{}%disable hyper at arxiv
{}%use pdflatex at arxiv

\documentclass[12pt,a4paper]{article}

\newif\ifpublic\publictrue

\usepackage{amsmath,amssymb}
\ifpublic\else\usepackage{showkeys}\fi
\usepackage[bookmarks=true,hyperfigures=true]{hyperref}
\usepackage{graphicx}
\usepackage[nosort]{cite}
\usepackage[bulletsep]{collref}
\usepackage{multirow}

\newcommand{\foreign}[1]{\textit{#1}}

\def\showkeysrefformat#1{{\normalfont\tiny\ttfamily#1}}
\makeatletter
\def\SK@@ref#1>#2\SK@{%
 {\@inlabelfalse\leavevmode\vbox to\z@{%
 \vss\SK@refcolor\rlap{\vrule\raise .75em%
  \hbox{\showkeysrefformat{#2}}}}}}
\makeatother

\usepackage[a4paper,text={160mm,247mm},centering]{geometry}
\allowdisplaybreaks[3]

\numberwithin{equation}{section}

\usepackage[font=small,labelfont=bf,width=0.85\textwidth]{caption}

\expandafter\def\expandafter\bfseries\expandafter{\bfseries\ifmmode\else\boldmath\fi}
\expandafter\def\expandafter\mdseries\expandafter{\mdseries\ifmmode\else\unboldmath\fi}
\expandafter\def\expandafter\normalfont\expandafter{\normalfont\ifmmode\else\unboldmath\fi}

\RequirePackage{verbatim}

\makeatletter
\newwrite\bibinl@out
\newenvironment{bibtex}[1][\jobname]{%
  \immediate\openout\bibinl@out #1.bib
  \immediate\write\bibinl@out{\@percentchar generated from `\jobname' starting line \the\inputlineno^^J}%
  \def\verbatim@processline{\immediate\write\bibinl@out{\the\verbatim@line}}%
  \@bsphack\let\do\@makeother\dospecials\catcode`\^^M\active\verbatim@start
}%
{\immediate\closeout\bibinl@out\@esphack}
\makeatother

\usepackage{fixmath}

\newcommand{\sfrac}[2]{{\textstyle\frac{#1}{#2}}}
\newcommand{\half}{\sfrac{1}{2}}

\newcommand{\Complex}{\mathbb{C}}
\newcommand{\Integer}{\mathbb{Z}}
\newcommand{\Natural}{\mathbb{N}}

\newcommand{\id}{\mathrm{id}}
\newcommand{\hopf}[1]{\mathrm{#1}}
\newcommand{\env}{\hopf{U}}
\newcommand{\double}{\hopf{D}}
\newcommand{\alg}[1]{\mathfrak{#1}}

\newcommand{\pro}{\mu}
\newcommand{\copro}{\mathrm{\Delta}}
\newcommand{\cop}{\text{cop}}

\newcommand{\aut}{{\text{A}}}
\newcommand{\exta}{\tilde{a}}

\newcommand{\counit}{\epsilon}
\newcommand{\antipode}{\mathrm{S}}

\newcommand{\rmat}{\mathcal{R}}

\newcommand{\asymL}{\tilde{L}}

\newcommand{\charfn}{\theta}

\RequirePackage{amsmath}
\makeatletter
\newcommand{\brk@ord}{\bBigg@{0}}
\newcommand{\brk@ordl}{\mathopen\brk@ord}
\newcommand{\brk@ordr}{\mathclose\brk@ord}
\newcommand{\brk@ordm}{\mathrel\brk@ord}
\newcommand{\brk@var}{\brk@ord}
\newcommand{\brk@varl}{\left}
\newcommand{\brk@varr}{\right}
\newcommand{\brk@varm}{\mathrel\brk@var}
\newcommand{\brk@altname}[3]{\expandafter\def\csname#2\expandafter\@gobble\string#1\endcsname{#1[#3]}}
\newcommand{\brk@usearg}[3]{%
  \def\brk@star{*}\def\brk@blank{}\def\brk@arg{#1}%
  \ifx\brk@arg\brk@blank\def\brk@arg{brk@ord}\fi%
  \ifx\brk@arg\brk@star\def\brk@arg{brk@var}\fi%
  \csname\brk@arg #2\endcsname#3}

\newcommand{\DeclareMathBrackets}[3]{
  \newcommand{#1}[2][]{\brk@usearg{##1}{l}{#2}##2\brk@usearg{##1}{r}{#3}}
  \brk@altname{#1}{big}{big}\brk@altname{#1}{lr}{*}}
\newcommand{\DeclareMathBiBrackets}[4]{
  \newcommand{#1}[3][]{\brk@usearg{##1}{l}{#2}##2#3##3\brk@usearg{##1}{r}{#4}}
  \brk@altname{#1}{big}{big}\brk@altname{#1}{lr}{*}}
\newcommand{\DeclareMathBiMBracketsStar}[4]{
  \newcommand{#1}[3][]{\brk@usearg{##1}{l}{#2}##2\brk@usearg{##1}{m}{#3}##3\brk@usearg{##1}{r}{#4}}
  \brk@altname{#1}{big}{big}}
\newcommand{\DeclareMathBiBracketsStar}[4]{
  \newcommand{#1}[3][]{\brk@usearg{##1}{l}{#2}##2\brk@usearg{##1}{}{#3}##3\brk@usearg{##1}{r}{#4}}
  \brk@altname{#1}{big}{big}}

\makeatother

\DeclareMathBrackets{\brk}{(}{)}
\DeclareMathBrackets{\sbrk}{[}{]}
\DeclareMathBrackets{\set}{\{}{\}}
\DeclareMathBiBracketsStar{\rset}{\{}{|}{\}}
\DeclareMathBrackets{\abs}{|}{|}
\DeclareMathBrackets{\eval}{.}{|}
\DeclareMathBrackets{\spn}{\langle}{\rangle}
\DeclareMathBiBrackets{\comm}{[}{,}{]}
\DeclareMathBiBrackets{\pair}{\langle}{,}{\rangle}
\DeclareMathBrackets{\ideal}{\langle}{\rangle}

\DeclareMathOperator{\Li}{Li}
\DeclareMathOperator{\ad}{ad}

\newcommand{\order}{\mathcal{O}}

\newcommand{\nln}{\nonumber\\}

\def\[{\begin{equation}}
\def\]{\end{equation}}

\providecommand{\href}[2]{#2}

\makeatletter
\def\mr@ignsp#1 {\ifx\:#1\@empty\else #1\expandafter\mr@ignsp\fi}%
\newcommand{\multiref}[1]{\begingroup%\let\protect\string%
\xdef\mr@no@sparg{\expandafter\mr@ignsp#1 \: }%
\def\mr@comma{}%
\@for\mr@refs:=\mr@no@sparg\do{\mr@comma\def\mr@comma{,}\ref{\mr@refs}}%
\endgroup}
\renewcommand{\eqref}[1]{(\multiref{#1})}
\makeatother

\makeatletter
\newcommand{\namedref}[2]{\hyperref[#2]{#1~\ref*{#2}}}
\newcommand{\secref}{\@ifstar{\namedref{Section}}{\namedref{Sec.}}}
\newcommand{\appref}{\@ifstar{\namedref{Appendix}}{\namedref{App.}}}
\newcommand{\tabref}{\@ifstar{\namedref{Table}}{\namedref{Tab.}}}
\newcommand{\figref}{\@ifstar{\namedref{Figure}}{\namedref{Fig.}}}
\makeatother

\usepackage{amsthm}
\newtheorem{prop}{Proposition}[section]
\newtheorem{lemma}{Lemma}[section]
\let\oldbib=\thebibliography
\def\thebibliography{\phantomsection\addcontentsline{toc}{section}{\refname}\oldbib}

\let\oldtoc=\tableofcontents
\def\tableofcontents{\phantomsection\addcontentsline{toc}{section}{\contentsname}\oldtoc}

\RequirePackage{verbatim}

\makeatletter
\newwrite\mpi@out
\def\mpi@write#1{\immediate\write\mpi@out{#1}}
\immediate\openout\mpi@out\jobname.mp
\mpi@write{\@percentchar generated from `\jobname'}%
\AtEndDocument{\mpostdone}

\def\mpostdone{
  \immediate\closeout\mpi@out%
  \ifpublic\else%
    \immediate\write18{mpost -tex=latex \jobname.mp}
  \fi%
  \gdef\mpostdone{}
}

\newcommand{\mpi@putlineno}{%
  \mpi@write{\@percentchar---------------------------------------}%
  \mpi@write{\@percentchar l.\the\inputlineno}%
}

\newcommand{\mpi@verbatim}{
  \@bsphack
  \let\do\@makeother\dospecials
  \catcode`\^^M\active
  \def\verbatim@processline{\mpi@write{\the\verbatim@line}}%
  \verbatim@start
}

\newenvironment{mpostcmd}{%
  \mpi@putlineno%
  \mpi@verbatim%
}%
{\mpi@write{}\@esphack}

\newenvironment{mpostfile}[1]{%
  \mpi@putlineno%
  \mpi@write{filenametemplate "#1";}%
  \mpi@write{beginfig(0)}%
  \mpi@verbatim%
}%
{\mpi@write{endfig;}\@esphack}

\newcommand{\includegraphicsex}[2][]{%
  \xdef\mpi@tmp{#2}%
  \IfFileExists{\mpi@tmp}%
    {\includegraphics[#1]{\mpi@tmp}}%
    {\textbf{??}\typeout{file \mpi@tmp{} missing}}%
}

\makeatother

\makeatletter
\newsavebox{\apb@box}\newlength{\apb@width}
\newcommand{\autoparbox}[2][c]{\sbox{\apb@box}{#2}%
 \settowidth{\apb@width}{\usebox{\apb@box}}%
 \parbox[#1]{\apb@width}{\usebox{\apb@box}}}
\newcommand{\includegraphicsbox}[2][]{\autoparbox{\includegraphicsex[#1]{#2}}}
\makeatother

\providecommand{\hypersetup}[1]{}
\providecommand{\texorpdfstring}[2]{#1}

\hypersetup{plainpages=false}
\hypersetup{pdfpagemode=UseNone}
\hypersetup{bookmarksnumbered=true}
\hypersetup{pdfstartview=FitH}
\hypersetup{colorlinks=false}
\hypersetup{citebordercolor={.5 1 .5}}
\hypersetup{urlbordercolor={.5 1 1}}
\hypersetup{linkbordercolor={1 .7 .7}}
\makeatletter
\let\@keywords\@empty
\let\@subject\@empty
\providecommand{\keywords}[1]{\gdef\@keywords{#1}}
\providecommand{\subject}[1]{\gdef\@subject{#1}}
\def\thetitle{\@title}
\def\theauthor{\@author}
\def\thesubject{\@subject}
\def\thedate{\@date}
\def\thekeywords{\@keywords}
\makeatother
\AtBeginDocument{
\hypersetup{pdftitle={\thetitle}}%
\hypersetup{pdfauthor={\theauthor}}%
\hypersetup{pdfsubject={\thesubject}}%
\hypersetup{pdfkeywords={\thekeywords}}%
}

\newif\ifshownote
\shownotefalse

\ifpdf\else\RequirePackage[active]{srcltx}\fi
\RequirePackage{color}

\newcommand{\remark}[2][]{{\normalfont\sffamily\hspace{1ex}%
  \def\tmparga{#1}\def\tmpargb{}\ifx\tmparga\tmpargb\else \textbf{#1:} \fi%
  #2\hspace{1ex}}}

\ifshownote\else\renewcommand{\remark}[2][]{}\fi

\begin{mpostcmd}
prologues:=2;

verbatimtex
\documentclass{article}
\usepackage{amsfonts,amssymb,latexsym,amsmath}
\newcommand{\alg}{\mathfrak}
\newcommand{\aut}{{\text{A}}}
\newcommand{\Complex}{\mathbb{C}}
\begin{document}
etex
\end{mpostcmd}

\begin{mpostcmd}
picture copyrightline,copyleftline;
copyrightline := btex \copyright\ \textsf{2015 Niklas Beisert} etex;
copyleftline := btex $\circledast$ etex;
def putcopyspace =
label.bot(btex \vphantom{gA} etex scaled 0.1, lrcorner(currentpicture));
enddef;
def putcopy =
label.ulft(copyrightline scaled 0.1, lrcorner(currentpicture)) withcolor 0.9white;
label.urt(copyleftline scaled 0.1, llcorner(currentpicture)) withcolor 0.9white;
currentpicture:=currentpicture shifted (10.5cm,14cm);
enddef;
\end{mpostcmd}

\begin{mpostcmd}
def pensize(expr s)=withpen pencircle scaled s enddef;
def fillshape(expr p,ci,tb,cb)=
  fill p withcolor ci;
  draw p pensize(tb) withcolor cb;
enddef;
def filldot(expr z,s,ci)=
  fillshape(fullcircle scaled s shifted z, ci, 0.5pt, 0.0white);
enddef;
def filldotcross(expr z,s,ci)=
  fillshape(fullcircle scaled s shifted z, ci, 0.5pt, 0.0white);
  draw (point 1 of fullcircle scaled s shifted z)--(point 5 of fullcircle scaled s shifted z) pensize(0.5pt) withcolor 0.0white;
  draw (point 3 of fullcircle scaled s shifted z)--(point 7 of fullcircle scaled s shifted z) pensize(0.5pt) withcolor 0.0white;
enddef;
xu:=1cm;
pair pos[];
path paths[];
def midarrow (expr p, t) =
fill arrowhead subpath(0,arctime(arclength(subpath (0,t) of p)+0.5ahlength) of p) of p;
enddef;
\end{mpostcmd}

\begin{mpostfile}{FigGenerators.mps}
pos[1]:=(2xu,0);
pos[2]:=(0,1.5xu);
pos[3]:=(0,0.8xu) slanted 0.5;
fill (2.5pos[1]-(0,0.8xu))--(-2.5pos[1]+(0,0.8xu))--(-2.5pos[1]-2.5pos[2])--(+2.5pos[1]-2.5pos[2])--cycle withcolor 0.95white;
draw (-2pos[1])--(+2pos[1]) pensize(0.3pt) dashed evenly;
draw (-2pos[3])--(+2pos[3]) pensize(0.3pt) dashed evenly;
draw (-2pos[2])--(+2pos[2]) pensize(0.3pt) dashed evenly;
draw (-pos[1]+pos[2]-pos[3])--(+pos[1]+pos[2]-pos[3])--(+pos[1]+pos[2]+pos[3])--(-pos[1]+pos[2]+pos[3])--cycle 
      pensize(0.3pt) dashed evenly;
draw (-pos[1]-pos[2]-pos[3])--(+pos[1]-pos[2]-pos[3])--(+pos[1]-pos[2]+pos[3])--(-pos[1]-pos[2]+pos[3])--cycle 
      pensize(0.3pt) dashed evenly;
draw (-pos[1]-pos[2]-pos[3])--(-pos[1]+pos[2]-pos[3]) pensize(0.3pt) dashed evenly;
draw (+pos[1]-pos[2]-pos[3])--(+pos[1]+pos[2]-pos[3]) pensize(0.3pt) dashed evenly;
draw (-pos[1]-pos[2]+pos[3])--(-pos[1]+pos[2]+pos[3]) pensize(0.3pt) dashed evenly;
draw (+pos[1]-pos[2]+pos[3])--(+pos[1]+pos[2]+pos[3]) pensize(0.3pt) dashed evenly;
filldot (+2pos[1], 7.5pt, white); label.rt(btex $E_1$ etex, +2pos[1]+(3pt,0));
filldot (-2pos[1], 7.5pt, 0.95white); label.lft(btex $F_1$ etex, -2pos[1]-(3pt,0));
filldot (+2pos[3], 7.5pt, white); label.top(btex $E_3$ etex, +2pos[3]+(0,3pt));
filldot (-2pos[3], 7.5pt, 0.95white); label.bot(btex $F_3$ etex, -2pos[3]-(0,3pt));
filldotcross (-pos[1]+pos[2]-pos[3], 7.5pt, white); label.ulft(btex $E_2$ etex, -pos[1]+pos[2]-pos[3]+(-2pt,2pt));
filldotcross (+pos[1]+pos[2]-pos[3], 5pt, white); label.ulft(btex $E_{12}$ etex, +pos[1]+pos[2]-pos[3]);
filldotcross (-pos[1]+pos[2]+pos[3], 5pt, white); label.ulft(btex $E_{32}$ etex, -pos[1]+pos[2]+pos[3]);
filldotcross (+pos[1]+pos[2]+pos[3], 5pt, white); label.ulft(btex $E_{132}$ etex, +pos[1]+pos[2]+pos[3]);
filldotcross (-pos[1]-pos[2]-pos[3], 5pt, 0.95white); label.lrt(btex $F_{213}$ etex, -pos[1]-pos[2]-pos[3]);
filldotcross (+pos[1]-pos[2]-pos[3], 5pt, 0.95white); label.lrt(btex $F_{23}$ etex, +pos[1]-pos[2]-pos[3]);
filldotcross (-pos[1]-pos[2]+pos[3], 5pt, 0.95white); label.lrt(btex $F_{21}$ etex, -pos[1]-pos[2]+pos[3]);
filldotcross (+pos[1]-pos[2]+pos[3], 7.5pt, 0.95white); label.lrt(btex $F_2$ etex, +pos[1]-pos[2]+pos[3]-(-2pt,2pt));
filldot (+2pos[2]+(-5pt,0), 7.5pt, 0.75white); label.ulft(btex $L$ etex, +2pos[2]+(-7pt,0));
filldot (+2pos[2]+(+3pt,0), 5pt, 0.75white); label.urt(btex $P$ etex, +2pos[2]+(+3pt,0));
filldot (-2pos[2]+(-5pt,0), 7.5pt, 0.75white); label.llft(btex $M$ etex, -2pos[2]+(-7pt,0));
filldot (-2pos[2]+(+3pt,0), 5pt, 0.75white); label.lrt(btex $K$ etex, -2pos[2]+(+3pt,0));
filldot ((+3pt,+3pt), 5pt, white); 
filldot ((+3pt,-3pt), 5pt, 0.75white); 
filldot ((-3pt,+3pt), 5pt, white); 
filldot ((-3pt,-3pt), 5pt, 0.75white); 
draw (2.5pos[1]-(0,0.8xu))--(-2.5pos[1]+(0,0.8xu)) dashed (withdots scaled 0.5) pensize(1pt);
label.ulft(btex $\alg{b}^+$ etex, (2.5pos[1]-(0,0.8xu)+(0,3pt)));
label.llft(btex $\alg{b}^-$ etex, (2.5pos[1]-(0,0.8xu)));
label.ulft(btex $H_1$ etex, (-3pt,1pt));
label.llft(btex $H_2$ etex, (-3pt,-1pt));
label.urt(btex $H_3$ etex, (3pt,1pt));
label.lrt(btex $H_\aut$ etex, (3pt,-1pt));
putcopyspace;putcopy;
\end{mpostfile}

\begin{mpostfile}{FigInclusions.mps}
pos[1]:=(4xu,0);
pos[2]:=(1xu,0);
pos[3]:=(0,-0.5xu);
pos[4]:=(0,-0.75xu);
pos[5]:=(0,-3.00xu);
label(btex $K_{}$ etex scaled 1.5, pos[1]+1pos[2]);
label(btex $P_{}$ etex scaled 1.5, pos[1]+0pos[2]);
label(btex $C_{}$ etex scaled 1.5, pos[1]-1pos[2]);
label(btex $E_k$ etex scaled 1.5, 0pos[1]+pos[2]);
label(btex $H_{\smash{1,3}}$ etex scaled 1.5, 0pos[1]+0pos[2]);
label(btex $F_k$ etex scaled 1.5, 0pos[1]-pos[2]);
label(btex $H_\aut$ etex scaled 1.5, -pos[1]+pos[2]);
label(btex $L_{}$ etex scaled 1.5, -pos[1]+0pos[2]);
label(btex $M_{}$ etex scaled 1.5, -pos[1]-pos[2]);

draw ((0pos[1]-1.25pos[2])--(0pos[1]+1.25pos[2])) shifted (-pos[4]-0pos[3]) pensize(5pt) withcolor 0.5white;
draw ((-1pos[1]+0.75pos[2])--(0pos[1]+1.25pos[2])) shifted (-pos[4]-1pos[3]) pensize(5pt) withcolor 0.5white;
draw ((0pos[1]-1.25pos[2])--(1pos[1]-0.75pos[2])) shifted (-pos[4]-2pos[3]) pensize(5pt) withcolor 0.5white;
draw ((-1pos[1]+0.75pos[2])--(1pos[1]-0.75pos[2])) shifted (-pos[4]-3pos[3]) pensize(5pt) withcolor 0.5white;
label.lft(btex $\alg{psl}(2|2)$ etex, -pos[4]-0pos[3]-pos[1]+0.5pos[2]);
label.lft(btex $\alg{pgl}(2|2)$ etex, -pos[4]-1pos[3]-pos[1]+0.5pos[2]);
label.lft(btex $\alg{sl}(2|2)$ etex, -pos[4]-2pos[3]-pos[1]+0.5pos[2]);
label.lft(btex $\alg{gl}(2|2)$ etex, -pos[4]-3pos[3]-pos[1]+0.5pos[2]);

draw ((0pos[1]-1.25pos[2])--(0pos[1]+1.25pos[2])) shifted (pos[4]+0pos[3]) pensize(5pt) withcolor 0.5white;
draw ((-1pos[1]-1.25pos[2])--(0pos[1]+1.25pos[2])) shifted (pos[4]+1pos[3]) pensize(5pt) withcolor 0.5white;
draw ((0pos[1]-1.25pos[2])--(1pos[1]+1.25pos[2])) shifted (pos[4]+2pos[3]) pensize(5pt) withcolor 0.5white;
draw ((-1pos[1]-1.25pos[2])--(1pos[1]+1.25pos[2])) shifted (pos[4]+3pos[3]) pensize(5pt) withcolor 0.5white;
label.lft(btex $\alg{psg}$ etex, +pos[4]-0pos[3]-0pos[1]-1.5pos[2]);
label.lft(btex $\alg{pg}$ etex, +pos[4]+1pos[3]-pos[1]-1.5pos[2]);
label.lft(btex $\alg{sg}$ etex, +pos[4]+2pos[3]-pos[1]-1.5pos[2]);
label.lft(btex $\alg{g}$ etex, +pos[4]+3pos[3]-pos[1]-1.5pos[2]);

draw ((1pos[1]-1.25pos[2])--(1pos[1]+1.25pos[2])) shifted (pos[4]+0pos[3]) pensize(5pt) withcolor 0.5white;
draw ((-1pos[1]-1.25pos[2])--(-1pos[1]+1.25pos[2])) shifted (pos[4]+0pos[3]) pensize(5pt) withcolor 0.5white;
label.lft(btex $\alg{sl}(2)$ etex, +pos[4]-0pos[3]-1pos[1]-1.5pos[2]);
label.lft(btex $\Complex^3$ etex, +pos[4]-0pos[3]+1pos[1]-1.5pos[2]);

label(btex $-$ etex,   +1pos[1]+1pos[2]+0.6pos[4]);
label(btex $+$ etex,   +1pos[1]+0pos[2]+0.6pos[4]);
label(btex $\pm$ etex, +1pos[1]-1pos[2]+0.6pos[4]);
label(btex $+$ etex,    0pos[1]+1pos[2]+0.6pos[4]);
label(btex $\pm$ etex,  0pos[1]+0pos[2]+0.6pos[4]);
label(btex $-$ etex,    0pos[1]-1pos[2]+0.6pos[4]);
label(btex $\pm$ etex, -1pos[1]+1pos[2]+0.6pos[4]);
label(btex $+$ etex,   -1pos[1]+0pos[2]+0.6pos[4]);
label(btex $-$ etex,   -1pos[1]-1pos[2]+0.6pos[4]);
label.rt(btex $\alg{b}^\pm$ etex, 1pos[1]+1.5pos[2]+0.6pos[4]);

putcopyspace;putcopy;
\end{mpostfile}

\begin{mpostcmd}
verbatimtex
\end{document}
etex

end;
\end{mpostcmd}

\mpostdone

%%%%%%%%%%%%%%%%%%%%%%%%%%%%%%%%%%%%%%%%%%%%%%%%%%%%%%%%%%%%%%%%%%%%%%%%%%%%%%%%
%%%%%%%%%%%%%%%%%%%%%%%%%%%%%%%%%%%%%%%%%%%%%%%%%%%%%%%%%%%%%%%%%%%%%%%%%%%%%%%%
\title{Maximally extended \texorpdfstring{$\alg{sl}(2|2)$}{sl(2|2)} as a quantum double}
\author{Niklas Beisert, Marius de Leeuw and Reimar Hecht}

%%%%%%%%%%%%%%%%%%%%%%%%%%%%%%%%%%%%%%%%%%%%%%%%%%%%%%%%%%%%%%%%%%%%%%%%%%%%%%%%
\begin{document}

\pdfbookmark[1]{Title Page}{title}
\thispagestyle{empty}

%\begingroup\raggedleft\footnotesize\ttfamily
%\arxivlink{1204.4406}
%\par\endgroup

\vspace*{2cm}
\begin{center}%
\begingroup\Large\bfseries\thetitle\par\endgroup
\vspace{1cm}

%\begingroup\scshape\theauthor\par\endgroup
%\vspace{5mm}%

\begingroup\scshape
Niklas Beisert${}^{1}$, Marius de Leeuw${}^{1,2}$ and
Reimar Hecht${}^{1}$
\endgroup
\vspace{5mm}

\textit{${}^{1}$ Institut f\"ur Theoretische Physik,\\
Eidgen\"ossische Technische Hochschule Z\"urich,\\
Wolfgang-Pauli-Strasse 27, 8093 Z\"urich, Switzerland}
\vspace{0.1cm}

\begingroup\ttfamily\small
\verb+{+nbeisert,hechtr\verb+}+@itp.phys.ethz.ch\par
\endgroup
\vspace{5mm}

\textit{${}^{2}$ Niels Bohr Institute, \\ Copenhagen University,\\
Blegdamsvej 17, 2100 Copenhagen \O, Denmark}\\[0.1cm]
\begingroup\ttfamily\small
deleeuwm@nbi.ku.dk\par
\endgroup
\vspace{5mm}

\vfill

\textbf{Abstract}\vspace{5mm}

\begin{minipage}{12.7cm}
We derive the universal R-matrix of the quantum-deformed enveloping algebra 
of centrally extended $\alg{sl}(2|2)$ using Drinfeld's quantum
double construction. 
We are led to enlarging the algebra by additional generators 
corresponding to an $\alg{sl}(2)$ automorphism. For this maximally extended algebra 
we construct a consistent Hopf algebra structure 
where the extensions exhibit several uncommon features. 
We determine the corresponding universal R-matrix containing
some non-standard functions.
Curiously, this Hopf algebra has one extra deformation parameter
for which the R-matrix does not factorize into products of exponentials.
\end{minipage}

\vspace*{4cm}

\end{center}

\newpage

%%%%%%%%%%%%%%%%%%%%%%%%%%%%%%%%%%%%%%%%%%%%%%%%%%%%%%%%%%%%%%%%%%%%%%%%%%%%%%%%
%%%%%%%%%%%%%%%%%%%%%%%%%%%%%%%%%%%%%%%%%%%%%%%%%%%%%%%%%%%%%%%%%%%%%%%%%%%%%%%%

\section{Introduction}

Integrable models are usually characterized by an invertible finite-dimensional solution $
R:\mathbb{C}^n\otimes \mathbb{C}^n \rightarrow \mathbb{C}^n\otimes \mathbb{C}^n$ of the 
so-called Yang--Baxter equation
\begin{align}
R_{12}R_{13}R_{23}
=
R_{23}R_{13}R_{12}.
\end{align}
For instance, the R-matrix corresponds to the scattering matrix in integrable field theories. In
the language of the algebraic Bethe Ansatz, the R-matrix describes the symmetry algebra
that underlies the integrable model. It also parameterizes the Hamiltonian. Alternatively, knowing 
the full symmetry algebra of the model usually allows one to derive the R-matrix. 

The intimate relation between algebra and R-matrices is made manifest in quasi-triangular
Hopf algebras. These Hopf algebras contain an operator $\rmat$, called the \textit{universal} R-matrix, 
which is an invertible operator that intertwines the Hopf algebra structure and its opposite counterpart.
One can show that the universal R-matrix satisfies the Yang--Baxter equation
\begin{align}
\rmat_{12}\rmat_{13}\rmat_{23}
=
\rmat_{23}\rmat_{13}\rmat_{12} .
\end{align}
In particular, R-matrices whose symmetry algebra is a quasi-triangular 
Hopf algebra can then be obtained by evaluating the universal R-matrix in the corresponding representation,
$R = (\rho\otimes \rho')(\rmat)$. 
For a large class of solutions of the former Yang--Baxter equation,
the associated quasi-triangular Hopf algebra is known and can be formulated very explicitly.
Prominent examples are q-deformed, quantum affine and Yangian algebras
based on simple Lie algebras and superalgebras.
However, there exist several peculiar R-matrices for which 
the question of the underlying algebra remains obscure.
In particular, despite many efforts, the algebraic structure 
that governs the R-matrix of the one-dimensional Hubbard model 
and of the AdS/CFT integrable system is only known to some extent.
In this paper we take 
some first steps towards understanding the universal structure of the R-matrix of 
these integrable systems.

\medskip

The R-matrix that underlies the integrability of the Hubbard model \cite{Hubbard}, 
see also \cite{HubbBook}, was found by Shastry \cite{Shastry} without knowledge of its algebraic origins.
Only parts of the underlying algebra were known. For instance, it is well known that the Hubbard model 
exhibits two $\alg{sl}(2)$  algebras,  that are associated with spin and charge. These algebras can even be 
extended to a full Yangian symmetry \cite{Uglov:1993jy}, but these symmetries are not 
sufficient to determine the R-matrix.

Later, the algebraic structure of the Hubbard model was elucidated by using 
input from a rather different area of theoretical physics.
It turned out that the Hubbard model has a remarkable relation to string
and gauge theory via the AdS/CFT correspondence.
The prime example of the gauge/string correspondence ---
the duality between $\mathcal{N}=4$ SYM and superstrings on $\mathrm{AdS}_5\times \mathrm{S}^5$ ---
proved to be an integrable system.
Moreover, the R-matrix that describes this system was found
to consist of two copies of the Hubbard model R-matrix \cite{Martins:2007hb,BAna}.

From string and gauge theory considerations it then became
clear that Shastry's R-matrix actually exhibits supersymmetry.
More precisely, there is an unusual Lie superalgebra underlying \cite{Bsu22,AFP}
the AdS/CFT R-matrix and hence also the Hubbard model R-matrix exhibits this symmetry algebra.
This Lie algebra is centrally extended $\alg{psl}(2|2)$
and the symmetry algebra of the R-matrix is given by a novel type of 
Yangian algebra \cite{BYang,dLYang,BdLHubbAlgebra} 
corresponding to this Lie superalgebra.

\medskip

The next question that arises is whether a universal R-matrix exists from which
the Hubbard model R-matrix can be derived. Answering this question is important for our 
understanding of the Hubbard and AdS/CFT integrable models. In particular, it would 
indicate whether the Hopf algebra is  quasi-triangular. Moreover, the unusual nature of 
the algebra might lead to some new algebraic structures that arise  in the construction of the 
universal R-matrix. A positive answer to the question of the existence of the R-matrix 
would potentially have important implications. For example, it should provide a proof
of the BES conjecture \cite{Beisert:2006ez}. Moreover, the universal R-matrix can be 
used to compute correlation functions \cite{LeClair:1991cf}, which will help solving the 
Hubbard model and the AdS/CFT integrable model. 

Hints of a universal algebraic structure can be found at the classical level \cite{TClassical,MTClassical,BSclassical}.
At the classical level, the R-matrix reduces to a classical r-matrix
that satisfies the classical Yang--Baxter equation.
In fact, it was shown that the universal classical r-matrix put forward in \cite{BSclassical}
indeed correctly describes the classical limit of the scattering matrices appearing
in the $\mathrm{AdS}_5\times \mathrm{S}^5$ superstring \cite{MdLClassicalR,MdLThesis}.
Remarkably, it was even found to describe the R-matrix to quadratic order \cite{MdLThesis}.
Furthermore, the R-matrices from these models already contain universal sub-structures \cite{AdLTuniversal}.
Nevertheless, despite all these indications a universal R-matrix has never been found.

\medskip

In the theory of quantum algebras there is a standard way to generate universal R-matrices,
which goes under the name of a quantum double  \cite{DQauntumGroups}, see also \cite{Tdouble}.
The idea behind it is to construct from a given Hopf algebra $\hopf{H}$ and its `dual' Hopf algebra $\hopf{H}^*$, 
a quasi-triangular Hopf algebra 
--- the quantum double $\double \hopf{H}$ ---
whose R-matrix is simply given by the sum over a pair of dual bases $\rmat = \sum_i e_i \otimes e_i^*$.
For a given R-matrix, whose symmetry algebra is known,
one can then endeavor to write this symmetry algebra as a quantum double or embed it into one.
In other words, we would like to construct the smallest Hopf algebra
that can be written as a double that contains the symmetry algebra of the Hubbard model.

In the present paper we will consider, as a starting point, the centrally extended $\alg{psl}(2|2)$ Lie superalgebra,
which is finite-dimensional, rather than the corresponding infinite-dimensional Yangian algebra. 
Not only is this a logical first step to take, but the R-matrix of the Hubbard model in the fundamental
representation is actually fixed by the finite-dimensional algebra \cite{Bsu22,AFP}. In other words, we 
might already gain insight into the structure of the Hubbard model R-matrix by restricting to this case. 

However, in order to get a non-trivial quantum algebra to which we can apply the quantum double
construction, we need to q-deform the algebra. In \cite{BK} this algebra, denoted by 
$\env_{q}(\alg{psl}(2|2)\ltimes \Complex^3)$, was
defined by considering the quantum deformation of the universal  enveloping algebra of 
$\alg{psl}(2|2)\ltimes \Complex^3$. Analogously to the presence of Yangian symmetry for 
the undeformed case, the symmetry algebra can be enlarged 
by an affine extension \cite{BGMaffine}. Also in the deformed case, the classical limit exhibits universal structures 
 \cite{BQclassical}. The undeformed model can be recovered by taking the rational limit $q\rightarrow 1$. 

In this paper we successfully construct the smallest double algebra that contains $\env_{q}(\alg{psl}(2|2)\ltimes \Complex^3)$.
To this end we need to introduce three additional boost operators that are dual to the central extensions. They 
form an $\alg{sl}(2)$ algebra. We find that the total algebra,
which we shall call the \emph{maximal extension} of (quantum-deformed) $\alg{psl}(2|2)$,
\begin{align}
\env_{q,\kappa}\bigbrk{\alg{sl}(2) \ltimes \alg{psl}(2|2)\ltimes \Complex^3}
\end{align}
is of a novel type.
It depends on an additional parameter $\kappa$ and has some unusual features. 
For instance, its commutation relations depend (polynomially) not only on $q$
but also on $\hbar := \log q$. We will derive all algebra and coalgebra relations 
that define this Hopf algebra. Moreover, we explicitly work out its universal R-matrix.

\medskip
%next: outline

This paper is organized as follows.
In \secref{sec:GenDouble} we will discuss some
general features of quantum deformed algebras as well as the construction of the
quantum double.
After this we work out the example of $\alg{sl}(3)$ in detail.
We then turn to the algebra of interest:
the quantum-deformed maximally extended $\alg{sl}(2|2)$ algebra.
First we summarize the algebra relations in \secref{sec:overview22}.
We then carry out the construction of a quantum double in \secref{sec:double}, 
which leads to a natural extension of the algebra. 
Subsequently in \secref{sec:Rmat} we derive the universal R-matrix for this extended algebra.
Finally, we perform the classical limit in \secref{sec:classical}.
The details of our computation are presented in the appendix.

%%%%%%%%%%%%%%%%%%%%%%%%%%%%%%%%%%%%%%%%%%%%%%%%%%%%%%%%%%%%%%%%%%%%%%%%%%%%%%%%
%%%%%%%%%%%%%%%%%%%%%%%%%%%%%%%%%%%%%%%%%%%%%%%%%%%%%%%%%%%%%%%%%%%%%%%%%%%%%%%%

\section{Hopf algebras as a quantum double}
\label{sec:GenDouble}

In this section, we briefly introduce the notion of a quantum double and quantum enveloping algebras.
We will work with superalgebras and to this end we introduce the corresponding $\Integer_2$ grading.
The degree of a generator $a$ is denoted by
\begin{align}
\left|a\right|:=\begin{cases}
0, & a\text{ is even},\\
1, & a\text{ is odd}.
\end{cases}
\end{align}
We furthermore use the graded tensor product
\begin{align}
(a \otimes b)(c \otimes d) = (-1)^{|b||c|}\, ac\otimes bd
\end{align}
and all commutators are to be understood in the graded sense, \foreign{i.e.}\ 
\[
[a,b] := ab -(-1)^{|a||b|} ba.
\]

%%%%%%%%%%%%%%%%%%%%%%%%%%%%%%%%%%%%%%%%%%%%%%%%%%%%%%%%%%%%%%%%%%%%%%%%%%%%%%%%
\subsection{The quantum double}

In the following, we will develop the general framework underlying
the construction of a quantum double.

%%%%%%%%%%%%%%%%%%%%%%%%%%%%%%%%%%%%%%%%
\paragraph{Hopf algebras.}

A Hopf algebra is a unital associative algebra $(\hopf{H},\pro,1)$
together with linear maps $\copro$, $\counit$ and $\antipode$,
called coproduct, counit and antipode,
\begin{align}
  \copro :& \; \hopf{H} \rightarrow \hopf{H} \otimes \hopf{H}, &\counit :&
\; \hopf{H} \rightarrow \Complex, &\antipode :& \; \hopf{H} \rightarrow \hopf{H},
\end{align}
that satisfy for all $a \in \hopf{H} $
\begin{align}
  (\copro \otimes \id)\circ \copro(a) &= (\id\otimes \copro) \circ \copro(a),
\\
  (\counit \otimes \id) \circ \copro(a) &\cong (\id \otimes \counit)\circ \copro(a) \cong a \label{eq:counit},
\\
  \pro \circ (\antipode \otimes \id) \circ \copro(a) &=\pro \circ (\id \otimes \antipode) \circ \copro(a)
= \counit(a) 1,\label{eq:antipode}
\end{align}
where the symbol $\cong$ in \eqref{eq:counit} denotes the canonical 
isomorphisms between $\hopf{H}$ and $\Complex \otimes \hopf{H}$ and $\hopf{H} 
\otimes \Complex$.
Furthermore algebra and coalgebra need to satisfy the compatibility relations for any $a,b \in \hopf{H} $
\begin{align}
  \copro(ab) &= \copro(a) \copro(b), &\counit (ab) &= \counit (a) \counit(b).
\end{align}
It is often convenient to write the coproduct using the Sweedler notation
\[
\copro(a)
= a_{(1)} \otimes a_{(2)}
:= \sum\nolimits_k a_{(1),k} \otimes a_{(2),k}.
\]
Here $(a_{(1),k},a_{(2),k})$ form a collection of pairs of elements
describing the coproduct of $a\in\hopf{H}$.
Usually we shall drop the sum and use the abbreviated middle form.

Within a Hopf algebra it is useful to define two bilinear compositions
called the left and right adjoint actions
\begin{align}
\label{eq:adjointaction}
a \triangleright b &:= (-1)^{|b||a_{(2)}|} a_{(1)} \,b\, \antipode(a_{(2)})
,&
b \triangleleft a &:= (-1)^{|a_{(1)}||b|} \antipode(a_{(1)})\,b\,a_{(2)}
.
\end{align}
These actions provide generalizations of conjugation and the commutator in the q-\hspace{0pt}deformed case,
as will be seen later. 
Note that $1\triangleleft a=a\triangleright 1=\counit(a)1$ is the Hopf algebra relation \eqref{eq:antipode}
and that the action obeys the composition rule $a\triangleright (b\triangleright c)=(ab)\triangleright c$.

Provided that the antipode of a Hopf algebra $\hopf{H}$ is invertible,
one can define another Hopf algebra $\hopf{H}^{\cop}$
with the opposite coproduct $\copro^{\cop} := \tau \circ \copro $
and antipode $\antipode^{\cop}:= \antipode^{-1}$.
Here, $\tau$ is the (graded) permutation map $\tau(a\otimes b)=(-1)^{|a||b|}b\otimes a $.

%%%%%%%%%%%%%%%%%%%%%%%%%%%%%%%%%%%%%%%%
\paragraph{Quasi-triangular Hopf algebras.}

Integrable systems are closely related to quasi-triangular Hopf algebras.
These algebras constitute a special class of Hopf algebras for which
the coproduct and opposite coproduct are related by a similarity transformation.

More precisely, a quasi-triangular Hopf algebra $(\hopf{H},\rmat)$
is a Hopf algebra $\hopf{H}$ together with an invertible element $\rmat\in \hopf{H} \otimes \hopf{H}$,
called the universal R-matrix.
It relates the coproduct and the opposite coproduct 
for any $a\in \hopf{H}$ in the following way
\[\label{eq:intertwine}
\copro^{\cop}(a) \rmat=\rmat \copro(a),
\]
and furthermore has to satisfy the so-called fusion relations
\begin{align}
\label{eq:fusion}
(\copro \otimes \id)\rmat &= \rmat_{13}\rmat_{23}
,&
(\id\otimes \copro)\rmat &= \rmat_{13}\rmat_{12}
.
\end{align}
If we write $\rmat=\sum  \rmat^{(1)} \otimes \rmat^{(2)}$, then $\rmat_{ij}$
is the element of $\hopf{H} \otimes \hopf{H} \otimes \hopf{H}$ with $\rmat^{(1)}$
in the $i$-th factor of the tensor product, $\rmat^{(2)}$ in the $j$-th factor,
and $1$ in the remaining factor.

The above axioms directly imply the Yang--Baxter equation
\begin{align}\label{eq:YBE}
  \rmat_{12}\rmat_{13}\rmat_{23} &=
  \rmat_{23}\rmat_{13}\rmat_{12},
\end{align}
which is the key relation in the theory of integrable systems.
The universal R-matrix describes the scattering in an integrable model
from an algebraic point of view.

%%%%%%%%%%%%%%%%%%%%%%%%%%%%%%%%%%%%%%%%
\paragraph{Dual Hopf algebra.}

\label{sec:dual}
We call a Hopf algebra $\hopf{H}^*$ the dual
\unskip\footnote{Note that, as a vector space, this definition agrees in the 
finite-dimensional case with the usual definition of the dual space 
as the space of all linear maps $\hopf{H}\rightarrow \Complex$.
However, in the infinite-dimensional case the dual space of our definition
will only be a subspace of the actual algebraic dual space.}
of a Hopf algebra $\hopf{H}$,
if there exists a non-degenerate bilinear pairing
$\pair{\cdot}{\cdot} : \hopf{H}^* \otimes \hopf{H} \rightarrow \Complex $,
satisfying for all $f,g \in \hopf{H}^*$ and $a,b\in \hopf{H}$
\begin{align}\label{eq:pairing1}
  &\pair{fg}{a} = (-1)^{|a_{(1)}||g|} \pair{f}{a_{(1)}} \pair{g}{a_{(2)}}, &&
  \pair{f}{ab} = (-1)^{|a||f_{(2)}|} \pair{f_{(1)}}{a} \pair{f_{(2)}}{b}
\end{align}
and
\begin{align}\label{eq:pairing2}
& \pair{f}{1} = \counit(f),
&&\pair{1}{a} = \counit(a),
&& \pair{\antipode(f)}{a}=\pair{f}{\antipode(a)}.
\end{align}

Given a basis $\{e_i\}_{i \in I}$ of a Hopf algebra $\hopf{H}$ with $e_0=1$ 
and $\counit(e_i)=0$, for all $i \neq 0$,
we formally construct the dual Hopf algebra  $\hopf{H}^*$
and the pairing by defining the dual basis $\{ e_i^* \}_{i\in I}\in \hopf{H}^*$
such that $\pair{e_i^*}{e_j} = \delta_{ij}$. 
The dual Hopf structure is then found from the pairing relations \eqref{eq:pairing1}.
The product of two elements $f,g\in \hopf{H}^*$ can be expanded in the dual basis as
\[\label{eq:dualproduct}
fg=\sum_{i\in I} \pair{fg}{e_i} e_i^* 
=\sum_{i \in I} (-1)^{|(e_i)_{(1)}||g|} \pair{f}{(e_i)_{(1)}}\pair{g}{(e_i)_{(2)}} e_i^*,
\]
and the coproduct of an element $f\in \hopf{H}^*$
can be expanded in the dual tensor basis as
\[\label{eq:dualcoproduct}
\copro(f)=\sum_{i,j\in I} (-1)^{|e_i||e_j|} \pair{\copro(f)}{e_i \otimes e_j} e_i^* \otimes e_j^*
= \sum_{i,j \in I} (-1)^{|e_i||e_j|} \pair{f}{e_i e_j} e_i^* \otimes e_j^*.
\]
Notice that the (co)algebra structure of $\hopf{H}^*$
is completely fixed in terms of the (co)algebra structure of $\hopf{H}$.
Specifically, the coalgebra structure on $\hopf{H}$
determines the algebra structure on $\hopf{H}^*$ and \foreign{vice versa}. 
It already follows from the requirement $\counit (e_i)=\delta_{i0}$ that the dual of the unit $1^*$ is also the unit of the dual. 
We will omit the star on the unit for convenience $1^*=1$. 

%%%%%%%%%%%%%%%%%%%%%%%%%%%%%%%%%%%%%%%%
\paragraph{Quantum double.}

For any given Hopf algebra $\hopf{H}$ we can construct its quantum double $\double\hopf{H}$,
which is a Hopf algebra with a quasi-triangular structure.
It is generated by $\hopf{H}$ and $\hopf{H}^{*\cop}$ as Hopf sub-algebras 
\unskip\footnote{There exists also a version of the quantum double where instead 
of $\hopf{H}^{*\cop}$ the dual with the opposite product is used $\hopf{H}^{*op}$.}
and can be built on $\hopf{H} \otimes \hopf{H}^{*\cop}$ as a vector space.
We need to specify the algebra relations
that deal with elements from both $\hopf{H}$ and $\hopf{H}^{*\cop}$.
These so-called cross-relations are defined by
\begin{align}\label{eq:gencross}
\sum(-1)^{|x_{(1)}|\left(|f_{(1)}|+|f_{(2)}|\right)}x_{(1)}f_{(1)}\bigpair{f_{(2)}}{x_{(2)}}
& =\sum(-1)^{|f_{(1)}||f_{(2)}|}\bigpair{f_{(1)}}{x_{(1)}} f_{(2)}x_{(2)}
\end{align}
for $x\in \hopf{H}$, $f\in \hopf{H}^{*\cop}$.
Since the coproduct on the dual was transposed,
the pairing is now a skew pairing, \foreign{i.e.}\
the pairing relations \eqref{eq:pairing1} and \eqref{eq:pairing2} are replaced by
\[
\pair{f}{ab} =(-1)^{|a||f_{(1)}|}\pair{f_{(2)}}{a} \pair{f_{(1)}}{b},
\qquad \bigpair{\antipode^{-1}(f)}{a}=\bigpair{f}{\antipode(a)}.
\]
One of the virtues of the quantum double is
that there is an explicit formula for the universal R-matrix
\[\label{eq:R-matrix}
\rmat=\sum_{i\in I} e_i \otimes e_i^* \in \double\hopf{H}\otimes \double\hopf{H},
\]
where $\{e_i\}_{i\in I}\subset \hopf{H}$ and $\{e_i^*\}_{i\in I}\subset \hopf{H}^{*\cop}$ are dual bases.
The cross-relations \eqref{eq:gencross}
can in fact be found by the condition that the R-matrix
of this form has to satisfy \eqref{eq:intertwine}.

%%%%%%%%%%%%%%%%%%%%%%%%%%%%%%%%%%%%%%%%%%%%%%%%%%%%%%%%%%%%%%%%%%%%%%%%%%%%%%%%
\subsection{Quantum enveloping algebra}
\label{sec:enveloping algebra}

In this paper we will consider q-deformed universal enveloping algebras $\env_q(\alg{g})$
of a Lie (super)algebra $\alg{g}$.
The quantum enveloping algebra $\env_q(\alg{g})$
is the unital associative algebra
over the ring of formal power series $\Complex[[\hbar]]$,
where $q=e^\hbar$, freely generated by $1$ and the generators of $\alg{g}$
satisfying q-deformed commutation relations,
which we will define specifically later.
For simple Lie algebras $\alg{g}$, the quantum enveloping algebra $\env_q(\alg{g})$
is quasi-triangular.
The R-matrix can be obtained by writing $\env_q(\alg{g})$
as the quantum double of the positive Borel sub-algebra $\env_q(\alg{b}^+)$.
The positive and negative Borel sub-algebras $\alg{b}^{\pm}$
of a Lie algebra $\alg{g}$ are defined in terms of the positive
and negative root space $\alg{g}^{\pm}$ and the Cartan sub-algebra $\alg{h}$ as
\begin{align}
&\alg{b}^+ = \alg{g}^+ \oplus \alg{h},
&& \alg{b}^- = \alg{g}^- \oplus \alg{h}.
\end{align}
In order to relate the quantum double
$\double\env_q(\alg{b}^+)=\env_q(\alg{b}^+)\otimes \env_q(\alg{b}^+)^{*\cop}$ to $\env_q(\alg{g})$
we should identify
\[\label{eq:isob+*cop=b-}
\env_q(\alg{b}^+)^{*\cop}\cong \env_q(\alg{b}^-).
\]
In this way we obtain two copies of the Cartan algebra $\alg{h}$.
Taking this fact into account, we can write
\[\label{eq:isodouble}
\env_q (\alg{g})\cong \frac{\env_q (\alg{b}^+ )\otimes \env_q (\alg{b}^-)}{\ideal{H-\widehat{H}}}\,,
\]
where we have to quotient out by some ideal $\ideal{H-\widehat{H}}$
such that the Cartan generators $H$ and $\widehat{H}$ of both Borel halves are identified correctly.
The R-matrix is then given via the formula \eqref{eq:R-matrix}
with both copies of the Cartan generators identified.

We will go through this procedure for $\alg{g}=\alg{sl}(3)$
as a guideline and in order to illustrate the calculations.
Then we will focus on the actual algebra of interest $\alg{psl}(2|2) \ltimes \Complex^3$.
For this algebra \eqref{eq:isob+*cop=b-} does not hold true
and we will need to extend it in a consistent way
such that the extended algebra satisfies \eqref{eq:isob+*cop=b-}
and consequently can be written as a double.

%%%%%%%%%%%%%%%%%%%%%%%%%%%%%%%%%%%%%%%%%%%%%%%%%%%%%%%%%%%%%%%%%%%%%%%%%%%%%%%%

\subsection{\texorpdfstring{$\env_q(\alg{sl}(3))$}{Uq(sl(3))} as a quantum double}
\label{sec:sl3ex}

Next we will apply the techniques discussed above to
describe the dual structure of $\env_q(\alg{sl}(3))$.
We refer to \cite{Bsl3double} for additional details.
While this example is considerably
simpler than extended $\alg{psl}(2|2)$, it still exhibits most features
that we will encounter later on.
%We will use the derivation in this section as a guideline.

%%%%%%%%%%%%%%%%%%%%%%%%%%%%%%%%%%%%%%%%
\paragraph{Algebra.}

We begin by specifying the algebra structure of $\env_q(\alg{sl}(3))$.
The algebra is most conveniently defined in terms of Chevalley--Serre generators.
These are the positive and negative simple-root vectors $E_i$ and $F_i$
as well as the Cartan generators $H_i$, $i=1,2$.
The commutation relations among these are given by
\begin{align}\label{eq:DefAlg}
[H_i , E_j ]&= a_{ij} E_j, 
&
[H_i, F_j ] &=-a_{ij} F_j, 
&
[E_i , F_j ]=\delta_{ij} \frac{q^{H_i}-q^{-H_i}}{q-q^{-1}}\,,
\end{align}
where the Cartan matrix is
\begin{align}
a =
\begin{pmatrix}
2 & -1 \\
-1 & 2 \\
\end{pmatrix}.
\end{align}
In addition, the simple-root vectors need to satisfy the Serre relations ($i\neq j$)
\begin{align}\label{eq:SerreSl3}
E_i E_i E_j -(q+q^{-1}) E_i E_j E_i +E_j E_i E_i &=0, \\
F_i F_i F_j -(q+q^{-1}) F_i F_j F_i +F_j F_i F_i &=0.
\end{align}
%

%%%%%%%%%%%%%%%%%%%%%%%%%%%%%%%%%%%%%%%%
\paragraph{Coalgebra.}

The coproduct of the simple-root generators is defined as
\begin{align}
\label{eq:coproE0}
\copro E_i &=E_i \otimes 1 + q^{-H_i} \otimes E_i,
\\
\label{eq:coproF0}
\copro F_i &=F_i \otimes q^{H_i} + 1 \otimes F_i,
\\
\copro H_i &=H_i \otimes 1 + 1\otimes H_i.
\end{align}
The expressions for the counit and antipode can be easily derived from
the coproduct via their defining properties \eqref{eq:counit} and \eqref{eq:antipode}.

%%%%%%%%%%%%%%%%%%%%%%%%%%%%%%%%%%%%%%%%
\paragraph{Basis of the Borel sub-algebra.}

In order to deal with the cubic Serre relations \eqref{eq:SerreSl3}
and to define a Poincar\'e--Birkhoff--Witt basis, it is convenient
to define additional generators, corresponding to non-simple roots.
To that end we observe that the Serre relations can be expressed
in terms of the adjoint actions \eqref{eq:adjointaction} as
\begin{align}
&E_i \triangleright (E_i \triangleright E_j) = 0 =
F_j \triangleleft (F_i \triangleleft F_i), \qquad i\neq j.
\end{align}
It is therefore natural to define non-simple-root vectors
\begin{align}
E_{12} &:= E_1 \triangleright E_2 = E_1 E_2 - q E_2 E_1 , 
\\
F_{21} &:=  F_2 \triangleleft F_1 =  F_2 F_1 - q^{-1} F_1 F_2  .
\end{align}
The Serre relations are then expressed as
\unskip\footnote{The relations on the left hand side have a convenient
formulation in terms of the adjoint actions
$E_1\triangleright E_{12}=0=F_{12}\triangleleft F_1$,
but the ones on the right hand side do not.}
\begin{align}
E_1 E_{12}-q^{-1} E_{12} E_1&=0
,
& E_{12} E_2- q^{-1} E_2 E_{12} &=0
,\\
F_{12} F_1-q F_1 F_{12}&=0
,
& F_2 F_{12}- q F_{12} F_2 &=0
.
\end{align}
We can now define a convenient Poincar\'e--Birkhoff--Witt basis
for the positive Borel sub-algebra $\env_q (\alg{b}^+)$
spanned by the basis
\begin{align}\label{eq:PBWsl3}
\mathcal{B} = \bigrset{E_2^{n_2}E_{12}^{n_{12}}E_1^{n_1} H_1^{m_1} H_2^{m_2} }{n_i,m_i \in\Natural_0 }.
\end{align}
With the definition of this basis we made a particular choice regarding
the ordering of the generators and the definition of the non-simple-root vector $E_{12}$. 
\foreign{A priori}, it would have also been possible to define 
for instance $E_{21}=E_2 \triangleright E_1$ as the non-simple-root vector.
The basis \eqref{eq:PBWsl3} is however chosen such that later calculations,
especially of the R matrix, are rather simple. What this exactly means 
and how the ordering of the simple-root vectors is connected to the definition 
of non-simple-root vectors will be discussed in a later chapter.

%%%%%%%%%%%%%%%%%%%%%%%%%%%%%%%%%%%%%%%%
\paragraph{Dual of the Borel sub-algebra.}

Let us now consider the dual of the positive Borel sub-algebra $\env_q(\alg{b}^+)^*$
as defined in \eqref{eq:pairing1}.
We will explicitly calculate the Hopf structure of the dual generators.
The dual Hopf algebra $\env_q(\alg{b}^+)^*$ is, by definition,
spanned by the dual basis
\begin{align}
\label{eq:DualPBWsl3}
\mathcal{B}^* = \bigrset{(E_2^{n_2}E_{12}^{n_{12}}E_1^{n_1} H_1^{m_1}H_2^{m_2})^* }{n_i,m_i\in\Natural_0 }.
\end{align}
The product of two dual generators expressed in this basis is given by \eqref{eq:dualproduct}.
From that we can find the algebra relations on the dual.
For example let us calculate the product $E_1^* E_2^*$.
We need to find all basis elements in $\mathcal{B}$ whose coproduct has an $E_1 \otimes E_2$ term.
The product is then expanded in the basis $\mathcal{B}^*$
\[
E_1^* E_2^*=q\left(E_2E_1\right)^*+\left(1-q^2\right)E_{12}^*.
\]
Similarly we find
\[
E_2^* E_1^*=\left(E_2E_1\right)^*,
\]
which leads to the commutator
\[
E_1^*E_2^*-qE_2^*E_1^*=\left(1-q^2\right)E_{12}^*.
\]
In the same fashion one can obtain all commutation relations on the dual.
The non-trivial ones are
\begin{align}
\left[H_i^*,E_j^*\right]&=-\delta_{ij}\hbar E_j^*,\\
E_1^*E_{12}^*-q^{-1}E_{12}^*E_1^*&=0,
\\
E_{12}^*E_2^*-q^{-1}E_2^*E_{12}^*&=0.
\end{align}
The coproducts on the dual are found through relation \eqref{eq:dualcoproduct}.
For example, to obtain the coproduct $\copro E_i^*$,
one has to consider contributions coming from the unordered products 
$H_i^n E_j=E_j (H_i +a_{ij})^n $.
Thus the coproducts on the dual are
\begin{align}
\copro E_j^*&=E_j^* \otimes 1+e^{\sum_{i=1}^2 a_{ij} H_i^*}\otimes E_j^*, 
&
\copro H_i^*&=H_i^* \otimes 1 + 1\otimes H_i^*.
\end{align}
Remember that we omit the star at the dual unit $1^*=1$ for convenience.

Let us notice at this point that the positive Borel sub-algebra of 
$\env_q(\alg{sl}(3))$ is self-dual.
\[
\env_q(\alg{b}^+)\cong \env_q (\alg{b}^+)^*.
\]
This can be seen directly from the identifications
\begin{align}
\label{eq:sl3selfdualmap}
 H_j&\equiv -\frac{1}{\hbar} \sum_{i=1}^2 a_{ij}H_i^* , 
&E_j&\equiv \frac{1}{q-q^{-1}} E_j^*, 
&E_{12}&\equiv - \frac{q }{q-q^{-1}}E_{12}^* .
\end{align}
%

%%%%%%%%%%%%%%%%%%%%%%%%%%%%%%%%%%%%%%%%
\paragraph{$\env_q(\alg{sl}(3))$ as a quantum double.}

We can now proceed to construct the quantum double 
$\double\env_q(\alg{b}^+)=\env_q(\alg{b}^+)\otimes \env_q(\alg{b}^+)^{*\cop}$.
To that end we need to calculate the cross-relations \eqref{eq:gencross}. Explicitly we find  
\begin{align}
\comm{H_i^*}{E_j}&=\hbar\delta_{ij}E_j, &\comm{H_i}{E_j^*}&=-a_{ij}E_j^*,\\
\comm{H_i}{H_j^*}&=0, &\comm{E_j^*}{E_i}&=\delta_{ij}(q^{-H_i}-e^{\sum_{k=1}^2 a_{kj} H_k^*}).
\end{align}
The identification 
\begin{align}
 \widehat{H}_j&\equiv \frac{1}{\hbar} \sum_{i=1}^2 a_{ij}H_i^* , 
&F_j&\equiv \frac{1}{q-q^{-1}} E_j^*, 
&F_{21}&\equiv  \frac{1 }{q-q^{-1}}E_{12}^* .
\end{align}
shows the isomorphism 
\[
\env_q(\alg{b}^+)^{*\cop}\cong \env_q(\alg{b}^-).
\]
Note that each Borel half contains a copy of the Cartan sub-algebra.
To distinguish them in the double $\env_q(\alg{b}^+)\otimes \env_q(\alg{b}^-)$ 
we denote the Cartan generators of $\env_q(\alg{b}^-)$
by $\widehat{H}_i$. We can, however, identify the two copies by quotienting 
out the ideal generated by $\widehat{H}_i-H_i$ and thereby recover the Hopf algebra $\env_q(\alg{sl}(3))$.
Thus we find that we can write $\env_q(\alg{sl}(3))$ as the quantum double
\[
\env_q(\alg{sl}(3))\cong \frac{\env_q(\alg{b}^+) \otimes \env_q(\alg{b}^-)}{\bigideal{\widehat{H}_i-H_i}}\,.
\]
%

%%%%%%%%%%%%%%%%%%%%%%%%%%%%%%%%%%%%%%%%
\paragraph{The R-matrix.}

Having written $\env_q(\alg{sl}(3))$ as a quantum double,
it is now straight-forward to find the underlying universal R-matrix from formula \eqref{eq:R-matrix}.
It requires fixing a basis of the positive Borel sub-algebra which we already did in \eqref{eq:PBWsl3}.
Its dual basis \eqref{eq:DualPBWsl3} 
can be expressed in terms of the generators of the negative Borel sub-algebra as
\begin{align}\label{eq:DualPBWsl3sim}
(E_2^{n_2}E_{12}^{n_{12}}E_1^{n_1} H_1^{m_1}H_2^{m_2})^*
&=
\frac{[(q-q^{-1})F_2]^{n_2}}{[n_2;q^{-2}]!}\,
\frac{[(q-q^{-1})F_{21}]^{n_{12}}}{[n_{12};q^{-2}]!}\,
\frac{[(q-q^{-1})F_1]^{n_1}}{[n_1;q^{-2}]!}
\nln&\qquad
\cdot
\frac{\bigsbrk{\frac{\hbar}{3}(2H_1+H_2)}^{m_1}}{m_1!}
\frac{\bigsbrk{\frac{\hbar}{3}(H_1+2H_2)}^{m_2}}{m_2!}\,,
\end{align}
where, following \cite{MBook}, we introduced q-deformed factorials
\unskip\footnote{Another popular choice for q-numbers is $\lfloor n\rfloor_q :=(q^n - q^{-n})/(q-q^{-1})$. 
Both forms are related by $[n;q^2]=q^{n-1} \lfloor n \rfloor_q$ 
and the respective q-factorials by $[n;q^{2}]!=q^{n(n-1)/2} \lfloor n \rfloor _q!$.}
\begin{align}
 &
 [n;q]! :=[n;q][n-1;q]\cdots [1;q],
&&
[n;q] :=\frac{1-q^{n}}{1-q}\,.
\end{align}
Note that this basis transformation takes a rather simple form. This is due 
to the particular choice of generators and their ordering in the PBW basis 
\eqref{eq:PBWsl3}. We will discuss this issue and show the calculation of the 
basis transformation in detail in \secref{sec:Rmat} in the case of 
extended $\alg{sl}(2|2)$.

The factorized form of the basis transformation applied to \eqref{eq:R-matrix} 
leads immediately to a factorized R-matrix
\begin{align}
\rmat &=
e_{q^{-2}}^{(q-q^{-1})E_1\otimes F_1}
e_{q^{-2}}^{(q-q^{-1})E_{12}\otimes F_{21}}
e_{q^{-2}}^{(q-q^{-1})E_2\otimes F_2} 
q^{H_1\otimes (2H_1+H_2)/3}\,
q^{H_2\otimes (2H_2+H_1)/3},
\end{align}
where the q-deformed exponentials are defined as 
\[\label{eq:qexp}
e_{q}^X=\exp_q(X):=\sum_{n=0}^\infty \frac{X^n}{[n;q]!}\,.
\]

\subsection{Presentations and deformations}
\label{sec:presdef}

%%%%%%%%%%%%%%%%%Begin Suggestions
%%%%%%%%%%%%%%%%%%%%%%%%%%%%%%%%%%%%%%%%%%%%%%%%%%%%%%%%%%%%%%%%%%%%%%%%%%%%%%%%

In this paper we will construct a novel quantum algebra
by making a general ansatz for the algebra relations 
and requiring consistency in order to constrain the parameters of the ansatz. 
The parameters can be of different types with different implications
for the structure of the algebra. In particular one should distinguish between 
two classes of parameters.

One class of coefficients is related to
the presentation of the algebra, \foreign{i.e.}\ how to write the algebra in terms
of symbols that form a generating set of the algebra. 
Changing the labeling does not actually change the algebra, 
hence these presentation parameters have no significance,
yet some work is needed to identify their nature. 

The second class is formed by the remaining coefficients that are actual
parameters of the algebra. There are some standard deformations which can be 
applied to general quantum groups. The parameters that are associated with these 
deformations are under good control. However, the parameters that do not have 
such an explanation are the most interesting ones 
because they signal the presence of non-standard deformations.
One may view the quantum parameter $q=e^\hbar$ to be among them
because there is no deformation procedure (along the lines discussed below) 
to derive it. Nevertheless we will basically not consider this parameter and 
restrict our attention to the novel parameters appearing in the construction of 
our particular algebra.

Let us therefore discuss some standard manipulations of quantum algebras that 
will be needed later.

%%%%%%%%%%%%%%%%%%%%%%%%%%%%%%%%%%%%%%%%
\paragraph{Change of basis.}

A quantum algebra is usually presented in terms of a 
set of symbols, \foreign{e.g.}\ $X_i$, and relations among them.
We can redefine the symbols $X_i'=f(X_i,\epsilon_k)$ as functions of the original 
symbols and potentially some parameters $\epsilon_k$.
The algebra relations will take a different form and the presentation parameters 
may change. Yet they will still represent the same algebra.
Of particular interest are transformations that set the presentation 
parameters to special values. This makes most sense if there is a canonical choice
to reduce the complexity or to make the resulting expressions more symmetric. 

In q-deformed quantum algebras, the Cartan sub-algebra of the underlying
Lie algebra plays a central role.
While it is undeformed, it largely determines the deformations of the remaining algebra.
Therefore, transformations of the basis should 
preserve the weights (charges under the Cartan elements) in order not to obscure 
the algebra relations.

%%%%%%%%%%%%%%%%%%%%%%%%%%%%%%%%%%%%%%%%
\paragraph{Similarity transformations.}

Similarity transformations form a special class 
of basis changes. For an invertible element $G$ 
all basis elements are transformed according to
\[
X_i'= G X_i G^{-1}.
\]
Clearly this change of basis preserves 
the form of all algebra relations.
The form of the coalgebra relations usually changes 
unless the element $G$ is group-like.
The latter case will not affect the presentation parameters because no relations are changed. 
Even though one might ignore such similarity transformations right away, 
they are relevant when counting parameters of the algebra relations
\foreign{vs.}\ similarity transformations.

A standard similarity transformation uses a Cartan element $H$
\[
X_i'= e^{\alpha_m H_m} X_i e^{-\alpha_m H_m}.
\]
Since the conjugation element is group-like, this transformation has no effect
on any of the Hopf algebra relations.
By performing the commutators one can see that the similarity transformation
amounts to a rescaling of all generators 
\[
X_i'= e^{\alpha|X_i|_m} X_i
\]
with the exponent given by the weight $|X_i|_m$
defined by $\comm{H_m}{X_i}=|X_i|_m X_i$.

%%%%%%%%%%%%%%%%%%%%%%%%%%%%%%%%%%%%%%%%
\paragraph{Symmetric twist.}

One can also perform a similarity transformation with a quadratic
combination of the Cartan elements
\[
X_i'= e^{\gamma_{mn} H_m H_n/2} X_i e^{-\gamma_{mn} H_m H_n/2}
=
e^{\gamma_{mn} |X_i|_m (H_n-|X_i|_n/2)}
X_i .
\]
Here $\gamma_{mn}$ is a symmetric matrix of coefficients,
and the similarity transformation amounts to multiplying the generators 
by exponents of the Cartan elements.
The conjugation element is not group-like, and effectively 
only the form of the coproduct changes. Therefore, instead of transforming 
the generators, one can also take the different but equivalent 
point of view to only redefine the coproduct by the following twist 
\[
\label{eq:symtwist}
\copro' (X) = 
e^{\gamma_{mn} H_m\otimes H_n}
\copro (X)
e^{-\gamma_{mn} H_m\otimes H_n}.
\]
More explicitly, the conjugation of the coproduct acts 
by inserting various factors of exponentiated Cartan elements
\[
\label{eq:twisteffect}
e^{\gamma_{mn} H_m\otimes H_n}
(X\otimes Y)
e^{-\gamma_{mn} H_m\otimes H_n}
=
 X e^{\gamma_{mn} |Y|_m H_n }\otimes e^{\gamma_{mn} |X|_n H_m } Y.
\]

A noteworthy special case of the symmetric twist is the transformation
on the simple-root generators
\begin{align}
E_i' &= q^{\gamma H_i} E_i,
&
F_i' &= F_i q^{-\gamma H_i}.
\end{align}
It shifts the position of the exponential factors in the coproduct
\eqref{eq:coproE0,eq:coproF0}
\begin{align}
\copro E_i' &= E_i'\otimes q^{\gamma H_i} + q^{-(1-\gamma)H_i} \otimes E_i',
\\
\copro F_i' &= F_i' \otimes q^{(1-\gamma)H_i} + q^{-\gamma H_i} \otimes F_i' .
\end{align}
%

%%%%%%%%%%%%%%%%%%%%%%%%%%%%%%%%%%%%%%%%
\paragraph{Anti-symmetric twist.}

A standard deformation of the quantum algebra is given by the 
Reshetikhin twist of the coproduct \cite{RTwist}
\[
\label{eq:asymtwist}
\copro' (X)= 
e^{\beta_{mn} H_m\otimes H_n}
\copro (X)
e^{-\beta_{mn} H_m\otimes H_n},
\]
where in contradistinction to \eqref{eq:symtwist} $\beta_{mn}$ is 
an anti-symmetric matrix. As above in \eqref{eq:twisteffect}, 
the twist effectively inserts exponential Cartan elements into the coproduct.
In general this twist cannot be compensated by a basis transformation and will 
therefore lead to a different Hopf algebra. If the Hopf algebra was 
quasi-triangular then the twisted Hopf algebra is so as well and it has
\[
\rmat'=e^{\beta_{mn} H_n\otimes H_m}\rmat e^{-\beta_{mn} H_m\otimes H_n},
\] 
as its R-matrix.

%%%%%%%%%%%%%%%%%%%%%%%%%%%%%%%%%%%%%%%%%%%%%%%%%%%%%%%%%%%%%%%%%%%%%%%%%%%%%%%%
%%%%%%%%%%%%%%%%%%%%%%%%%%%%%%%%%%%%%%%%%%%%%%%%%%%%%%%%%%%%%%%%%%%%%%%%%%%%%%%%

\section{Maximally extended \texorpdfstring{$\env_q(\alg{psl}(2|2))$}{Uq(psl(2|2))}}
\label{sec:overview22}

In the following we will state the Hopf algebra structures of 
the maximal extension of $\env_q(\alg{psl}(2|2))$ which is one
of the central results of this paper.
This section is meant to provide an overview and summary 
of the structures and relationships of the algebra.
All derivations and proofs will be postponed to the following sections.

First, we will give an overview of the algebra and its generators, 
then we shall summarize the previously known relations of 
the central extension of $\env_q(\alg{psl}(2|2))$,
and finally state the results of the maximal extension
of $\env_q(\alg{psl}(2|2))$.

%%%%%%%%%%%%%%%%%%%%%%%%%%%%%%%%%%%%%%%%%%%%%%%%%%%%%%%%%%%%%%%%%%%%%%%%%%%%%%%%
\subsection{Overview of the algebras}
\label{sec:overview}

For conciseness, let us introduce abbreviations $\alg{[p][s]g}$
for the various extensions of $\alg{psl}(2|2)$ 
which we shall encounter.
They follow the naming conventions of the algebras $\alg{[p][s]u}(n|n)$ 
\begin{align}
\alg{psg}&:=\alg{psl}(2|2),
&
\alg{sg}&:=\alg{psl}(2|2)\ltimes\Complex^3,
\\
\alg{pg}&:=\alg{sl}(2)\ltimes\alg{psl}(2|2),
&
\alg{g}&:=\alg{sl}(2)\ltimes\alg{psl}(2|2)\ltimes\Complex^3.
\end{align}
The labeling of the corresponding Borel sub-algebras $\alg{[p][s]b}^\pm$ will
follow the same scheme.

\begin{figure}
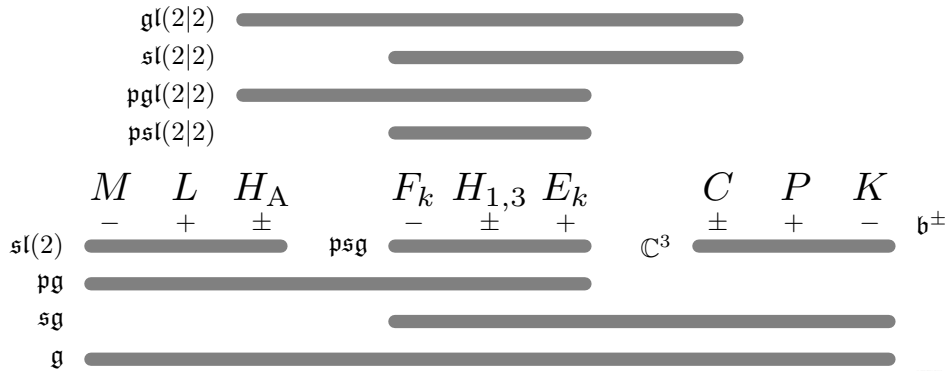
\centering
\includegraphicsbox{FigInclusions.mps}
\caption{Overview of the extended algebras, their inclusions
and generators.
The signs $+$/$-$ indicate to which of the Borel sub-algebras $b^+$/$b^-$
the generators belong; $\pm$ represents Cartan generators which belong to both.}
 \label{fig:inclusions}
\end{figure}

\begin{figure}
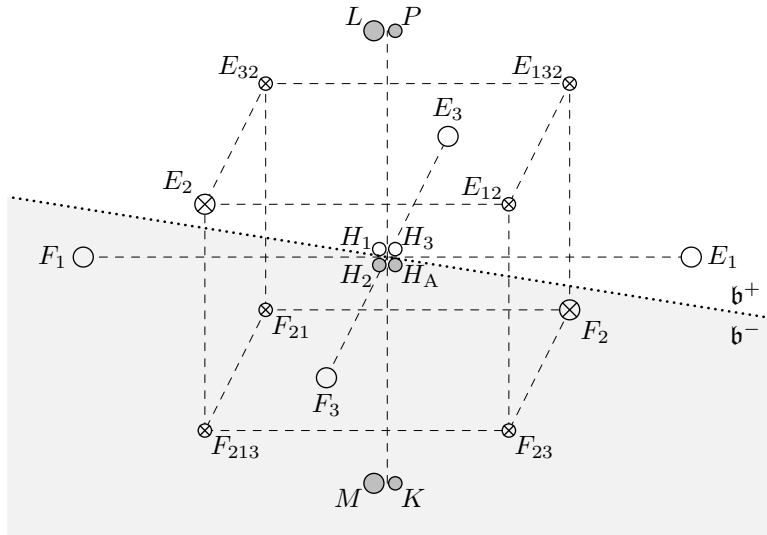
\centering
\includegraphicsbox{FigGenerators.mps}
\caption{Overview of the generators and their weights.
Big, crossed and shaded dots correspond to simple, fermionic and extended generators, respectively.}
\label{fig:generators}
\end{figure}

The algebras will be defined in terms of the Chevalley--Serre generators.
The simple algebra $\alg{psg}=\alg{psl}(2|2)$
has three pairs of positive and negative simple-root generators $E_i$, $F_i$
as well as the three Cartan generators $H_i$ (which are subject to one constraint).
Of these generators $E_2$ and $F_2$ are odd, while the other generators are even.
The central extension $\alg{sg}=\alg{psl}(2|2)\ltimes \Complex^3$ 
is obtained from $\alg{psg}=\alg{psl}(2|2)$ by relaxing three constraints. 
The resulting three additional generators are central, 
and they are denoted by $C,P,K$.
Dual to the central extension is the extension by an $\alg{sl}(2)$ outer 
automorphism algebra $\alg{pg}:=\alg{sl}(2)\ltimes\alg{psl}(2|2)$. We will 
denote the $\alg{sl}(2)$ automorphism generators by  $H_\aut,L,M$. 
The maximal extension
\unskip\footnote{We use the notation of $\ltimes$ freely. More 
precisely we could write depending on the point of view either $\alg{sl}
(2)\ltimes (\alg{psl}(2|2)\oplus_\chi \Complex^3)$ or $(\alg{sl}(2)\ltimes 
\alg{psl}(2|2))\ltimes_\chi \Complex^3$, where $\ltimes$ denotes the semidirect 
product, $\oplus_\chi$ denotes the central extension defined by the cocycle 
$\chi$, and $\ltimes_\chi$ denotes a combination of semidirect product and 
cocycle extension.}
$\alg{g}=\alg{sl}(2)\ltimes\alg{psl}(2|2)\ltimes \Complex^3$
finally combines both extensions into one algebra, where now the $\alg{sl}(2)$ 
automorphisms also act non-trivially on the $\mathbb{C}^3$ part.
Please refer to \figref{fig:inclusions} and \figref{fig:generators} 
for an overview of the generators and their weights.

In order to identify the additional generators unambiguously,
the extensions $C,P,K$ spanning $\Complex^3$ will be called \emph{momentum generators}
\unskip\footnote{These generators are not central in the maximally extended algebra
and hence they should not be called central elements.}
while the extensions $H_\aut,L,M$ spanning $\alg{sl}(2)$ will be called \emph{boost generators}.
These terms follow from the fact that the maximally extended algebra 
$\alg{sl}(2)\ltimes\alg{psl}(2|2)\ltimes \Complex^3$
can be viewed as a peculiar supersymmetric Poincar\'e algebra in three dimensions. 
In this case $\Complex^3$ serves as the ideal of momentum
generators whereas $\alg{sl}(2)$ is the sub-algebra of Lorentz rotations;
the simple algebra $\alg{psl}(2|2)$ contains 8 supercharges 
along with two further internal $\alg{sl}(2)$ symmetry algebras.

Finally, let us mention two relevant relationships for the elements of $\alg{g}$.
The invariant quadratic form of $\alg{g}$ induces a dual pairing of the (qualitative) form:
\[\label{eq:duals}
\begin{array}{c||ccc|ccc|ccc}
& \multicolumn{3}{c|}{\alg{sl}(2)} &  \multicolumn{3}{c|}{\alg{psl}(2|2)} & \multicolumn{3}{c}{\Complex^3}
\\\hline
\alg{g} & M&L&H_\aut & E_k&H_{1,3}&F_k & C&K&P
\\
\alg{g}^* & P^*&K^*&C^* & F_k^*&H_{1,3}^*&E_k^* & H_\aut^*&L^*&M^*
\\\hline
& \multicolumn{3}{c|}{(\Complex^3)^*} &  \multicolumn{3}{c|}{\alg{psl}(2|2)^*} & \multicolumn{3}{c}{\alg{sl}(2)^*}
\end{array}
\]
This paring is needed to relate the double of
the Borel sub-algebra to the full algebra.
Note in particular that the boosts are dual to the momenta.
The other relationship is the algebra automorphism 
which interchanges the Borel sub-algebras:
\[\label{eq:involution}
\begin{array}{c||ccc|ccc|ccc}
& \multicolumn{3}{c|}{\alg{sl}(2)} &  \multicolumn{3}{c|}{\alg{psl}(2|2)} & \multicolumn{3}{c}{\Complex^3}
\\\hline
 & M&L&H_\aut & E_k&H_{1,3}&F_k & C&K&P
\\
\mapsto & L&M&H_\aut & F_k&H_{1,3}&E_k & C&P&K
\end{array}
\]
The combination of the two above relationships 
relates each Borel sub-algebra to its dual
(as a bi-algebra).

%%%%%%%%%%%%%%%%%%%%%%%%%%%%%%%%%%%%%%%%%%%%%%%%%%%%%%%%%%%%%%%%%%%%%%%%%%%%%%%%
\subsection{Hopf structure of the centrally extended algebra}
\label{sec:centralext}

We start by reviewing the q-deformed universal enveloping algebra $\env_q(\alg{sg})$
as provided in \cite{BK}.

%%%%%%%%%%%%%%%%%%%%%%%%%%%%%%%%%%%%%%%%
\paragraph{Algebra.}

The commutation relations of the simple-root generators take the standard form
\begin{align}
\label{eq:defsu22A}
[H_i,H_j] &= 0,
& 
[E_i,F_j] &= d_i \delta_{ij}\, \frac{q^{H_i}-q^{-H_i}}{q-q^{-1}}\,,
\\
[H_i,E_j] &= a_{ij} E_j,
& 
[H_i,F_j ] &= -a_{ij} F_j
\end{align}
expressed in terms of the symmetric Cartan matrix and the vector of signs
\begin{align}\label{eq:CartanSU22}
& a_{ij} 
:=
\begin{pmatrix}
+2 & -1 & 0\\
-1 & 0 & +1\\
0 & +1 & -2
\end{pmatrix},
&  d_i:=( +1, -1, -1 ).
\end{align}
Note that the Cartan matrix has non-maximal rank 2. 
Correspondingly there is a central element within the Cartan sub-algebra, given by 
\unskip\footnote{Notice that we use a different sign convention for $C$ than in,
\foreign{e.g.}\ \cite{BK}.}
\begin{align}
\label{eq:central}
C&:=\sum_{i=1}^3 c_i H_i=\sfrac{1}{2}H_1+H_2+\sfrac{1}{2} H_3,
&
c_i&:=(\half,1,\half).
\end{align}

In addition, the simple-root generators satisfy the Serre relations
\begin{align}
0 & =\bigcomm{E_1}{E_3}=E_2E_2, 
\label{eq:E1E3}\\
0 & =\bigcomm{F_1}{F_3}=F_2F_2, 
\\
0 & =E_iE_iE_2-(q+q^{-1})E_iE_2E_i+E_2E_iE_i, \qquad i=1,3,
\\
0 & =F_iF_iF_2
-\left(q+q^{-1}\right)F_iF_2F_i
+F_2F_iF_i, \qquad i=1,3,
\end{align}
which can also be expressed much more compactly 
using the adjoint action \eqref{eq:adjointaction} as
\[
\label{eq:serretriangle}
0=E_1 \triangleright E_3 
  =F_3 \triangleleft F_1 
= E_i\triangleright (E_i\triangleright E_2 )
  =(F_2 \triangleleft F_i)\triangleleft F_i,
\qquad i=1,3.
\]
It is straightforward to show that there are two further central elements  
$P$ and $K$
\begin{align}
P & :=E_1E_2E_3E_2+E_2E_1E_2E_3
- (q+q^{-1})E_2E_1E_3E_2
+ E_3E_2E_1E_2+E_2E_3E_2E_1,
\label{eq:defP}\\
K & :=F_1F_2F_3F_2
+F_2F_1F_2F_3
- (q+q^{-1})F_2F_1F_3F_2
+ F_3F_2F_1F_2
+F_2F_3F_2F_1.
\label{eq:defK}
\end{align}
Setting them to zero reduces the algebra to $\env_q(\alg{sl}(2|2))$,
in which case the relations $P=K=0$ serve as the quartic Serre relations 
common to Lie superalgebras.
Furthermore setting $C=0$ 
leads to the simple algebra $\env_q(\alg{psl}(2|2))=\env_q(\alg{psg})$.
For completeness, let us state the centrality relations
\[
\label{eq:PKCentral}
[H_i,X]=[E_i,X]=[F_i,X]=[X,X'] = 0,
\qquad 
i=1,2,3,\quad
X,X'= C,P,K.
\]
%

%%%%%%%%%%%%%%%%%%%%%%%%%%%%%%%%%%%%%%%%
\paragraph{Coalgebra.}

We define the q-deformed coproduct as 
\begin{align}
\label{eq:coproE}
\copro E_i &= E_i\otimes 1+q^{-H_i}\otimes E_i, 
\\
 \copro F_i &=F_i\otimes q^{H_i} + 1\otimes F_i, 
\\
\copro H_i &= H_i \otimes 1 + 1\otimes H_i. 
\end{align}
The coproduct of the central elements $C,P,K$ 
follows from their definitions \eqref{eq:central}, \eqref{eq:defP} and \eqref{eq:defK}
via the compatibility condition and takes the form
\begin{align}
\copro C&=C\otimes 1+1\otimes C,
\\
\copro P&=P\otimes 1+q^{-2C}\otimes P,
\\
\copro K&=K \otimes q^{2C} + 1 \otimes K.
\end{align}
The counit and the antipode follow from the coproduct 
by their defining property \eqref{eq:counit} and \eqref{eq:antipode}
and are given by $\counit(X)=0, \; X=H_i,E_i,F_i,P,K$ and
\begin{align}
\antipode (H_i) &= -H_i,
&\antipode (E_i) &= -q^{H_i}E_i,
&\antipode (F_i) &= -F_i q^{-H_i}.
\end{align}

%%%%%%%%%%%%%%%%%%%%%%%%%%%%%%%%%%%%%%%%
\paragraph{Non-simple generators.}

For later usage we introduce non-simple-root generators
as polynomials in the simple roots.
The positive non-simple-root generators read
\begin{align}
E_{12} & :=E_1\triangleright E_2=E_1E_2-qE_2E_1
,\\
E_{32} & :=E_3\triangleright E_2=E_3E_2-q^{-1}E_2E_3
,\\
E_{132} & :=(E_1E_3)\triangleright E_2
=E_1E_{32}-qE_{32}E_1
=E_3E_{12}-q^{-1}E_{12}E_3
,\label{eq:defE123}\\
P&:=\comm{ E_1\triangleright E_2}{E_3\triangleright E_2 }.
\label{eq:defP-PBW}
\end{align}
The corresponding negative ones read
\begin{align}
F_{21} & := F_2 \triangleleft F_1 = F_2 F_1 - q^{-1} F_1 F_2
,\\
F_{23} & := F_2 \triangleleft F_3 = F_2 F_3 - q F_3 F_2
,\\
F_{213} & := F_2 \triangleleft (F_1F_3) 
=F_{23} F_1-q^{-1} F_1 F_{23}
=F_{21} F_3-q F_3F_{21}
,\\
K&:= \comm{ F_2\triangleleft F_1}{F_2\triangleleft F_3 }
.
\end{align}
Note that the central elements $P$ and $K$,
which have been introduced above, are naturally among the non-simple-root generators. 

The Serre relations \eqref{eq:serretriangle} are now expressed as
\begin{align}
\label{eq:E12E32}
E_1 \triangleright E_{12}&=0, & E_3 \triangleright E_{32}&=0,\\
F_{21}\triangleleft F_1&=0, & F_{23}\triangleleft F_3&=0.
\end{align}

Other algebraic relations of the non-simple-root generators follow from their definitions,
and we shall not write them out. 
Due to the special role of $P$ and $K$, we shall nevertheless provide
many of their relations.

%%%%%%%%%%%%%%%%%%%%%%%%%%%%%%%%%%%%%%%%%%%%%%%%%%%%%%%%%%%%%%%%%%%%%%%%%%%%%%%%
\subsection{Hopf structure of the maximally extended algebra}
\label{sec:extalg}

In the following we present the maximally extended algebra $\env_{q,\kappa}(\alg{g})$.
This is understood as the smallest quantum algebra 
which has the form of a quantum double $\double \env_{q,\kappa}(\alg{b}^+)$
and which contains centrally extended $\env_q(\alg{sg})$ 
as a proper sub-algebra. 
It turns out there exists a one-parameter family 
$\env_{q,\kappa}(\alg{g})$ of such algebras
labeled by the parameter $\kappa$. 
The algebraic relations we present here have two parameters $\kappa,\omega$ (apart from 
the conventional quantum parameter $q=e^{\hbar}$). The first is a true parameter
of the Hopf algebra, while the second parameter $\omega$ is merely a parameter of the presentation. 

We will first simply state the defining relations of said Hopf algebra. 
Compared to the above algebra, it suffices to specify 
the relations involving any of the boost generators $H_\aut,L,M$.
In the subsequent section we will provide its construction. 

%%%%%%%%%%%%%%%%%%%%%%%%%%%%%%%%%%%%%%%%
\paragraph{Algebra.}

The algebra has one additional Cartan generator $H_{\aut}$.
It is therefore convenient to extend the Cartan matrix $a$ of $\alg{psl}(2|2)$  
\eqref{eq:CartanSU22} by one row and one column and define a new matrix $\exta$ as follows
\[
\label{eq:CartanU22}
\exta_{ij}
=\begin{pmatrix}
\zeta & 0 & +1 & 0 \\
0 & +2 & -1 & 0\\
+1 & -1 & 0 & +1\\
0 & 0 & +1 & -2
\end{pmatrix}.
\]
We added the new elements at the top and on the left of the Cartan matrix $a$ 
so that the indices run now through ($i,j=\aut,1,2,3$).
\unskip\footnote{The matrix $\exta$ is not in the usual sense the Cartan matrix 
of the extended algebra, since there is no fourth simple-root generator $E_\aut$. 
Instead we have the boost $L$, yet the adjoint action of $H_2$ is not diagonalizable,
so the $\aut$-column is of no use to define commutation relations of $L$ with the Cartan sub-algebra, 
but will be useful in another context.}
The extended Cartan matrix $\exta$ now has full rank $4$.
There is some freedom to choose the top-left element $a_{\aut\aut}$,
and we will parametrize this freedom by the variable 
\[\zeta:= -\kappa-2\omega.\] 

The commutation relations of the Cartan generators and the simple-root generators 
can now be written in terms of the extended Cartan matrix
\[
[H_i,E_j]=\exta_{ij}E_j,\qquad [H_i,F_j]=-\exta_{ij}F_j,\quad i=\aut,1,2,3,\; j=1,2,3.
\]
The centrally extended algebra $\alg{sg}$ is contained as a sub-algebra in the bigger algebra. 
Thus the commutation relations \eqref{eq:defsu22A}, 
the Serre relations \eqref{eq:E1E3} and
the centrality relations \eqref{eq:PKCentral}
carry over to the maximally extended algebra.
Note, however, 
that the momentum generators $C,P,K$ are no longer central in the maximally extended algebra. 
For instance, $P$ and $K$ have a non-trivial charge under $H_\aut$
\begin{align}
[H_\aut, P] &= 2P,
&
[H_\aut,K] &= -2K.
\end{align}

The algebra relations involving the positive boost $L$ read
\begin{align}\label{eq:Lrel}
[H_\aut, L] &= 2L +  \omega \,\frac{q-q^{-1}}{2\hbar}\, P ,
\\
[H_2, L] &=  -  \frac{q-q^{-1}}{2\hbar}\, P,
\\
[L,E_2]&= \half (q-q^{-1}) E_2P,
\\
[L,E_3]&=  q(q-q^{-1}) E_{32}E_{132},
\\
[L,F_2]&=  q\bigsbrk{E_{132}+(q-q^{-1}) E_{32}E_1}q^{-H_2},
\\
[L,F_3]&= q^{-1}(q-q^{-1})q^{H_3}E_2 E_{12},
\\
[L,X]&=0, \qquad X=H_1,H_3,E_1,F_1,
\end{align}
whereas those for the negative boost $M$ take the analogous form
\begin{align}
[H_\aut, M] &=
-2 M -\omega \,\frac{q-q^{-1}}{2 \hbar}\,  K,
\\
[H_2, M] &=  \frac{q-q^{-1}}{2 \hbar}\, K ,
\\
[M,E_2]&= q^{-1}\bigsbrk{-q^{H_2} F_{213}+(q-q^{-1})q^{H_2} F_1 F_{23} },
\\
[M,E_3]&= q^{-1} (q-q^{-1}) q^{-H_3} F_{21} F_2 ,
\\
[M,F_2]&= \half (q-q^{-1}) K F_2 ,
\\
[M,F_3]&= q^{-1}(q-q^{-1}) F_{213} F_{23},
\\
[M,X]&=0, \qquad X=H_1,H_3,E_1,F_1.
\end{align}
Finally, the cross-relation for the boosts reads
\begin{align}
[L,M]&=
- \half\bigsbrk{q^{2C}+q^{-2C}} \bigsbrk{H_\aut + (\kappa+\omega)C }.
\end{align}
It is convenient to note the algebra relations 
between the boost and momentum extensions
\begin{align}
[L,P] &= 0,
&
[M,P]&=- \frac{q^{2C}-q^{-2C}}{q-q^{-1}}\, ,
\\
[L,C] &=  \frac{q-q^{-1}}{2\hbar} P,
&
[M,C] &=- \frac{q-q^{-1}}{2\hbar} K,
\\
[L,K]&= \frac{q^{2C}-q^{-2C}}{q-q^{-1}}\,,
&
[M,K]&= 0.
\end{align}

%%%%%%%%%%%%%%%%%%%%%%%%%%%%%%%%%%%%%%%%
\paragraph{Coalgebra.} 

The coproduct of the simple-root vectors \eqref{eq:coproE} is unchanged,
and also the boost element $H_\aut$ of the Cartan sub-algebra follows the standard trivial form.
For the boosts $L$ and $M$ we find the following expressions 
\begin{align}
\label{eq:Lcopro}
\copro L
&= L\otimes 1 +q^{-2C}\otimes L 
\nonumber \\ &\qquad
+\half (q-q^{-1}) \bigsbrk{ H_\aut + (\kappa+\omega) C } q^{-2C} \otimes P
\nonumber \\ &\qquad
 -  q^{-1}(q-q^{-1})^2 E_{3} q^{-H_1-2H_2} \otimes E_2 E_{12}
\nonumber \\ &\qquad
-  (q-q^{-1}) E_{32}q^{-H_1-H_2} \otimes E_{12} 
\nonumber \\ &\qquad
+   q(q-q^{-1})\bigsbrk{E_{132} +(q-q^{-1}) E_{32}E_{1}}q^{-H_2} \otimes E_2 
 ,
\\
\copro M &= 
M \otimes q^{2C} 
+ 1 \otimes M 
\nonumber \\ &\qquad
-\half (q-q^{-1}) K \otimes q^{2C} \bigsbrk{H_\aut +(\kappa+\omega)C}
\nonumber  \\ &\qquad
- q(q-q^{-1})^2 F_{21} F_2 \otimes q^{H_1+2H_2}F_3 
\nonumber  \\ &\qquad
+ (q-q^{-1}) F_{21}\otimes q^{H_1+H_2}F_{23}
\nonumber \\ &\qquad
+ q^{-1}(q-q^{-1}) F_2 \otimes q^{H_2} \bigsbrk{-F_{213}+(q-q^{-1}) F_1 F_{32}}
.
\end{align}
The antipode reads
\begin{align}
 \antipode(L)&= 
-L q^{2C}
+\half (q-q^{-1})  \bigsbrk{ PH_\aut + (\kappa+\omega) PC }q^{2C}
\nonumber \\ &\qquad
+  (q-q^{-1})\bigsbrk{E_{12}E_{32}-q^{-1}E_2E_{132}+q^{-1}(q-q^{-1})^2E_2E_{12}E_3}q^{2C}
,\\
\antipode (M) &= 
-q^{-2C}M 
-\half (q-q^{-1})  q^{-2C}\bigsbrk{ H_\aut K + (\kappa+\omega) C K }
\nonumber \\ &\qquad
+  (q-q^{-1})q^{-2C}\bigsbrk{-F_{23} F_{21}+qF_{213} F_2 + q(q-q^{-1})^2F_3 F_{21} F_2}
.
\end{align}

%%%%%%%%%%%%%%%%%%%%%%%%%%%%%%%%%%%%%%%%%%%%%%%%%%%%%%%%%%%%%%%%%%%%%%%%%%%%%%%%
\subsection{Special features}
\label{sec:features}

Finally, we collect and discuss various salient and unusual features that our algebra exhibits.

%%%%%%%%%%%%%%%%%%%%%%%%%%%%%%%%%%%%%%%%
\paragraph{Combinations of generators.}

First, let us comment on the appearance of exponential functions: 
conventional q-deformed algebras can be formulated in terms of 
exponentiated Cartan generators $K_i := q^{H_i}$
and the quantum parameter  $q:=e^\hbar$
without the need to resort to $\log K_i$ or $\log q$
(merely the R-matrix requires these in one factor).
In this sense, the new Cartan generator $H_\aut$ appears in a non-standard form
because it is never exponentiated in the algebra or coalgebra relations. 
If $H_\aut$ was replaced by its exponent $K_\aut := q^{H_\aut}$, 
many relations would have to be formulated in terms of $\log K_\aut$.

Conversely, the Cartan generator $C$ can almost always be exponentiated,
except for a few terms which vanish upon setting 
the presentation parameter $\omega = -\kappa$. 
In other words, this is an artifact of our presentation rather than a feature of the algebra itself. 
On a related note, some plain factors of $\hbar=\log q$ appear in the Hopf algebra relations, 
for instance in $[H_2, L]$. 
However, this factor cancels neatly for exponentiated Cartan generators, 
\foreign{e.g.}\ $q^{H_2} L =  L q^{H_2} -\half (q-q^{-1}) P q^{H_2}$.

Another unusual feature is the appearance of non-trivial products 
of generators in both the algebra and coalgebra structures,
see \foreign{e.g.}\ \eqref{eq:Lrel,eq:Lcopro}.

%%%%%%%%%%%%%%%%%%%%%%%%%%%%%%%%%%%%%%%%
\paragraph{Undeformed automorphisms.}

While the q-deformation for the $\alg{psl}(2|2)$ generators 
and the momenta $C,P,K$ is rather standard,
the deformation for the boosts $H_\aut,L,M$ is faint. 
For instance, when dropping all other generators, 
the boosts obey the algebra $\env(\alg{sl}(2))$ 
rather than $\env_q(\alg{sl}(2))$.
Moreover, we can even remove the appearance of the momenta $C,P,K$ 
by a redefinition (with $b+c=\pm 1$)
\begin{align}
J_+ &= q^{2b C}\bigbrk{L + \sfrac{1}{2}c (q-q^{-1})P (H_\aut+\omega C)},
\\
J_- &= q^{2c C}\bigbrk{M + \sfrac{1}{2}b (q-q^{-1})K (H_\aut+\omega C)},
\\
J_0 &= H_\aut + \omega C.
\end{align}
Their algebra then reads
\[
\comm{J_0}{J_\pm}=\pm 2 J_\pm,
\qquad
\comm{J_+}{J_-}=-J_0 - \half \kappa \bigbrk{1+q^{4(b+c)C}} C,
\]
which is undeformed $\env(\alg{sl}(2))$
up to the term proportional to $\kappa$.

This feature is related to the absence of exponentials of the type $q^{H_\aut}$
noted above.
It can be attributed to the coefficients $d_i$
governing the norm of simple roots in non-simply laced Lie algebras.
While the coefficients $d_i$, $i=1,2,3$ for 
the simple algebra $\alg{psl}(2|2)$ all equal $\pm1$,
the coefficient $d_\aut$ for the boost generators
is (in some sense) infinitesimally small.
Therefore the corresponding exponential
$q^{d_\aut H_\aut}=1+\hbar d_\aut H_\aut+\order(d_\aut^2)$
is approximated well by the constant and linear term,
and consequently the algebra of the boosts is undeformed.

%%%%%%%%%%%%%%%%%%%%%%%%%%%%%%%%%%%%%%%%
\paragraph{Symmetry of the presentation.}

Whereas the centrally extended algebra is 
symmetric w.r.t.\ the interchange of simple-root generators $1\leftrightarrow 3$,
the automorphisms appear to break this discrete symmetry, \eqref{eq:Lrel}. 
However, the breaking is due to our choice of basis.
There is an equivalent presentation which makes 
$[L,E_1]$ rather than $[L,E_3]$ non-trivial;
there is also a presentation which makes the $1\leftrightarrow 3$
symmetry manifest (see \cite{BdLH2}), 
but this choice will not be convenient to calculate the universal R-matrix.
Note that the asymmetry $1\leftrightarrow 3$ 
also shows up in the q-deformed secret symmetry \cite{deLeeuw:2011fr}.

%%%%%%%%%%%%%%%%%%%%%%%%%%%%%%%%%%%%%%%%
\paragraph{Momentum invariant.}

Note that there is a quadratic invariant $X$ involving the momentum generators $C,P,K$,
whose form was already observed in the shortening condition 
for representations in \cite{BK}
\[\label{eq:mominv}
X= P K - \lrbrk{\frac{q^{C} -q^{-C}}{q-q^{-1}} }^2.
\]
Since  $C,P,K$ are central in $\env_q(\alg{sg})$, 
it suffices to check that $[M,X]=[L,X]=0$
to ensure centrality in $\env_{q,\kappa}(\alg{g})$.
The latter follows from the algebra relations presented in \secref{sec:extalg}.

%%%%%%%%%%%%%%%%%%%%%%%%%%%%%%%%%%%%%%%%
\paragraph{Deformation parameters.}

A final note is that our algebra has two non-trivial deformation parameters
$\hbar$ and $\kappa$ (besides the standard ones discussed in \secref{sec:presdef}).
The additional parameter $\omega$ has no significance for the Hopf algebra because it merely
deforms the presentation.
In particular, it can be absorbed completely by a redefinition 
\[\label{eq:omegaredef}
H_\aut' = H_\aut +\omega C
\]
It is nevertheless instructive to keep it in the presentation rather than fixing it to a specific value. 
The existence of the parameter $\kappa$ 
can be attributed to the unconstrained element $a_{\aut\aut}$ 
in the extended Cartan matrix. 

A curious fact is that the parameter $\kappa$ 
can be removed from all algebra relations as well 
(but not from the coalgebra relations)
by a redefinition
\begin{align}\label{eq:kapparedef}
L'&= L+\sfrac{1}{4}\kappa f(C,X)P,
&
M'&= M+\sfrac{1}{4}\kappa f(C,X)K.
\end{align}
Here $f(C,X)$ is a function of $C$ and 
the momentum invariant $X$ in \eqref{eq:mominv},
and it should obey the differential equation
\[
\frac{q^{2C}-q^{-2C}}{(q-q^{-1})C}\,f
+\frac{q-q^{-1}}{2\hbar C}\,\frac{\partial f}{\partial C}
\lrsbrk{X +  \lrbrk{\frac{q^{C} -q^{-C}}{q-q^{-1}} }^2}
=q^{2C}+q^{-2C}\,.
\]
This equation can be solved by a deformation function (with $Y:=\sfrac{1}{2}(q-q^{-1})\sqrt{X}$)
\begin{align}
f&=
\frac{q-q^{-1}}{\hbar}
\lrsbrk{-\frac{1}{2}+\frac{\hbar \bigbrk{q^{2C}-q^{-2C}}C+4Y\sqrt{1-Y^2}\arcsin Y}
{\bigbrk{q^{C}-q^{-C}}^2+4Y^2}}
\nln
&=1+\hbar^2\lrbrk{\sfrac{1}{6}+\sfrac{4}{3}C^2-\sfrac{2}{3}PK} +\order(\hbar^4).
\end{align}
We refrain from implementing this transformation because it would
mess up the coalgebra.

%%%%%%%%%%%%%%%%%%%%%%%%%%%%%%%%%%%%%%%%%%%%%%%%%%%%%%%%%%%%%%%%%%%%%%%%%%%%%%%%
%%%%%%%%%%%%%%%%%%%%%%%%%%%%%%%%%%%%%%%%%%%%%%%%%%%%%%%%%%%%%%%%%%%%%%%%%%%%%%%%

\section{Extending the algebra}
\label{sec:double}

We aim to express
the centrally extended algebra $\env_q(\alg{psl}(2|2)\ltimes\Complex^3)$
as a quantum double.
The procedure will be analogous to the example of $\alg{sl}(3)$ discussed in \secref{sec:sl3ex}.
However, the presence of the momentum ideal $\Complex^3$ will turn out
to cause the addition of an $\alg{sl}(2)$ sub-algebra to our algebra.
In the end we find that the centrally extended algebra
can be embedded in the larger algebra
$\env_q(\alg{sl}(2)\ltimes \alg{psl}(2|2)\ltimes\Complex^3)$ presented above
and that the latter takes the form of a quantum double.

In this section we will derive the aforementioned Hopf algebra 
by enlarging $\env_q(\alg{sg})$
with additional generators $L,H_\aut,M$ 
such that this enlarged algebra can be identified 
with the quantum double of its Borel sub-algebra.
This is done in several steps:
First, we shall construct the dual of 
the positive Borel sub-algebra $\env_q(\alg{sb}^+)$.
This is isomorphic to the algebra $\env_q(\alg{pb}^+)$ 
which contains some of the relations of the additional generators.
Some of their relations follow from the fact that they are
dual to the momentum generators $C,P,K$, respectively, see \eqref{eq:duals}.
The dual of the established central extension $\env_q(\alg{sg})$
thus contains relations involving the boosts (but not the momenta).
This statement can be made exact at the level of Borel sub-algebras
\[ \env_q(\alg{sb}^+)^* \cong \env_q(\alg{pb}^+). \]

Next, we enlarge the positive Borel sub-algebra $\env_q(\alg{sb}^+)$ 
by an additional Cartan generator $H_\aut$ and by a new generator $L$ 
to $\env_{q,\beta}(\alg{b}^+)$. 
We will do this in a very general way which leaves us with four freely adjustable parameters
$\beta_i$, $i=\aut,1,2,3$.
Finally, we construct the quantum double of $\env_{q,\beta}(\alg{b}^+)$,
and find that the matching of the Cartan sub-algebras 
of the dual $\env_{q,\beta}(\alg{b}^+)^*$ 
with the negative sub-algebra $\env_{q,\beta}(\alg{b}^-)$
imposes further restrictions on the $\beta_i$.
The resulting double is nevertheless not unique, 
but it forms a one-parameter family of consistent Hopf algebras $\env_{q,\kappa}(\alg{g})$
which contain the centrally extended algebra $\env_{q}(\alg{sg})$. 

%%%%%%%%%%%%%%%%%%%%%%%%%%%%%%%%%%%%%%%%%%%%%%%%%%%%%%%%%%%%%%%%%%%%%%%%%%%%%%%%%%%%%%%%%
\subsection{Dual of the centrally extended Borel sub-algebra}
\label{sec:dualnonext}

In the following we will construct the relations of the dual Hopf algebra 
of the positve Borel sub-algebra $\env_q(\alg{sb}^+)$ 
of the centrally extended algebra as defined in \secref{sec:centralext}.
As a result of this construction, we will observe explicitly that 
$\env_q(\alg{sb}^+)$ is not dual to 
the negative Borel sub-algebra 
\[ \env_q(\alg{sb}^+)^*\ncong \env_q(\alg{sb}^-). \]
Consequently we cannot write the centrally extended algebra $\env_q(\alg{sg})$ 
as the quantum double of its positive Borel sub-algebra $\double\env_q(\alg{sb}^+)$ 
which motivates the introduction of the boosts in the next chapter.

%%%%%%%%%%%%%%%%%%%%%%%%%%%%%%%%%%%%%%%%
\paragraph{Dual algebra.}

The calculation of the commutation relations on the dual algebra is completely
analogous to the calculation in the $\alg{sl}(3)$ case.
We first need to fix a basis of $\env_q(\alg{sb}^+)$. 
We choose the PBW basis with following ordering
\begin{equation}
\label{eq:PBW}
\bigrset{  E_2^{n_2}E_{12}^{n_{12}} P^{n_P} E_{32}^{n_{32}} E_{132}^{n_{132}} E_1^{n_1}E_3^{n_3}
H_1^{m_1}  H_2^{m_2} H_3^{m_3}
}{n_i,m_j \in \Natural_0} ,
\end{equation}
and define the dual vector space $\env_q(\alg{sb}^+)^*$
as the span of the dual basis
\begin{equation}
\label{eq:PBW*}
\bigrset{(  E_2^{n_2}E_{12}^{n_{12}} P^{n_P} E_{32}^{n_{32}} E_{132}^{n_{132}}E_1^{n_1}E_3^{n_3}
H_1^{m_1} H_2^{m_2} H_3^{m_3})^*
}{n_i,m_j \in \Natural_0 } .
\end{equation}
The product of two dual generators expressed in the dual basis 
is found as prescribed by $\eqref{eq:dualproduct}$. 
For instance, the commutator $[H_i^*,E_j^*]$ follows from the basis expansions
\begin{align}
H_i^*E_j^*
&=
(E_jH_i)^*
-\hbar \delta_{ij} E_j^*,
\\
E_j^*H_i^*&=(E_j H_i)^*.
\end{align}
In this way all commutators and Serre relations are calculated. 
We find the commutators
\unskip\footnote{The fact that $\comm{P^*}{E_3^*}$
is different from the other relations, 
in particular from $\comm{P^*}{E_1^*}$, 
follows from our choice of PBW basis.}
\begin{align}
\bigcomm{H_i^*}{H_j^*}&=0, 
&
\bigcomm{H_i^*}{E_j^*}&=-\hbar \delta_{ij} E_j^* 
\label{eq:dualCartan},\\
\bigcomm{H_i^*}{P^*} 
& = -2\hbar c_iP^*,
& 
\bigcomm{P^*}{E_j^*} 
& =\delta_{j3}(q-q^{-1})E_{32}^*E_{132}^* ,
\label{eq:PstartComRel3}
\end{align}
where $c_i$ is the null vector of the Cartan matrix defined in \eqref{eq:central}.
The dual Serre relations read
\begin{align}
0 & =\left[E_1^*,E_3^*\right]=E_2^*E_2^*,
\\
0 & =E_i^*E_i^*E_2^*-(q+q^{-1})E_i^*E_2^*E_i^*+E_2^*E_i^*E_i^*,\qquad i=1,3,
\\
0&=E_1^*E_2^*E_3^*E_2^* +
E_2^*E_1^*E_2^*E_3^* -
(q+q^{-1})E_2^*E_1^*E_3^*E_2^* +
E_3^*E_2^*E_1^*E_2^* +
E_2^*E_3^*E_2^*E_1^*,
\label{eq:dualSerre}
\end{align}
and the duals of the non-simple generators are related to the dual simple generators by
\begin{align}
E_{2}^{*}E_{1}^{*}-q^{-1}E_{1}^{*}E_{2}^{*}&=(q-q^{-1})E_{12}^{*},
\\
E_{2}^{*}E_{3}^{*}-qE_{3}^{*}E_{2}^{*}&=-(q-q^{-1})E_{32}^{*},
\\
E_{12}^{*}E_{3}^{*}-qE_{3}^{*}E_{12}^{*}&=-(q-q^{-1})E_{132}^{*}.
\end{align}

%%%%%%%%%%%%%%%%%%%%%%%%%%%%%%%%%%%%%%%%
\paragraph{Dual coproduct.}

The coproduct of dual generators can be expressed in the dual basis using \eqref{eq:dualcoproduct}.
As an example let us consider the coproduct $\copro P^*$.
To find all contributions to that coproduct, we need to consider all pairs of basis elements $(x,y)$
such that their product $xy$ expressed in the basis \eqref{eq:PBW} 
contains a contribution of $P$.
This happens in the following cases
\[
\pair{ P^*}{xy}=\left\{
\begin{array}{ l l l}
\prod_{i=1}^3\left(a_{i1}+a_{i2}\right)^{n_i}, & x=E_{32}\prod_{i=1}^3H_i^{n_i}, & y=E_{12},
\\
-q\prod_{i=1}^3a_{i2}^{n_i}, & x=E_{132}\prod_{i=1}^3H_i^{n_i}, & y=E_2,
\\
\prod_{i=1}^3\left(a_{i1}+2a_{i2}\right)^{n_i}, & x=E_3\prod_{i=1}^3H_i^{n_i}, & y=E_2E_{12},
\\
\prod_{i=1}^3\left(a_{i2}\right)^{n_i}, & x=E_{32}E_1\prod_{i=1}^3H_i^{n_i},& y=E_2,
\\
1, & x=P,& y=1,
\\
\prod_{i=1}^3 (a_{i1}+2a_{i2}+a_{i3})^{n_i}, & x=\prod_{i=1}^3 H_i^{n_i}, & y=P.
\end{array}
\right.
\]
Performing the analogous consideration for all dual generators 
we find the dual coproducts, noting that $(H_i^{n_i})^*=(H_i^*)^{n_i}/n!$
\begin{align}
\label{eq:coprodual2}
\copro H_i^* & = H_i^*\otimes1+1\otimes H_i^*
,\\
\label{eq:coprodual1}
\copro E_j^*&=E_j^*\otimes 1 +
\exp\lrbrk{\sum_{i=1}^3a_{ij}H_i^*} \otimes E_j^*
,\\
\copro P^* 
& =P^*\otimes 1+ 1 \otimes P^*
+(qE_{132}^*  -E_{32}^*E_1^*)\exp\lrbrk{\sum_{i=1}^3a_{i2}H_i^*} \otimes E_2^*\label{eq:DP*_1}
\nonumber \\  &\qquad 
-E_{32}^* \exp\lrbrk{\sum_{i=1}^3 \left(a_{i1}+a_{i2}\right) H_i^*} \otimes E_{12}^*
\nonumber \\  &\qquad 
-E_3^*\exp\lrbrk{\sum_{i=1}^3 \left(a_{i1}+2a_{i2}\right) H_i^*} \otimes E_2^*E_{12}^*
.
\end{align}
For details on the calculation of the coproduct $\copro P^*$ 
see also \appref{app:copro}.

%%%%%%%%%%%%%%%%%%%%%%%%%%%%%%%%%%%%%%%%
\paragraph{Dual Hopf algebra structure.}

The above dual Hopf algebra relations show explicitly 
that $\env_q(\alg{sb}^+)$ is neither dual to itself
nor to $\env_q(\alg{sb}^-)$.

This fact can be noticed in several of the algebra relations:
First of all \eqref{eq:dualCartan} shows 
that there is no element in the dual Cartan sub-algebra that is central.
Therefore it is impossible to identify the dual Cartan sub-algebra 
with the Cartan sub-algebra of $\env_q(\alg{sb}^\pm)$.
Furthermore, the dual quartic Serre relation \eqref{eq:dualSerre} 
has no generator on the left hand side, and is not related to $P^*$. 
So again, we cannot make an identification with $\env_q(\alg{sb}^\pm)$
because there is no element in the dual we could identify with $K$.
Finally, $P^*$ is a non-central element \eqref{eq:PstartComRel3}
which has no analogue in $\env_q(\alg{sb}^\pm)$.

Alternatively, this fact follows from the dual coproduct:
To identify $E_j$ or $F_j$ with $E_j^*$ 
one would also need to identify $H_j$ with $\hbar^{-1}\sum_{i=1}^3a_{ij}H_i^*$ 
to make the exponent in the coproduct match. 
This is, however, not possible since the Cartan matrix $a_{ij}$ is degenerate. 
Furthermore, the unusual form of the coproduct of $P^*$ 
makes it clear that we cannot identify it with any element in $\env_q(\alg{sb}^\pm)$.

All in all we find that, unlike in the $\alg{sl}(3)$ example, 
we cannot identify the dual of the positive Borel sub-algebra 
with the negative Borel sub-algebra $\env_q(\alg{sb}^+)^{*\cop}\ncong \env_q(\alg{sb}^-)$.
This is solely due to the presence of the central elements which 
fail to be central upon dualization.

We will eventually fix this issue by the introduction 
of three additional boost generators $H_\aut, L , M$ to the algebra, 
such that the Borel sub-algebras of that extended algebra satisfy the duality relation
\[
\env_q(\alg{b}^+)^{*\cop}\cong \env_q(\alg{b}^-).
\]
The idea is that the dual generators of the boosts $H_\aut^*,L^*$ 
shall be identified with the (almost) central generators $C,K$
while the the duals of the (almost) central generators $C^*,P^*$ 
shall be identified with the boosts $H_\aut,M$. 
In total, the situation is depicted in \tabref{tab1}.

\begin{table}
\centering
\begin{tabular}{c || c  c  c c c || c }
    & $\env_q(\alg{b}^+)$ & dualization & $\env_q(\alg{b}^+)^*$ & identification & $\env_q(\alg{b}^-)$  & 
\\ \hline
\multirow{ 2}{*}{$\alg{psb}^+$}& $H_{1,3}$ & $\stackrel{*}{\longrightarrow}$ &  $H^*_{1,3}$  
&  $\stackrel{\sim}{\longrightarrow}$ & $H_{1,3}$ &  \multirow{ 2}{*}{$\alg{psb}^-$} 
\\
 & $E_i$ & $\stackrel{*}{\longrightarrow}$ &  $E_i^*$  & $\stackrel{\sim}{\longrightarrow}$ & $F_i$&
\\
 \color{red} &  \color{red}$C$ & $\stackrel{*}{\longrightarrow}$ & \color{blue} $C^*$  
& $\stackrel{\sim}{\longrightarrow}$ &\color{blue} $H_\aut$ &  \color{blue}  
\\
 \multirow{-2}{*}{ \color{red} $(\Complex^3)^+$} &  \color{red} $P$ &  $\stackrel{*}{\longrightarrow}$ 
&\color{blue} $P^*$  & $\stackrel{\sim}{\longrightarrow}$ & \color{blue} $M$ &  \multirow{-2}{*}{\color{blue} $\alg{sl}(2)^-$}
\\
\color{blue} & \color{blue}$H_\aut$ &  $\stackrel{*}{\longrightarrow}$ & \color{red} $H_\aut^*$  
& $\stackrel{\sim}{\longrightarrow}$ & \color{red} $C$ &     \color{red}
\\
 \multirow{-2}{*}{\color{blue} $\alg{sl}(2)^+$} &  \color{blue} $L$ &  $\stackrel{*}{\longrightarrow}$ 
& \color{red} $L^*$  & $\stackrel{\sim}{\longrightarrow}$ & \color{red} $M$ & 
\multirow{-2}{*}{ \color{red} $(\Complex^3)^-$}
\\
 \hline
\end{tabular}
\caption{The relations between the positive sub-algebra, its dual and the negative sub-algebra. 
Under dualization, the boost generators $C,P,K$ (blue) 
are mapped to the momentum generators $H_\aut,M,L$ (red) and \foreign{vice versa}.}
\label{tab1}
\end{table}

Furthermore the introduction of the boost generators shall be such that 
it keeps centrally extended $\env_q(\alg{psl}(2|2)\ltimes \Complex^3)$ unchanged 
as a Hopf sub-algebra of the enlarged Hopf algebra.

%%%%%%%%%%%%%%%%%%%%%%%%%%%%%%%%%%%%%%%%%%%%%%%%%%%%%%%%%%%%%%%%%%%%%%%%%%%%%%%%
\subsection{Extending the positive Borel sub-algebra}

We extend the Borel sub-algebra $\env_q(\alg{sb}^+)$
to $\env_q(\alg{b}^+)$ by adding the two boost generators $H_\aut$ and $\asymL$ 
such that its quantum double $\double \env_q(\alg{b}^+)$ contains a sub-algebra 
that can be identified with $\env_q(\alg{sg})$. 
This requirement will fix the Hopf structure of the boost generators.
We denote $\asymL$ with a tilde to leave the plain $L$ for a redefined version of it later.
We will make a general ansatz with a couple of free parameters
 which we subsequently constrain to ensure a consistent Hopf algebra structure.

%%%%%%%%%%%%%%%%%%%%%%%%%%%%%%%%%%%%%%%%
\paragraph{PBW basis.}

We have to include the new generators in our basis and define a PBW basis 
for the positive Borel sub-algebra $\env_q(\alg{b}^+)$
with the following ordering of generators
\begin{align}
\label{eq:extendedPBW}
\bigrset{ H_\aut^{n_\aut}  E_2^{n_2}E_{12}^{n_{12}} 
\asymL^{n_L} P^{n_P} E_{32}^{n_{32}} E_{132}^{n_{132}}E_1^{n_1}E_3^{n_3}
H_1^{m_1} H_2^{m_2} H_3^{m_3} }{n_i,m_i \in\Natural_0 }.
\end{align}
This ordering will turn out to be a convenient choice for calculating the R-matrix. 
The reason for that will be explained later in \secref{sec:Rmat}.
Given this basis we define the dual space as the span of the dual basis 
\begin{align}
\label{eq:extendedPBW*}
\bigrset{ (H_\aut^{n_\aut}  E_2^{n_2}E_{12}^{n_{12}} 
\asymL^{n_L} P^{n_P} E_{32}^{n_{32}} E_{132}^{n_{132}}E_1^{n_1}E_3^{n_3}
H_1^{m_1} H_2^{m_2} H_3^{m_3})^* }{n_i,m_i \in\Natural_0 }.
\end{align}

%%%%%%%%%%%%%%%%%%%%%%%%%%%%%%%%%%%%%%%%
\paragraph{Extending the Cartan sub-algebra.}

First we focus on the Cartan sub-algebra and the additional boost generator $H_\aut$. 
We have seen before that the ranks of the Cartan matrix and the dual Cartan matrix did not match.
Now let us make the Cartan sub-algebra self-dual by adding an additional generator $H_\aut$.
For it to be part of the Cartan sub-algebra we require
\begin{align}
[H_i,H_\aut]&=0, &\copro H_\aut &=H_\aut\otimes 1 + 1\otimes H_\aut.
\end{align}
For the commutators with the simple-root vectors we extend the Cartan matrix by a fourth row
\[
\label{eq:CartanAut}
[H_\aut,E_j]=\exta_{\aut j} E_j.
\]
The new entries $\exta_{\aut j}$ have to be such that the rank of the extended Cartan matrix 
is equal to the rank of the extended dual Cartan matrix which will turn out to be $3$. 
Thus we require $\exta_{\aut 1}+2\exta_{\aut 2}+\exta_{\aut 3}\neq 0$.  
By a redefinition of $H_\aut$ we can always set $\exta_{\aut i}=\delta_{2i}$. 
So without loss of generality, we have the extended Cartan matrix 
\begin{equation}\label{eq:extCartan0}
\exta_{ij} 
=\begin{pmatrix}
 0 & 1 & 0 \\
 2 & -1 & 0\\
 -1 & 0 & +1\\
 0 & +1 & -2
\end{pmatrix}.
\end{equation} 

Now, let us repeat the calculation of the dual of the extended Cartan sub-algebra 
to see whether the Cartan sub-algebra has become self-dual.
Actually, hardly anything changes compared to the 
non-extended case in \secref{sec:dualnonext}.
The new dual generator $H_\aut^*$ commutes with all simple-root generators
since $H_\aut$ does not appear in the coproduct of any of the simple-root generators $E_i$
\[\label{eq:HAalmostcentral}
\bigcomm{H_\aut^*}{H_i^*}=\bigcomm{H_\aut^*}{E_i^*}=0,\qquad i=1,2,3.
\]
The commutator \eqref{eq:dualCartan} remains unchanged
\[
\label{eq:dualCartan2}
\bigcomm{H_i^*}{E_j^*}=-\hbar \delta_{ij} E_j^* ,\qquad i,j=1,2,3.
\]
Furthermore, the coproduct of the dual Cartan generators \eqref{eq:coprodual2} 
is  not touched and also the same for $H_\aut^*$
\[
\copro H_i^* = H_i^*\otimes1+1\otimes H_i^*,\qquad i=\aut,1,2,3.
\]
We observe that now we have a central element in both the Cartan ($C$) 
and the dual Cartan sub-algebra ($H_\aut^*$) so that its possible to identify them with each other. 

Before we continue with the introduction of the additional boost generators $\asymL$ 
we note that the coproduct of the simple-root generators \eqref{eq:coprodual1} 
now also has an exponential in  $H_\aut^*$ appearing in the \emph{right} tensor factor. 
This is due to \eqref{eq:CartanAut} and our choice of PBW basis \eqref{eq:extendedPBW} 
\begin{align}
\label{eq:Ecoprodual}
\copro E_j^* =E_j^*\otimes \exp \lrbrk{-\exta_{\aut j}H_\aut^*} +
\exp\lrbrk{\sum_{i=1}^{3}\exta_{ij}H_i^*} \otimes E_j^*.
\end{align}
%

%%%%%%%%%%%%%%%%%%%%%%%%%%%%%%%%%%%%%%%%
\paragraph{Introducing the positive boost.}

Next we introduce the additional boost generator $\asymL$ to our Borel sub-algebra. 
We denote it with a tilde to leave the plain $L$ for a redefined version of it later.
The strategy that we are following is this:
The Hopf structure is completely fixed once we know the coproduct of all generators and their duals. 
However, we do not immediately know what the coproduct of the new element $\asymL$ should be. 
Also the coproduct of $P^*$ calculated above might get additional terms through the introduction of $\asymL^*$. 
So in order to find the coproduct of $\asymL$ and $P^*$ we first consider the commutators. 
Only commutators that produce a single factor of $\asymL^*$ or $P$ 
\foreign{i.e.}\ $[\cdot,\cdot]=\asymL^*+\ldots$ or $[\cdot,\cdot]=P+\ldots$ 
can give rise to contributions of the coproduct of $\asymL$ and $P^*$, respectively.
Therefore, we first focus on commutators of such form and try to determine them. 
The requirement to leave the $\alg{sg}$ sub-algebra unchanged 
indeed fixes them so far that we only need to use 8 parameters to make the most general ansatz. 
Subsequently, we can calculate the coproduct of all generators 
and their dual generators. This fixes also all other commutators.
Finally the parameters are constrained by the requirement of compatibility between coproduct and commutators. 

In the case of non-trivial $P$ 
we have shown above that the dual quartic Serre relation remains trivial, 
$\comm{E_{12}^*}{E_{32}^*} =0$. 
In order to accommodate for the momentum extension in the dual,
we modify this relation by the dual generator $\asymL^*$
in analogy to the definition of the momentum 
$P=\comm{E_{12}}{E_{32}}$ in \eqref{eq:defP-PBW}
\[
\label{eq:DefL*}
\bigcomm{E_{12}^*}{E_{32}^*}= (q-q^{-1}) \asymL^* .
\]
Here, we fixed the prefactor corresponding to a rescaling of $\asymL^*$ 
for later convenience. 
From this new relation and the coproduct of the $E^*_i$ in \eqref{eq:Ecoprodual}, 
the coproduct of $\asymL^*$ follows straight-forwardly
\[\label{eq:coproL*}
\copro \asymL^* = \asymL^*\otimes e^{-2 H_\aut^*} + 1\otimes \asymL^*.
\]
Equivalently, the commutators with the Cartan sub-algebra 
follow from \eqref{eq:HAalmostcentral} and \eqref{eq:dualCartan2} as
\begin{align}
\bigcomm{H_\aut^*}{\asymL^*} &= 0,
&
\bigcomm{H_i^*}{\asymL^*} &= -2\hbar c_i \asymL^*,
\quad i=1,2,3.
\end{align}
However, we have some freedom to modify the
commutators of the Cartan sub-algebra with $P^*$ given in \eqref{eq:PstartComRel3} 
along with $\comm{H_\aut^*}{P^*}=0$
by the introduction of $\asymL^*$ as follows%
\begin{align}
\label{eq:H*P*}
\bigcomm{H_\aut^*}{P^*} &= \hbar\beta_\aut \asymL^*,
&
\bigcomm{H_i^*}{P^*} &= -2\hbar c_i P^*+\hbar\beta_i \asymL^*,
\quad i=1,2,3.
\end{align}
The four new parameters $\beta_{\aut,1,2,3}$ parametrize our ignorance.
One could also allow for additional product terms such as $\asymL^* H_j^{*n}$. 
They, however, will not affect the calculation of the coproduct,
and, eventually, consistency of the Hopf structure will rule them out. 

Similarly, we can now construct some of the algebra relations of $\asymL$
which will be needed for the coproduct $\copro P^*$.
From the dual coproduct $\copro \asymL^*$ in \eqref{eq:coproL*}
the algebra relations of the Cartan sub-algebra with $\asymL$ are determined to some extent
by dualization \eqref{eq:dualcoproduct}. 
In analogy to \eqref{eq:H*P*} we can extend the resulting relations by the introduction of $P$
\begin{align}
\label{eq:HL}
\bigcomm{H_\aut}{\asymL} &= 2\asymL + \alpha_\aut P,
&
\bigcomm{H_i}{\asymL} &= \alpha_i P,
\quad i=1,2,3.
\end{align}
This adds four more free parameters $\alpha_{\aut,1,2,3}$ to our algebra.

At this stage, the remaining coproducts $\copro P^*$ and $\copro \asymL^*$ are fixed
from the algebra relations.
Note that we do not yet know the relation $\comm{\asymL}{E_j}$ and $\comm{P^*}{E^*_j}$,
but due to the weights of the involved generators
we know that they cannot contain a term proportional to the basis elements $P$ and $\asymL^*$.
By dualization \eqref{eq:dualcoproduct} of the above algebra relations we obtain
\begin{align}\label{eq:copro_asymL}
\copro \asymL&= \asymL\otimes 1 +q^{-2C}\otimes \asymL - \hbar\beta_\aut P\otimes H_\aut
+ \hbar\sum_{i=1}^3 \beta_i H_iq^{-2C}\otimes P 
\nonumber\\ &\qquad 
+   q(q-q^{-1})\bigsbrk{E_{132} +(q-q^{-1}) E_{32}E_{1}}q^{-H_2} \otimes E_2 
\nonumber\\ &\qquad 
 - (q-q^{-1}) E_{32}q^{-H_1-H_2} \otimes E_{12}- q^{-1}(q-q^{-1})^2 E_{3} q^{-H_1 - 2 H_2} \otimes E_2 E_{12}
,\\
\copro P^* &=P^*\otimes e^{-2H_\aut^*} + 1 \otimes P^* 
-\alpha_\aut \asymL^* \otimes H^*_\aut  e^{-2H_\aut^*} +\sum_{i =1}^3 \alpha_i H_i^* \otimes \asymL^* 
\nonumber \\ &\qquad
 +\bigbrk{qE_{132}^*  -E_{32}^*E_1^*}e^{H_3^*-H_1^*} \otimes e^{-H_\aut^*} E_2^* 
\nonumber \\ &\qquad
-E_{32}^*e^{H_1^* -H_2^* + H_3^*} \otimes e^{-H_\aut^*}E_{12}^*
 -E_3^*e^{2H_3^*-H_2^*} \otimes  E_2^*E_{12}^*
,
\end{align}
where the latter equation extends the relation \eqref{eq:DP*_1}.
This completes the structure of the coalgebras. 

The remaining algebra relation follow by dualizing once more
\begin{align}
\label{eq:LEj}
\bigcomm{\asymL}{E_j}&=\lrsbrk{\delta_{j2}\sum_{i=1}^3 \exta_{i2}\beta_i+(\delta_{j1}+\delta_{j3})\alpha_j  } \hbar P E_j
+\delta_{j3}q (q-q^{-1}) E_{32}E_{132},
\\
\bigcomm{P^*}{E_j^*}&= \lrsbrk{\delta_{j2}(\beta_\aut - \alpha_2) 
-(\delta_{j1}+\delta_{j3})\sum_{i=1}^3 \exta_{ij}\beta_i} \hbar \asymL^*E_j^* + \delta_{j3}(q-q^{-1}) E_{32}^*E_{132}^*.
\end{align}
Let us also derive two noteworthy commutators
\begin{align}
\bigcomm{\asymL}{P}&=\hbar\lrsbrk{\sum_{i=1}^3 2 \exta_{i2}\beta_i+\alpha_1+\alpha_3 -\hbar^{-1}(q-q^{-1})} P^2,
\\
%\bigcomm{P^*}{\asymL^*}&=\hbar\lrsbrk{\beta_\aut- \half \sum_{i=1}^3 \alpha_i c_i}\asymL^{*2}.
\bigcomm{P^*}{\asymL^*}
&=\hbar\lrsbrk{2\beta_\aut-\sum_{i=1}^3 (\exta_{i1}+\exta_{i3})\beta_i-2\alpha_2 -\hbar^{-1}(q-q^{-1})}\asymL^{*2}.
\end{align}

It remains to be seen if our ansatz (in terms of 8 parameters $\alpha_i,\beta_i$) 
gives indeed a consistent Hopf algebra, 
\foreign{i.e.}\ we need to check the compatibility of product and coproduct. 
In particular, compatibility of the commutators $[\asymL,E]$ and $[P^*,E^*]$ with the coproduct
induce relations between the parameters $\alpha$ and $\beta$. 
By considering the terms $\asymL\otimes E, E\otimes \asymL$ and $P^*\otimes E^*, E^*\otimes P^*$ 
we find that the coproduct is only compatible with the commutators if
\begin{align}\label{eq:constraint}
\delta_{j2} \hbar^{-1}(q-q^{-1}) + \alpha_j &=\sum_{i=\aut,1,2,3} \exta_{ij} \beta_i , \qquad j=1,2,3.
\end{align}
Thus, we find that the extended Hopf structure is consistent 
if and only if \eqref{eq:constraint} is satisfied. 
This provides three constraints leaving us with five free parameters. 
This concludes the construction of the Hopf relations 
of the added boost generators in terms of 8 parameters $\alpha_i,\beta_j$. 
Before we continue to construct 
the quantum double of the enlarged Borel sub-algebra 
let us try to understand the parameters of our ansatz.  

%%%%%%%%%%%%%%%%%%%%%%%%%%%%%%%%%%%%%%%%
\paragraph{Presentations and deformations.}

We have derived a quantum algebra for $\alg{b}^+$ 
along with its dual in terms of 8 additional parameters 
$\alpha_{\aut,1,2,3}$, $\beta_{\aut,1,2,3}$
subject to 3 constraints.
Let us investigate the deformations of the algebra and of its presentation
along the lines of \secref{sec:presdef}
in order to understand the roles of these parameters better.

We can perform similarity transformations by conjugating with
exponentiated Cartan elements. 
As usual these merely rescale the generators by numerical factors,
and change neither the algebra relations nor their presentation. 
One noteworthy similarity transformation is by $e^{\epsilon C}$:
It leaves all elements unchanged, but transforms $\asymL$ according to 
\[
e^{\epsilon C} \asymL e^{-\epsilon C} = \asymL + \epsilon \comm{C}{\asymL} 
= \asymL + \epsilon \bigsbrk{\beta_\aut-\hbar^{-1}(q-q^{-1})} P.
\]
A shift of $\asymL$ by $P$ is thus inconsequential.

Similarity transformations by exponentiated quadratic combinations
of the Cartan elements lead to symmetric twists of the coalgebra.
Most of these modify the presentation of the centrally extended sub-algebra,
and thus we do not want to consider them here.
There remains one admissible symmetric twist by $e^{\epsilon C^2/2}$
which merely transforms $\asymL$ according to
\[
e^{\epsilon C^2/2} \asymL e^{-\epsilon C^2/2} = \asymL + \epsilon C\comm{C}{\asymL} 
= \asymL + \epsilon \bigsbrk{\beta_\aut-\hbar^{-1}(q-q^{-1})} CP.
\]
This similarity transformation changes the coefficient of
$C\otimes P$ in the coproduct $\copro \asymL$
and it introduces an additional term $P\otimes C$.

The anti-symmetric twists also modify the centrally extended algebra structures,
hence it remains to consider redefinitions of the generators. 
In particular, we will focus on the redefinitions of $\asymL$
in order to preserve the centrally extended algebra manifestly:
We can rescale $\asymL$ (without rescaling $P$ at the same time). 
This transformation amounts to changing the normalization chosen 
in \eqref{eq:DefL*}.
We can also shift $\asymL$ by $E_2E_{132}$, $E_2E_{12}E_3$,
$E_2E_{32}E_1$, $E_{12}E_{32}$ or $H_\aut P$.
This changes the presentations of the Hopf algebra relations substantially.
These transformations can be used to lift the distinguished
role of $E_3$ in \eqref{eq:LEj} and instead let $E_1$ take this role.
Similarly, one can find a more democratic prescription
where $E_1$ and $E_3$ are on equal footing.
These presentations, however, will not be convenient for 
our discussion of the R-matrix, and we shall not consider them here.

Finally, we can shift the boost $H_\aut$ by the momentum $C$.
This transformation can be seen to introduce a term $P\otimes C$ in $\copro \asymL$
and at the same time shift the commutator $[H_\aut,\asymL]$ by $P$.
This transformation combines nicely with the above similarity transformation 
such that the additional contributions $C\otimes P$ in the coproduct 
$\copro \asymL$ cancel. The relevant combination reads
\begin{align}
\label{eq:presentationshiftgen}
\asymL' &= \asymL + \hbar \epsilon \beta_\aut CP,
&
H_\aut'&= H_\aut - \epsilon C.
\end{align}
Thus only the set of established parameters changes by 
\begin{align}
\label{eq:presentationshiftpar}
\beta_i' &= \beta_i+ c_i \beta_\aut\epsilon,
&
\alpha_\aut' &= \alpha_\aut+ \bigsbrk{\hbar^{-1}(q-q^{-1})-\beta_\aut}\epsilon.
\end{align}
Two Hopf algebras related by this change of parameters
are actually identical.
This fact can be used to fix one of the parameters $\beta_i$
or $\alpha_\aut$ to any convenient value,
\foreign{e.g.}\ $\alpha_\aut =0$.
\unskip\footnote{This is possible unless $\beta_\aut=\hbar^{-1}(q-q^{-1})$,
a special case which will not be of further interest to us.}

Altogether this implies that the above Borel sub-algebra algebra 
$\env_{q,\beta}(\alg{b}^+)$
can be specified in terms of the four parameters $\beta_{\aut,1,2,3}$
along with the quantum parameter $q$. The parameter
$\alpha_\aut$ merely serves as a deformation of the presentation 
and can be fixed at convenience.

%%%%%%%%%%%%%%%%%%%%%%%%%%%%%%%%%%%%%%%%
\paragraph{Duality relationship.}

Finally, let us discuss the structure of the dual algebra.
In our construction we imposed a similar set of relations 
on the algebra and its dual. Therefore it is likely that
the algebra is structurally self-dual. Indeed by identifying the
generators with the dual generators as follows
\unskip\footnote{Note that the combination of terms 
appearing in the identification of $\asymL$ is reminiscent of 
the conjugation of $P^*$ by $e^{H^*_\aut H^*_2/\hbar}$.
However, 
$ e^{-H^*_\aut H^*_2/\hbar}  P^* e^{H^*_\aut H^*_2/\hbar}
= e^{2H^*_\aut}
\sbrk{ P^* -\beta'_\aut \asymL^*H^*_2+\ldots}$
where $\beta'_\aut=-\beta_\aut+ \hbar^{-1}(q-q^{-1})$,
\foreign{i.e.}\ the two expressions are unrelated.}
\begin{align}
 H_j &\equiv 
-\frac{1}{\hbar}\sum_{i=\aut,1,2,3} \exta_{ij}H_i^*,
\qquad
j=\aut,1,2,3,
\\
\asymL&\equiv 
\frac{1}{q-q^{-1}}\, e^{2H_\aut^*} \bigsbrk{ P^* - \beta_2\, \asymL^*H_\aut^*  + \beta_\aut  \asymL^* H_2^* },
\\
 E_j &\equiv \frac{(-1)^{\delta_{j3}}}{q-q^{-1}}\, E_j^*, \qquad j=1,3,
\\
 E_j &\equiv \frac{\lambda_j}{q-q^{-1}} \, e^{H_\aut^* } E_j^*, \qquad j=2,12,32,132,
\\
 C &\equiv  -\frac{1}{\hbar}\, H_\aut^*,
\\
P&\equiv \frac{1}{q-q^{-1}}\, e^{2H_\aut^*}\asymL^*,
\end{align}
where $\lambda_2=1$, $\lambda_{12}=-q$, $\lambda_{32}=-q^{-1}$ and $\lambda_{132}=1$,
the dual Hopf algebra has the same structure as the original one. 

Even though the Hopf algebra structure is the same, 
the parameters $\beta_{\aut,1,2,3}$ and $\alpha_{\aut,1,2,3}$ 
change between the algebra and its dual according to
\unskip\footnote{One can observe that the undetermined parameter $\zeta$
in the above identification translates between the particular choices of $\alpha_\aut$ 
and $\alpha'_\aut$ in each of the algebras.}
\begin{align}
\beta'_i &= -\beta_i   
+ c_i\sbrk{-\alpha_\aut+\beta_2+\zeta\beta_\aut-\zeta \hbar^{-1}(q-q^{-1})} ,
\\
\beta'_\aut &=-\beta_\aut+\hbar^{-1}(q-q^{-1}),
\\
\alpha'_i &= - \alpha_i  - \exta_{\aut i} \hbar^{-1}(q-q^{-1}) ,
\\
 \alpha'_\aut &=  -(\beta_2 +\zeta \beta_\aut).
\end{align}
Thus the formal duality statement is 
\[\label{eq:selfduality}
\env_{q,\beta}(\alg{b}^+) \cong  
\env_{q,\beta'}(\alg{b}^+)^*.
\]
Only for the special choice of parameters
\begin{align}\label{eq:parduality}
\beta_\aut &=\frac{q-q^{-1}}{2\hbar}\,,
&
\beta_i &=  c_i (\omega+\kappa)\, \frac{q-q^{-1}}{2\hbar}\,,
\\
 \alpha_\aut &=  \omega \,\frac{q-q^{-1}}{2\hbar}\,,
&
\alpha_i &= - \exta_{\aut i} \, \frac{q-q^{-1}}{2\hbar}\,,
\end{align}
the Hopf algebra becomes self-dual.
Here, the choice $\zeta = - 2\omega-\kappa$ 
of the undetermined element $\exta_{\aut\aut}$ of the Cartan matrix 
\eqref{eq:CartanU22} 
ensures that the duality transformation 
maps between equal presentations of the algebra.
This algebra has one degree of freedom $\kappa$ 
whereas $\omega$ merely describes a degree of freedom of its presentation.

%%%%%%%%%%%%%%%%%%%%%%%%%%%%%%%%%%%%%%%%%%%%%%%%%%%%%%%%%%%%%%%%%%%%%%%%%%%%%%%%
\subsection{Doubling the extended sub-algebra}

In this section we compute the quantum double corresponding to the extended positive sub-algebra. 
We find that identifying the dual with the negative sub-algebra 
puts additional restrictions on our parameters. 
From now on we use the dual with the opposite coproduct $\env_q (\alg{b}^+)^{*\cop}$ 
as required by the quantum double construction.

%%%%%%%%%%%%%%%%%%%%%%%%%%%%%%%%%%%%%%%%
\paragraph{Cross-relations.} 

Let us first calculate the cross-relations defined by \eqref{eq:gencross}. 
The commutation relations between the generators and their duals of the original sub-algebra are
\begin{align}
[H_i^*,E_j]&=\hbar \,\delta_{ij} E_j,  
&
[H_i,E_j^*]&=-\exta_{ij}E_j^*, 
\\
[H_i,H_j^*]&=0,  
&
[E_j^*, E_i]&= \delta_{ij}\bigbrk{q^{-H_i}e^{-\exta_{\aut j}H_\aut^*} - e^{\sum_{k=1}^3 \exta_{kj}H_k^*}}.
\end{align}
The commutators between the Cartan sub-algebra and the new generators are given by
\begin{align}
[\asymL^*,H_i]&=\delta_{i \aut} 2 \asymL^*,
\\
[P^*, H_i]&=\delta_{i \aut} 2 P^* +\alpha_i \asymL^*,
\\
[H_i^*,\asymL]&=2\hbar c_i \asymL- \hbar \beta_i P \label{eq:HstarL},
\end{align}
while the remaining commutation relations are finally
\begin{align}
[E_j,\asymL^*]&=0, 
\\
[E_j,P^*]&= \delta_{j3}q^{-H_3} E_2^*E_{12}^*+\delta_{j2}(qE_{132}^*-E_{32}^*E_1^*)e^{H_3^*-H_1^*},
\\
[\asymL,E_j^*]&=\delta_{j2} q (q-q^{-1}) \bigsbrk{E_{132}+(q-q^{-1})E_{32}E_1}q^{-H_2} e^{-H_{\aut}^*}
\nonumber\\&\qquad
-\delta_{j3} (q-q^{-1}) (1-q^{-2}) e^{H_2^*-2H_3^*}E_2 E_{12},
\\
[\asymL,\asymL^*]&=1-q^{-2C} e^{-2H_\aut^*},
\\
[P,P^*]&=1-q^{-2C} e^{-2H_\aut^*},
\\
[\asymL,P^*]&=\sum_{i=1}^3\bigsbrk{ \alpha_i H_i^* - \hbar \beta_i H_i q^{-2C} e^{-2H_\aut^*}}
 +\alpha_\aut q^{-2C}H_\aut^* e^{-2H_\aut^*} - \hbar \beta_\aut H_\aut.
\end{align}
%

%%%%%%%%%%%%%%%%%%%%%%%%%%%%%%%%%%%%%%%%
\paragraph{Identification.}

We have constructed the quantum double of the enlarged Borel sub-algebra 
$\double \env_{q,\beta} (\alg{b}^+)=\env_{q,\beta}(\alg{b}^+)\otimes \env_{q,\beta}(\alg{b}^+)^{*\cop}$. 
Instead of using the dual generators we would rather like to express the double 
with the generators of the negative Borel sub-algebra. 
Indeed, for the generators of the negative Borel half of $\alg{sg}$ the 
identification with respective dual generators is straight-forwardly 
found by comparing the commutators 
and the coproduct of $\env_{q}(\alg{b}^+)^{*\cop}$ and $\env_{q}(\alg{b}^-)$
\begin{align}
F_j&:= d_i\, %(-1)^{|E^*_j|} 
\frac{E_j^*}{q-q^{-1}}\, , \qquad j=1,3
,\\
F_j&:= \frac{e^{H^*_\aut} E_{j^*}^*}{q-q^{-1}}\, ,\qquad j=2,21,23,213
,\label{eq:identF_fer} \\
K&:= \frac{e^{2H^*_\aut}\asymL^*}{q-q^{-1}} \,
.
\end{align}
So far we have not yet defined the negative boost generator $M$. Therefore we define 
it essentially as the dual generator $P^*$. On the level of the algebra this 
means that we define the negative Borel half of the maximally extended algebra $\alg{g}$ via 
\[
\env_{q,\beta}(\alg{b}^-)\cong\env_{q,\beta}(\alg{b}^+)^{*\cop}. 
\]
However, we have a certain freedom in doing so and we use this freedom to choose 
a symmetric version between both Borel halves.
\[
M:=\frac{1}{q-q^{-1}}\bigsbrk{e^{2H^*_\aut}P^* 
    + \alpha_\aut e^{2H^*_\aut}\asymL^* H^*_\aut} \label{eq:defM} ,
\]
To that end we also redefine the boost 
\[
L:=\asymL + \half (q-q^{-1}) P H_\aut
.\label{eq:redefL}
\]
The Cartan generators $\widehat{H}_j$ of the negative Borel sub-algebra 
are identified as 
\[\label{eq:idCartan}
\widehat{H}_j := \frac{1}{\hbar} \sum_{i=\aut,1,2,3} \exta_{ij}H_i^* ,\qquad j=\aut,1,2,3 .
\]
This identification explains why it was useful to introduce the $\aut$-column 
in the extended Cartan matrix \eqref{eq:CartanU22}.  
Here, the new parameter $\exta_{\aut\aut}=\zeta$ 
represents the freedom to add the momentum generator $C$ to $H_\aut$.

%%%%%%%%%%%%%%%%%%%%%%%%%%%%%%%%%%%%%%%%
\paragraph{Reduction.}

This concludes the identification 
$\env_{q,\beta}(\alg{b}^+)^{*\cop}\cong \env_{q,\beta}(\alg{b}^-)$,
and we can thus write the quantum double as 
$\double \env_{q,\beta} (\alg{b}^+)=\env_{q,\beta}(\alg{b}^+)\otimes \env_{q,\beta}(\alg{b}^-)$. 
The quantum double, however, contains two copies of the Cartan generators, 
so that we would like to identify them by quotienting out 
the respective ideal as we have seen in the $\alg{sl}(3)$ case.
This identification of the two copies of Cartan generators 
provides another constraint on the parameters $\alpha_i,\beta_i$. 
Namely for the commutators \eqref{eq:HstarL} and  \eqref{eq:HL} 
to be consistent using the identification \eqref{eq:idCartan}, we require
\[\label{eq:constraintH}
\alpha_i=- \sum_{j=\aut,1,2,3} \exta_{ji}\beta_j, \qquad i=\aut,1,2,3.
\]
This provides an additional four relations on our \foreign{a priori} $8+1$ parameters
$\alpha_i,\beta_i$ and $\zeta$.
Together with the three constraints \eqref{eq:constraint} 
we are left with two degrees of freedom. We express the family of solutions 
in terms of two free parameters $\kappa,\omega$
\begin{align}
\alpha_j&=-\delta_{j2} \frac{q-q^{-1}}{2\hbar}\,,\quad j=1,2,3, 
& 
 \alpha_\aut &= \omega\, \frac{q-q^{-1}}{2\hbar}\,,
\\
\beta_j&=c_j(\omega+\kappa)\frac{q-q^{-1}}{2\hbar}\,,\quad j=1,2,3, 
&
\beta_\aut& = \frac{q-q^{-1}}{2 \hbar}\,,
\\
\zeta &= -2\omega-\kappa.
\end{align}
This set of parameters is exactly the same set of parameters \eqref{eq:parduality} 
that is required for a self-dual Borel sub-algebra. 
Therefore, self-duality is naturally required by the identification 
of the quantum double with $\env_{q,\kappa}(\alg{g})$.
This concludes our derivation of the algebra relations for the maximally extended algebra $\alg{g}$ 
presented in \secref{sec:extalg}.

At this point it makes sense to discuss the remaining parameters.
The requirement that the Hopf structure satisfies the compatibility relation 
between coproduct and product together with the requirement 
that we can identify the two copies of the Cartan sub-algebra 
in the quantum double fixes all but two parameters 
of our ansatz $\alpha_i,\beta_i$ and $\zeta$. 
Furthermore, the redefinition in \eqref{eq:presentationshiftpar}
reduces in terms of the generators \eqref{eq:defM} and \eqref{eq:redefL} to
\[\label{eq:shiftLM}
L'= L, \qquad M' = M , \qquad H_\aut' = H_\aut -\epsilon C
\]
and acts on the remaining parameters $\omega,\kappa$ as
\[
\omega'=\omega+\epsilon,
\qquad
\kappa' = \kappa.
\]
%\rhnote{in terms of the dual generators the relevant shift is:
%\begin{align*}
%\asymL^{\omega+\epsilon}&=\asymL^\omega +\half \epsilon (q-q^{-1})PC\\
%P^{*\omega+\epsilon}&=P^{*\omega}-\epsilon \frac{q-q^{-1}}{2\hbar}\asymL^* H_\aut^*\\
%H_\aut^{\omega+\epsilon}&=H_\aut^\omega-\epsilon C\\
%H_2^{*\omega+\epsilon}&=H_2^{*\omega}+\epsilon H_\aut^*
%\end{align*}
%}
This shows that the resulting Hopf algebra has merely one
degree of freedom $\kappa$ whereas 
$\omega$ serves as a parameter of the presentation.
We can thus set $\omega$ to any desired value 
such as $\omega=0$ or $\omega=-\kappa$.

In conclusion, we have found a one-parameter family of Hopf algebras
\[
\env_{q,\kappa}(\alg{g})
=
\frac{\double\env_{q,\kappa}(\alg{b}^+)}{\spn{\widehat{H}-H}}
\]
that contain q-deformed centrally extended $\alg{sl}(2|2)$ as a Hopf sub-algebra.

%%%%%%%%%%%%%%%%%%%%%%%%%%%%%%%%%%%%%%%%%%%%%%%%%%%%%%%%%%%%%%%%%%%%%%%%%%%%%%%%
%%%%%%%%%%%%%%%%%%%%%%%%%%%%%%%%%%%%%%%%%%%%%%%%%%%%%%%%%%%%%%%%%%%%%%%%%%%%%%%%

\section{R-matrix}
\label{sec:Rmat}

Having constructed the quantum double of our extended algebra, 
we are left with the construction of the corresponding R-matrix. 
It follows from the general formula for the universal R-matrix of a quantum double
\[
\rmat=\sum_i e_i\otimes e_i^*.
\]
The above sum runs over a complete basis $\set{e_i}_i \subset \env_q(\alg{b}^+)$
and its dual basis. 

%%%%%%%%%%%%%%%%%%%%%%%%%%%%%%%%%%%%%%%%%%%%%%%%%%%%%%%%%%%%%%%%%%%%%%%%%%%%%%%%
\subsection{Basis}

In order to get a compact expression for the R-matrix
it is important to make a good choice for the basis. 
Therefore let us first briefly explain what we consider a good basis 
and whether such a basis exists for our algebra $\alg{g}$.

%%%%%%%%%%%%%%%%%%%%%%%%%%%%%%%%%%%%%%%%
\paragraph{General considerations.}

Since we are dealing with a universal enveloping algebra a convenient basis
will be of PBW type $e_1^{n_1}e_2^{n_2}\cdots e_l^{n_l}$
in terms of some generators $\set{ e_i}_{1\leq i\leq l}$.
In addition it should also satisfy that its dual basis can be expressed as
PBW type basis of the dual generators $\set{ e^*_i }_{1\leq i\leq l}$. 
In other words we would like that the pairing relation factorizes such that
\begin{equation}
 e_1^{*n_1} \cdots e_l^{*n_l}
=(-1)^{\sum_{i=1}^l\sum_{j=i+1}^l n_i n_j |e_i||e_j|} \pair{e_1^{*n_1}}{e_1^{n_1} }
\cdots \pair{ e_l^{*n_l}}{e_l^{n_l} }
\left(e_1^{n_1}\cdots e_l^{n_l}\right)^*. \label{eq:dualbasis1}
\end{equation}
The benefit is that then also the R-matrix factorizes which provides an easier expression
\begin{align}
\rmat&=\sum_{n_1,\cdots ,n_l}^{\infty}
\left(e_1^{n_1}\cdots e_l^{n_l}\right)
\otimes \left(e_1^{n_1}\cdots e_l^{n_l}\right)^*
\nonumber\\
&= \sum_{n_1}^{\infty}
\frac{e_1^{n_1}
\otimes e_1^{*n_1} }{\pair{ e_1^{*n_1} }{ e_1^{n_1} }} \cdots \sum_{n_l}^{\infty}
\frac{e_l^{n_l}
\otimes e_l^{*n_l} }{{ e_l^{*n_l} }{ e_l^{n_1} }}\,.
\label{eq:R-matrixFactorized}
\end{align}
A sufficient condition for the paring to factorize is the following:

Given the unit $1$ and $l$ generators $e_i$, $i=1,\ldots,l$ with $\counit (e_i)=0$ for all $i$. 
Define for $1\leq i \leq j \leq l$ the sets
\[
\mathcal{B}_{ij}:= \bigrset{e_i^{n_i} e_{i+1}^{n_{i+1}} \cdots e_j^{n_j} }{n_k \in \Natural_0,\;1\leq k \leq j }.
\]
%

%%%%%%%%%%%%%%%%%%%%%%%%%%%%%%%%%%%%%%%%
\paragraph{Our algebra.}

Let us now assume that $\mathcal{B}_{1l}$ is a PBW basis of $\env_q (\alg{g})$.
Furthermore assume that the Hopf structure of the generators $e_i$ 
satisfies the following conditions regarding the linear spans $\spn{ \mathcal{B}_{ij} }$:
\begin{itemize}
\item The product respects the ordering of the basis
\[\label{eq:condprod}
e_i e_j \in \spn{ \mathcal{B}_{\min (i,j) \max (i,j)} }.
\]
\item The coproduct respects the ordering of the basis
\[\label{eq:condcopro}
\copro e_i \in \spn{ \mathcal{B}_{il} } \otimes
\spn{ \mathcal{B}_{1i} } .
\]
\end{itemize}
If these conditions are met then the pairing factorizes as given by \eqref{eq:dualbasis1}.
A proof of this statement is given in \appref{app:Basis}. 

For the quantum double of the enlarged algebra constructed above we can only 
find such a basis if $\omega=\kappa=\zeta=0$. In that case our basis choice \eqref{eq:extendedPBW}
satisfies the conditions above.  
To see this, consider first the commutators 
\begin{align}
[C,L]&=- \frac{q-q^{-1}}{2\hbar}\, P,
\\
[H_\aut ,L]&=2L+\omega\, \frac{q-q^{-1}}{2\hbar}\,P.
\end{align}
They tell us that to satisfy the condition \eqref{eq:condprod} 
we have to put $P$ between $C$ and $L$ and between $H_\aut$ and $L$ 
in the ordering of the basis; the latter, however, only if $\omega\neq 0$.
Now, consider the following part of the coproduct of $\asymL$
\begin{align}
\copro \asymL&= \asymL\otimes 1 +q^{-2C}\otimes \asymL - \half (q-q^{-1})P\otimes H_\aut
+ \half(\omega+\kappa)(q-q^{-1})\, C q^{-2C}\otimes P +\ldots\,.
\end{align}
The last two terms tell us that in order to satisfy condition \eqref{eq:condcopro} 
we have to choose the ordering $H_\aut LP$ and $PLC$;
the latter of course only if $\omega+\kappa\neq0$. 
It is now immediate to see that we can only find an ordering of generators 
satisfying conditions \eqref{eq:condprod} and \eqref{eq:condcopro} 
if $\omega=\kappa=0$. In that case our choice of PBW basis \eqref{eq:extendedPBW} 
satisfies these conditions.

%%%%%%%%%%%%%%%%%%%%%%%%%%%%%%%%%%%%%%%%%%%%%%%%%%%%%%%%%%%%%%%%%%%%%%%%%%%%%%%%
\subsection{Computation}

We will first calculate the universal R-matrix for the special case $\omega=\kappa=0$. 
Later we will extend the calculation to the general case;
this will take considerably more effort,
and it will not lead to the factorized form \eqref{eq:R-matrixFactorized}.

%%%%%%%%%%%%%%%%%%%%%%%%%%%%%%%%%%%%%%%%
\paragraph{R-matrix for $\kappa=\omega=0$.}

Henceforth we set $\omega=\kappa=0$. We have explicitly
\begin{align}
&\bigl\langle H_\aut^{*m_\aut} E_2^{*n_2}E_{12}^{*n_{12}}  
\asymL^{*n_L} P^{*n_P} E_{32}^{*n_{32}} E_{132}^{*n_{132}}E_1^{*n_1}E_3^{*n_3}
H_1^{*m_1} H_2^{*m_2} H_3^{*m_3} ,
\\
&\hspace{1cm}H_\aut^{m_\aut}  E_2^{n_2}E_{12}^{n_{12}} 
\asymL^{n_L} P^{n_P} E_{32}^{n_{32}} E_{132}^{n_{132}}E_1^{n_1}E_3^{n_3}
H_1^{m_1} H_2^{m_2} H_3^{m_3}\bigr\rangle 
\nonumber
\\
=\mathord{}&(-1)^{n_2(n_{12}+n_{32}+n_{132})+n_{12}(n_{32}+n_{132})+n_{32}n_{132}}
\bigpair{ H_\aut^{*m_\aut}}{H_\aut^{m_{\aut}} } 
\bigpair{ E_2^{*n_2} }{E_2^{n_2}} \cdots \bigpair{ H_3^{*m_3} }{ H_3^{m_3} }.
\nonumber
\end{align}
We only need to renormalize the PBW basis of dual generators by appropriate prefactors.
These prefactors are straight-forwardly 
obtained by means of the pairing relations
(see also \appref{app:copro}) 
\begin{align}
\bigpair{ H_i^{*n}}{H_i^m} &= \delta_{n,m} n!,
 \\
\bigpair{ E_i^{*n}}{ E_i^m} &= \delta_{n,m} [n;q^{-\exta_{ii}}] !, \quad i=1,3,
\\
\bigpair{ E_i^{*n}}{ E_i^m } &= \delta_{n,0}\delta_{m,0}+ \delta_{n,1}\delta_{m,1} , \quad i=2,12,32,132,
\\
\bigpair{ P^{*n}}{P^m } &=\bigpair{ \asymL^{*n}}{\asymL^{ m}}= \delta_{n,m}n!.
\end{align}

From what we have learned above, the R-matrix factorizes in our choice of basis
into powers of each generator
\begin{align}
&\sum_{n=0}^\infty H_i^{n}\otimes\left(H_i^{n}\right)^*
=\exp\bigsbrk{H_i \otimes H^*_i},
&&\sum_{n=0}^\infty E_i^{n}\otimes\left(E_i^{n}\right)^*
=\exp\bigsbrk{E_i\otimes E_i^*},\quad i=2,12,32,132,
\\
&\sum_{n=0}^{\infty} P^n\otimes \left(P^n\right)^*
=\exp\bigsbrk{ P \otimes  P^*},
&&
\sum_{n_1=0}^\infty E_1^{n_1}\otimes\left(E_1^{n_1}\right)^*
=\exp_{q^{-2}}\bigsbrk{ E_1\otimes E_1^{*}},
\\
& \sum_{n=0}^{\infty} \asymL^{ n}\otimes  (\asymL^{n} )^*
=\exp\bigsbrk{\asymL \otimes  \asymL^*},
&&
\sum_{n_3=0}^{\infty} E_3^{n_3}\otimes\left(E_3^{n_3}\right)^*
=\exp_{q^2}\bigsbrk{E_3\otimes E_3^{*}}.
\end{align}
Altogether the R-matrix of the quantum double $\double \env_{q,0}(\alg{b}^+)$ is
\unskip\footnote{The exponents for the odd terms terminate after the first term, 
\foreign{e.g.}\ $\exp(E_2\otimes E_2^*)=1\otimes 1+E_2\otimes E_2^*$. 
The q-exponentials were defined in \eqref{eq:qexp}.}
\begin{align}\label{eq:Rmat}
\rmat&=
\exp\bigsbrk{H_\aut \otimes H^*_\aut}
\exp\bigsbrk{E_2\otimes E_2^*}
\exp\bigsbrk{E_{12}\otimes E_{12}^*}
\exp\bigsbrk{ \asymL \otimes  \asymL^*}
\exp\bigsbrk{ P \otimes  P^*}
\nln&\qquad \cdot
\exp\bigsbrk{E_{32}\otimes E_{32}^*}
\exp\bigsbrk{E_{132} \otimes E_{132}^*}
\exp_{q^{-2}}\bigsbrk{ E_1\otimes E_1^{*}}
\exp_{q^2}\bigsbrk{E_3\otimes E_3^{*}}
\nln&\qquad \cdot
\exp\bigsbrk{H_1 \otimes H^*_1} 
\exp\bigsbrk{H_2 \otimes H^*_2} 
\exp\bigsbrk{H_3 \otimes H^*_3}.
\end{align}
%

%%%%%%%%%%%%%%%%%%%%%%%%%%%%%%%%%%%%%%%%
\paragraph{R-matrix for $\kappa \neq0$.}

For $\kappa\neq 0$ we cannot find a PBW basis that satisfies the conditions \eqref{eq:condprod} and \eqref{eq:condcopro}. 
One can see this for instance from the coproduct of $\asymL$ \eqref{eq:copro_asymL}: 
with the $C\otimes P$ term appearing we would need to choose an ordering $P \asymL C$ 
which is in violation with the ordering $\asymL P C$ demanded by the commutator $[\asymL,C]\propto P$. 

In particular, unlike the $\kappa=0$ case, the universal R-matrix does not factorize as nicely. 
The complication arises from the pairings $\pair{ \asymL^{*n}P^{*m}}{\asymL^{k}P^{l}}$ 
that are no longer proportional to $\delta_{n,k}\delta_{m,l}$. 
One can convince oneself that the introduction of $\kappa$ 
will only affect these pairings; the part of the R-matrix involving other generators will stay the same. 
In the following we set w.l.o.g.\ $\omega=0$ 
since it can be reintroduced by a simple redefinition of generators at the end. 

The pairing of arbitrary monomials is calculated by reducing 
it to pairings of single generators using multiple times \eqref{eq:pairing1}. 
The details of this rather lengthy calculation are found in \appref{app:r-matrix}. 
At the end (combining \namedref{Lemma}{lem:3}, \namedref{Lemma}{lem:4} and \namedref{Lemma}{lem:5})
we obtain the following expression for the relevant pairing:
\[\label{eq:kappapairing}
\bigpair{ \asymL^{*n}P^{*m}}{\asymL^{k}P^{l}}=
\delta_{m-l,k-n}\charfn_{m\geq l}\,k!\,m!\,(q-q^{-1})^{m-l}f_{m-l}.
\]
Here $\charfn_A$ denotes the characteristic function
\[
\charfn_A:=\begin{cases}
1, & \text{if condition $A$ holds},\\
0, & \text{otherwise},
\end{cases}
\]
and the sequence $f_{n}$ is generated by the function
\[
f(x):=
\sum_{n=0}^{\infty}f_{n}x^{n}
=\exp\lrsbrk{-\frac{\kappa}{4\hbar}\lrbrk{-\Li_2\lrsbrk{\frac{x}{x-1}}+\log(1-x)}}.
\]

Now, to get a nice expression for the R-matrix, the next step is to express the dual basis 
--- in particular $(\asymL^n P^m)^*$ --- in terms of the PBW basis of dual generators, \foreign{i.e.}\ $\asymL^{*k} P^{*l}$. 
The pairing provides the coefficients
\[
t^a_{nm}=\bigpair{ \asymL^{*a-n}P^{*n}}{\asymL^{a-m}P^{m}}
= \charfn_{n\geq m}\,(a-m)!\,n!\left(q-q^{-1}\right)^{n-m}f_{n-m}
\]
for the expansion ($0\leq n \leq a$)
\[
\asymL^{*a-n}P^{*n}=\sum_{m=0}^{a} t^a_{nm} \bigbrk{\asymL^{a-m}P^{m}}^{*}.
\]
We used the fact that only monomials with the same total number of generators contribute, 
as can be seen from \eqref{eq:kappapairing}. Therefore the basis transformation 
is a direct sum of basis transformations of finite-dimensional subspaces labelled by $a\geq 0 $. 

However, we are actually interested in the inverse transformation:
\[
\bigbrk{\asymL^{a-n}P^{n}}^{*}=\sum_{m=0}^{a}\tilde t^a_{nm} \asymL^{*a-m}P^{*m}.
\]
As shown in \namedref{Lemma}{lem:6} the inverse $\tilde t^a_{mk}$ is given by
\[
\tilde t^a_{mk}=\charfn_{m\geq k}\frac{\tilde f_{m-k}}{\left(a-m\right)!}\,\frac{1}{k!}\bigbrk{q-q^{-1}}^{m-k},
\]
where $\tilde f_{n}$ is generated by 
\[
\tilde f(x):=\frac{1}{f(x)}=\sum_{n=0}^{\infty}\tilde f_{n}x^{n}
=
\exp\lrsbrk{\frac{\kappa}{4\hbar}\lrbrk{-\Li_2\lrsbrk{\frac{x}{x-1}}+\log(1-x)}}.
\]

We have now found all ingredients for the R-matrix. 
The parts of it that do not contain the generator $\asymL$ or $P$ 
are just the same as in the $\kappa=0$ case. 
The term involving $\asymL$ and $P$ is:
\begin{align}
\sum_{m,n=0}^{\infty}&\bigbrk{\asymL^{m}P^{n}}\otimes\bigbrk{\asymL^{m}P^{n}}^{*} 
\nln
& =\sum_{a=0}^{\infty}\sum_{n=0}^{a}\bigbrk{\asymL^{a-n}P^{n}}\otimes\bigbrk{\asymL^{a-n}P^{n}}^{*}
\nln
 & =\sum_{a=0}^{\infty}\sum_{n=0}^{a}\sum_{m=0}^{a}
\frac{\charfn_{n\geq m}\tilde f_{n-m}}{(a-n)!\,m!}\left(q-q^{-1}\right)^{n-m}\asymL^{a-n}P^{n}\otimes \asymL^{*a-m}P^{*m}
\nln
 & =\sum_{k=0}^{\infty}\frac{\bigbrk{\asymL^{k}\otimes \asymL^{*k}}}{k!}\sum_{l=0}^{\infty}\tilde f_{l}\left(q-q^{-1}\right)^{l}
\bigbrk{P^{l}\otimes \asymL^{*l}}\sum_{m=0}^{\infty}\frac{P^{m}\otimes P^{*m}}{m!}
\nln
 & =\exp\bigsbrk{\asymL\otimes \asymL^{*}}\, 
\tilde f\bigsbrk{(q-q^{-1})P\otimes \asymL^{*}}  
\exp\bigsbrk{P\otimes P^{*}}.
\end{align}
Finally, the R-matrix for $\kappa\neq 0$, $\omega=0$ is given by
\begin{align}\label{eq:R-matrixKappa}
\rmat&=
\exp\bigsbrk{H_\aut \otimes H^*_\aut}
\exp\bigsbrk{E_2\otimes E_2^*}
\exp\bigsbrk{E_{12}\otimes E_{12}^*}
\nln&\qquad \cdot
\exp\bigsbrk{ \asymL \otimes  \asymL^*}\,
\tilde f\bigsbrk{(q-q^{-1})P\otimes \asymL^{*}}
\exp\bigsbrk{ P \otimes  P^*}
\exp\bigsbrk{E_{32}\otimes E_{32}^*}
\exp\bigsbrk{E_{132} \otimes E_{132}^*}
\nln&\qquad \cdot
\exp_{q^{-2}}\bigsbrk{ E_1\otimes E_1^{*}}
\exp_{q^2}\bigsbrk{E_3\otimes E_3^{*}}
\exp\bigsbrk{H_1 \otimes H^*_1} 
\exp\bigsbrk{H_2 \otimes H^*_2} 
\exp\bigsbrk{H_3 \otimes H^*_3}.
\end{align}
The generalization to $\omega\neq 0$ is straight-forward. We will not need it here, 
and we shall do it after transforming to the basis for $\env_{q,\kappa}(\alg{g})$ 
introduced in \secref{sec:overview22}.
%%
%\begin{align}
%\rmat&=
%\exp\bigsbrk{H_\aut \otimes H^*_\aut+\omega C \otimes H^*_\aut}
%\exp\bigsbrk{E_2\otimes E_2^*}
%\exp\bigsbrk{E_{12}\otimes E_{12}^*}
%\nln&\qquad \cdot
%\exp\bigsbrk{ \asymL \otimes  \asymL^*-\frac{\omega}{2}(q-q^{-1})PC\otimes  \asymL^* }\,
%\tilde f\bigsbrk{(q-q^{-1})P\otimes \asymL^{*}}
%\exp\bigsbrk{ P \otimes  P^*+\omega \frac{(q-q^{-1})}{2\hbar}\asymL^*H_\aut^* }
%\nln&\qquad \cdot
%\exp\bigsbrk{E_{32}\otimes E_{32}^*}
%\exp\bigsbrk{E_{132} \otimes E_{132}^*}
%\exp_{q^{-2}}\bigsbrk{ E_1\otimes E_1^{*}}
%\exp_{q^2}\bigsbrk{E_3\otimes E_3^{*}}
%\nln&\qquad \cdot
%\exp\bigsbrk{H_1 \otimes H^*_1} 
%\exp\bigsbrk{H_2 \otimes H^*_2-\omega H_2 \otimes H_\aut^*} 
%\exp\bigsbrk{H_3 \otimes H^*_3}.
%\end{align}
%% 

%%%%%%%%%%%%%%%%%%%%%%%%%%%%%%%%%%%%%%%%%%%%%%%%%%%%%%%%%%%%%%%%%%%%%%%%%%%%%%%%
\subsection{Chevalley--Serre form}

Instead of the dual generators we would like to express the R-matrix 
in terms of the Chevalley--Serre generators of the negative Borel sub-algebra. 
In the identification of the fermionic generators \eqref{eq:identF_fer} 
some factors of $e^{H_\aut^*}$ appear. 
Surprisingly, these are exactly 
the factors appearing if we commute the $\exp\sbrk{H_\aut\otimes H_\aut^*}$
term from the left to the right of the R-matrix
\begin{align}
\rmat&= 
\exp\bigsbrk{E_2\otimes E_2^* e^{H_\aut^*} }
\exp\bigsbrk{E_{12}\otimes E_{12}^*e^{H_\aut^*}} 
\exp\bigsbrk{ \asymL \otimes  \asymL^*e^{2H_\aut^*}}
\tilde f\bigsbrk{(q-q^{-1})P\otimes \asymL^{*}e^{2H_\aut^*}}
\nln & \qquad\cdot
\exp\bigsbrk{ 
  P \otimes e^{2H_\aut^*} P^*
  +\sfrac{1}{2}(q-q^{-1})PH_\aut \otimes \asymL^* e^{2H_\aut^*}
  }
\nln &\qquad \cdot
\exp\bigsbrk{E_{32}\otimes E_{32}^*e^{H_\aut^*}}
\exp\bigsbrk{ E_{132} \otimes E_{132}^*e^{H_\aut^*}}
\exp_{q^{-2}}\bigsbrk{E_1\otimes E_1^{*}}
\exp_{q^2}\bigsbrk{E_3\otimes E_3^{*}}
\nln &\qquad \cdot
\exp\lrsbrk{H_1 \otimes H^*_1+H_2 \otimes H^*_2+H_3 \otimes H^*_3+H_\aut \otimes H^*_\aut}.
\end{align}
So eventually in terms of the generators of the negative Borel sub-algebra 
and the redefined $L$ we have
\begin{align}
\rmat=\mathord{}&
\exp\bigsbrk{(q-q^{-1})E_2\otimes F_2}
\exp\bigsbrk{(q-q^{-1})E_{12}\otimes F_{21}}
\nln & \cdot
\exp\bigsbrk{(q-q^{-1})L\otimes K-\half(q-q^{-1})^{2}PH_{\aut}\otimes K}
\nln & \cdot
\tilde f\bigsbrk{(q-q^{-1})^{2}P\otimes K)}
\nln & \cdot
\exp\lrsbrk{(q-q^{-1})P\otimes M+\half(q-q^{-1})^{2}PH_{\aut}\otimes K}
\nln & \cdot
\exp\bigsbrk{(q-q^{-1})E_{32}\otimes  F_{23}}
\exp\bigsbrk{(q-q^{-1})E_{132} \otimes F_{213}}
\nln &\cdot
\exp_{q^{-2}}\bigsbrk{(q-q^{-1}) E_1\otimes F_1}
\exp_{q^2}\bigsbrk{-(q-q^{-1})E_3\otimes F_3}
\nln&\cdot
\exp\bigsbrk{\half \hbar H_1 \otimes H_1
  -\half \hbar H_3 \otimes H_3
  +\hbar C \otimes H_\aut 
  +\hbar H_\aut \otimes C 
  +\hbar \kappa C\otimes C}.
\end{align}
Now this expression contains two mixed exponentials each
with an unwanted term $PH_\aut\otimes K$. 
Interestingly, the unwanted terms come with the opposite sign.
Therefore it makes sense to combine these two exponents. 
Using \namedref{Lemma}{lem:expcombine} and its inverse with appropriately chosen 
$X$ and $Y$, and $Z=\half H_\aut\otimes 1$ we arrive at 
\begin{align}
\rmat=\mathord{}&
\exp\bigsbrk{(q-q^{-1})E_2\otimes F_2}
\exp\bigsbrk{(q-q^{-1})E_{12}\otimes F_{21}}
\nln & \cdot
\exp\bigsbrk{g_1\bigbrk{(q-q^{-1})^{2}P\otimes K}(q-q^{-1})L\otimes K}
\nln & \cdot
\exp\bigsbrk{-\half(q-q^{-1})^{2}PH_{\aut}\otimes K}
\tilde f\bigsbrk{(q-q^{-1})^{2}P\otimes K)}
\exp\bigsbrk{\half(q-q^{-1})^{2}PH_{\aut}\otimes K}
\nln & \cdot
\exp\lrsbrk{g_1\bigbrk{(q-q^{-1})^{2}P\otimes K}(q-q^{-1})P\otimes M}
\nln & \cdot
\exp\bigsbrk{(q-q^{-1})E_{32}\otimes  F_{23}}
\exp\bigsbrk{(q-q^{-1})E_{132} \otimes F_{213}}
\nln &\cdot
\exp_{q^{-2}}\bigsbrk{(q-q^{-1}) E_1\otimes F_1}
\exp_{q^2}\bigsbrk{-(q-q^{-1})E_3\otimes F_3}
\nln&\cdot
\exp\bigsbrk{\half \hbar H_1 \otimes H_1
  -\half \hbar H_3 \otimes H_3
  +\hbar C \otimes H_\aut 
  +\hbar H_\aut \otimes C 
  +\hbar \kappa C\otimes C},
\end{align}
where we defined the function
\[
g_1(x):=\frac{\log(1+x)}{x}
=
\sum_{n=0}^\infty \frac{(-1)^n}{n+1}\,x^n.
\]
The conjugation of the term $P\otimes K$ with $PH_\aut\otimes K$ yields
\begin{align}
&\exp\bigsbrk{-\half(q-q^{-1})^{2}PH_{\aut}\otimes K}
\tilde f\bigsbrk{(q-q^{-1})^{2}P\otimes K}
\exp\bigsbrk{\half(q-q^{-1})^{2}PH_{\aut}\otimes K}
\nln & \quad 
= \tilde f\lrsbrk{\frac{(q-q^{-1})^{2}P\otimes K}{1\otimes 1+(q-q^{-1})^{2}P\otimes K}}
=\exp\lrsbrk{-\frac{\kappa}{4\hbar}\,g_2\bigsbrk{(q-q^{-1})^2 P\otimes K}},
\end{align}
with the definition
\[
g_2(x):=\Li_2(-x)+\log(1+x)
=
\sum_{n=2}^\infty (-1)^{n+1}\,\frac{n-1}{n^2}\,x^n.
\]
It follows immediately from exponentiating the adjoint action
\[
\ad({PH_\aut\otimes K})^{n} (P\otimes K)=2^{n}n!\,P^{n+1}\otimes K^{n+1}.
\]
The R-matrix now takes the compact form
\begin{align}\label{eq:FinalRmat}
\rmat=\mathord{}&
\exp\bigsbrk{(q-q^{-1})E_2\otimes F_2}
\exp\bigsbrk{(q-q^{-1})E_{12}\otimes F_{21}}
\nln & \cdot
\exp\lrsbrk{g_1\bigsbrk{(q-q^{-1})^2 P\otimes K}(q-q^{-1})\bigbrk{P\otimes M+L\otimes K}
  -\frac{\kappa}{4\hbar}\,g_2\bigsbrk{(q-q^{-1})^2 P\otimes K}}
\nln & \cdot
\exp\bigsbrk{(q-q^{-1})E_{32}\otimes  F_{23}}
\exp\bigsbrk{(q-q^{-1})E_{132} \otimes F_{213}}
\nln &\cdot
\exp_{q^{-2}}\bigsbrk{(q-q^{-1}) E_1\otimes F_1}
\exp_{q^2}\bigsbrk{-(q-q^{-1})E_3\otimes F_3}
\nln&\cdot
\exp\bigsbrk{\half \hbar H_1 \otimes H_1
  -\half \hbar H_3 \otimes H_3
  +\hbar C \otimes H_\aut 
  +\hbar H_\aut \otimes C 
  +\hbar \kappa C\otimes C}.
\end{align}
We have verified the $\kappa$-dependence explicitly by 
means of the quasi-cocommutativity relation 
\eqref{eq:intertwine}. Note that some commutations
of exponents induce a derivative of $g_2$ which
cancels against a contribution from $g_1$ using the relation $g_2'=-g_1+(1+x)^{-1}$.

%%%%%%%%%%%%%%%%%%%%%%%%%%%%%%%%%%%%%%%%
\paragraph{R-matrix for $\omega\neq 0$.}

With the redefinition of $H_\aut$ in \eqref{eq:shiftLM}
we can reintroduce a non-trivial $\omega$ from the case $\omega=0$:
This is achieved by simply replacing $H_\aut \to H_\aut +\omega C$ in 
\eqref{eq:FinalRmat} leading to the fully complete universal R-matrix
\begin{align}
\rmat=\mathord{}&
\exp\bigsbrk{(q-q^{-1})E_2\otimes F_2}
\exp\bigsbrk{(q-q^{-1})E_{12}\otimes F_{21}}
\nln & \cdot
\exp\lrsbrk{g_1\bigsbrk{(q-q^{-1})^2 P\otimes K}(q-q^{-1})\bigbrk{P\otimes M+L\otimes K}
  -\frac{\kappa}{4\hbar}\,g_2\bigsbrk{(q-q^{-1})^2 P\otimes K}}
\nln & \cdot
\exp\bigsbrk{(q-q^{-1})E_{32}\otimes  F_{23}}
\exp\bigsbrk{(q-q^{-1})E_{132} \otimes F_{213}}
\nln &\cdot
\exp_{q^{-2}}\bigsbrk{(q-q^{-1}) E_1\otimes F_1}
\exp_{q^2}\bigsbrk{-(q-q^{-1})E_3\otimes F_3}
\nln&\cdot
\exp\bigsbrk{\half \hbar H_1 \otimes H_1
  -\half \hbar H_3 \otimes H_3
  +\hbar C \otimes H_\aut 
  +\hbar H_\aut \otimes C 
  +\hbar (\kappa+2\omega) C\otimes C}.
\end{align}
Note that the combination $-(\kappa+2\omega)$ is just 
the variable element $\exta_{\aut\aut}=\zeta$ of the extended Cartan matrix.
As such the terms on the latter line are precisely the
quadratic combination of the Cartan sub-algebra specified
by the inverse extended Cartan matrix.

%%%%%%%%%%%%%%%%%%%%%%%%%%%%%%%%%%%%%%%%%%%%%%%%%%%%%%%%%%%%%%%%%%%%%%%%%%%%%%%%
%%%%%%%%%%%%%%%%%%%%%%%%%%%%%%%%%%%%%%%%%%%%%%%%%%%%%%%%%%%%%%%%%%%%%%%%%%%%%%%%

\section{Classical limit}
\label{sec:classical}

Let us finally consider the classical limit $\hbar \rightarrow 0$. 
Our algebra $\env_{q,\kappa}(\alg{g})$ admits a well-defined $\hbar\rightarrow0$ limit and
we will consider the leading and the sub-leading order. 
To leading order, the algebra should simply 
reduce to the Lie superalgebra $\alg{g}$. 
In particular, we will see that the Lie superalgebra $\alg{g}$ does not depend on $\kappa$. 
The effects of the quantum deformation and $\kappa$ 
are seen in the next-to-leading order. At this order, our algebra $\env_{q,\kappa}(\alg{g})$
reduces to a Lie bialgebra with an interesting cobracket and classical r-matrix.

%%%%%%%%%%%%%%%%%%%%%%%%%%%%%%%%%%%%%%%%
\paragraph{Lie algebra.}

First, let us consider the commutation relations when $\hbar \rightarrow 0$.
The boost generators should form a standard $\alg{sl}(2)$ sub-algebra
whose commutation relations were already specified in \cite{BAna}.

Taking the classical limit of the commutation relations specified in \secref{sec:extalg} is 
straightforward and the the non-trivial commutation relations are given by
\begin{align}\label{eq:autclassical}
[H_\aut,L] &= 2L + \omega P,
&
[H_\aut,M] &= -2M -\omega K,
&
[L,M] &= -H_\aut - (\kappa+\omega) C,
\end{align}
together with
\begin{align}
[H_\aut, H_2 ] &= 0, 
&
[L, H_2 ] &= P  ,
&
[M, H_2 ] &= -K,
\\
[H_\aut, E_2 ] &= E_2 ,
&
[L, E_2 ] &= 0  ,
&
[M, E_2 ] &= F_{213},
\\
[H_\aut, F_2 ] &= -F_2 ,
&
[L, F_2 ] &= E_{132} ,
&
[M, F_2 ] &= 0,
\\
[H_\aut, P ] &= 2P, 
&
[L, P ] &= 0 , 
&
[M, P] &= -2C,
\\
[H_\aut, K ] &= -2K, 
&
[L, K ] &= 2C,  
&
[M, K] &= 0,
\end{align}
where, of course, $E_{132}$ and $F_{213}$ are understood as the classical limit
of \eqref{eq:defE123} and its analogue in the negative Borel sub-algebra. It is easy
to see that these relations agree with \cite{BAna} in the case $\kappa=\omega=0$.

The parameters $\kappa$ and $\omega$ only appear in then the commutation relations of the boost generators
\eqref{eq:autclassical}. They can be completely absorbed by the redefinition,
\foreign{cf.}\ \eqref{eq:omegaredef,eq:kapparedef}
\begin{align}\label{eq:redefclassical}
&H_\aut \rightarrow H_\aut + \omega C,
&& L \rightarrow L +\sfrac{1}{4}\kappa P,
&& M \rightarrow M +\sfrac{1}{4}\kappa K.
\end{align}
Finally, the coproducts of the boost generators trivialize
\begin{align}
\copro L &= L\otimes 1 + 1 \otimes L,
&
\copro M &= M\otimes 1  + 1 \otimes M.
\end{align}
Thus, we see that the algebra relations can be made $\kappa$ independent and
the algebra simply reduces to $\alg{g}$.

%%%%%%%%%%%%%%%%%%%%%%%%%%%%%%%%%%%%%%%%
\paragraph{Lie bialgebra.}

To study the effects of the quantization, 
we associate a quasi-triangular Lie bialgebra to our one-parameter family
of Hopf algebras. We will now work to first order in $\hbar$ 
and introduce the cobracket $\delta$ and classical r-matrix $r$
\begin{align}
\copro J- \copro^\cop J &= : 2 \hbar\, \delta(J) + \order(\hbar^2), 
\\
\rmat &= :1 + 2 \hbar\, r + \order(\hbar^2).
\end{align}
The cobrackets of the boost operators then directly follow from their coproducts
\begin{align}
%\delta(L) =&\,  (L-\half\kappa P) \wedge C - \half P\wedge(H_\aut + \omega C) 
\delta(L) =&\,  L \wedge C - \half P\wedge H_\aut -\half(\kappa+\omega) P \wedge C
+ E_{132} \wedge E_2 - E_{32}\wedge E_{12},
\\
%\delta(M) =&\, (M-\half\kappa K) \wedge C - \half K\wedge(H_\aut + \omega C)
\delta(M) =&\, M\wedge C - \half K\wedge H_\aut -\half(\kappa+\omega) K \wedge C
- F_{213} \wedge F_2 + F_{23}\wedge F_{21} .
\end{align}
Notice that the redefinition \eqref{eq:redefclassical} will only eliminate $\omega$ but not $\kappa$. 
In particular, one can make either the cobracket or the commutation relations independent of $\kappa$.

Similarly, one can derive the classical r-matrix directly from \eqref{eq:Rmat}
\begin{align}
r &=
 E_1 \otimes F_1 
+ E_2 \otimes F_2 
- E_3 \otimes F_3 
+ E_{32} \otimes F_{23} 
+ E_{12} \otimes F_{21} 
+ E_{132}\otimes F_{213}
\nonumber\\ &\qquad
+ P\otimes M 
+ L\otimes K 
+ \half  (\kappa+2\omega)  C\otimes C
\nonumber\\ &\qquad
+ \half C \otimes H_\aut 
+ \half H_\aut \otimes C 
+ \sfrac{1}{4} H_1\otimes H_1 
- \sfrac{1}{4} H_3\otimes H_3
. 
\end{align}
Again the $\omega$-dependence can be cancelled by \eqref{eq:redefclassical}. 
It satisfies the classical Yang--Baxter equation
\begin{align}
[r_{12},r_{13}] + [r_{12},r_{23}]  + [r_{13},r_{23}]  = 0.
\end{align}
Moreover, the r-matrix generates the cobracket via the so-called coboundary condition
\begin{align}
[J\otimes 1 + 1 \otimes J ,r] = \delta(J),
\end{align}
which is easily checked through direct computation.

%%%%%%%%%%%%%%%%%%%%%%%%%%%%%%%%%%%%%%%%%%%%%%%%%%%%%%%%%%%%%%%%%%%%%%%%%%%%%%%%
%%%%%%%%%%%%%%%%%%%%%%%%%%%%%%%%%%%%%%%%%%%%%%%%%%%%%%%%%%%%%%%%%%%%%%%%%%%%%%%%

\section{Conclusions and discussion}
\label{sec:conc}

In this paper we considered Drinfeld's quantum double construction for 
q-deformed centrally extended $\alg{psl}(2|2)$. 
We find that the dual elements corresponding to the 
central extensions are not central in the dual algebra. We are therefore 
led to the introduction of a new set of boost generators that form an 
$\alg{sl}(2)$ algebra to serve as the duals of the central extensions. 
By adjoining these generators to centrally extended
$\alg{psl}(2|2)$ we form a novel algebra which we call maximally extended $\alg{psl}(2|2)$. 
This algebra is defined as the smallest Hopf algebra that contains centrally extended 
$\alg{psl}(2|2)$ as a sub-algebra and that can be written as a double. 
These requirements lead to the algebra
\begin{align}
\env_{q,\kappa}\bigbrk{\alg{sl}(2) \ltimes \alg{psl}(2|2)\ltimes \Complex^3},
\end{align}
which depends on a free parameter $\kappa$. For convenience, its defining relations 
are summarized in \secref{sec:overview22}, in particular in \secref{sec:extalg}. 

This novel algebra displays a number of exciting features
that are not present for standard quantum algebras, 
\foreign{cf.}\ \secref{sec:features}.
For example, we observe the appearance of plain factors of $\hbar = \log q$ and 
parts of the extended algebra are actually not q-deformed.
Nevertheless, maximally extended 
$\alg{psl}(2|2)$ can be written as a quantum double and thus 
it has a universal R-matrix \eqref{eq:FinalRmat}.  
It turns out that, just like the maximally extended algebra, 
the R-matrix also displays some peculiar features. In particular,
it has a non-trivial functional form involving a dilogarithm function. 
Curiously, the R-matrix does not factorize into products of exponentials.
We have also computed the corresponding classical r-matrix, 
which yields a novel solution of the classical Yang--Baxter equation.

This is a first rigorous derivation of a universal R-matrix 
which is related to centrally extended $\alg{psl}(2|2)$. 
Yet the R-matrix \eqref{eq:FinalRmat} is not the universal R-matrix that describes 
the one-dimensional Hubbard model or the AdS/CFT integrable system.
Nevertheless, it should provide a first important 
step in the construction of the universal R-matrix of these models. 
In particular, for the Hubbard and AdS/CFT integrable models, the central extensions 
are identified with one braiding generator that deforms the coproduct. 
Moreover, these models also admit Yangian or quantum affine extensions, 
which we have not considered in the current paper.

Finally, the representation theory of this algebra is unexplored. It is not 
clear what kind of representations it admits. 
For instance, a (minimal) finite-dimensional representation
could be applied in the construction of transfer matrices
and the algebraic Bethe ansatz.
However, such a representation could not be unitary 
due to the structure of the algebra.
For purposes of physics, it would therefore be equally important to work out 
some unitarizable infinite-dimensional representation.
Last but not least, it would be useful to find some physical model
that exhibits the maximally extended algebra as a symmetry and for 
which the R-matrix would certainly play an important role.

%%%%%%%%%%%%%%%%%%%%%%%%%%%%%%%%%%%%%%%%
\paragraph{Acknowledgments.}

We would like to thank A.\ Torrielli and T.\ McLoughlin for useful discussions.
The work of NB and MdL is partially supported by grant no.\ 200021-137616 from the Swiss National Science Foundation
and through the NCCR SwissMAP.
The work of NB and RH is partially supported by grant no.\ 615203 from the European Research Council under the FP7.
MdL was also supported by FNU through grant number DFF--1323--00082.

%%%%%%%%%%%%%%%%%%%%%%%%%%%%%%%%%%%%%%%%%%%%%%%%%%%%%%%%%%%%%%%%%%%%%%%%%%%%%%%%
%%%%%%%%%%%%%%%%%%%%%%%%%%%%%%%%%%%%%%%%%%%%%%%%%%%%%%%%%%%%%%%%%%%%%%%%%%%%%%%%
%%%%%%%%%%%%%%%%%%%%%%%%%%%%%%%%%%%%%%%%%%%%%%%%%%%%%%%%%%%%%%%%%%%%%%%%%%%%%%%%

\appendix

%%%%%%%%%%%%%%%%%%%%%%%%%%%%%%%%%%%%%%%%%%%%%%%%%%%%%%%%%%%%%%%%%%%%%%%%%%%%%%%%
%%%%%%%%%%%%%%%%%%%%%%%%%%%%%%%%%%%%%%%%%%%%%%%%%%%%%%%%%%%%%%%%%%%%%%%%%%%%%%%%

\section{Serre relations and coproduct of \texorpdfstring{$P^*$}{P*}} \label{app:copro}

Here we list the explicit derivation of the terms in the quartic Serre relation
\begin{align}
E_1^*E_2^*E_3^*E_2^* & =
q(q-q^{-1})^2\left(E_2E_{132}\right)^*
+(q-q^{-1})^2\left(E_{12}E_{32}\right)^*
\nln&\qquad
-q(q-q^{-1})\left(E_2E_{32}E_1\right)^*
,\\
E_3^*E_2^*E_1^*E_2^* & =
q^{-1}(q-q^{-1})^2\left(E_2E_{132}\right)^*
-(q-q^{-1})^2\left(E_{12}E_{32}\right)^*
\nln&\qquad
+q^{-1}(q-q^{-1})\left(E_2E_{12}E_3\right)^*
,\\
E_2^*E_1^*E_2^*E_3^* & =
q(q-q^{-1})\left(E_2E_{12}E_3\right)^*
,\\
E_2^*E_3^*E_2^*E_1^* & =
-q^{-1}(q-q^{-1})\left(E_2E_{32}E_1\right)^*
,\\
E_2^*E_1^*E_3^*E_2^* & =
(q-q^{-1})^2\left(E_2E_{132}\right)^*
+(q-q^{-1})\left(E_2E_{12}E_3\right)^*
\nln&\qquad
-(q-q^{-1})\left(E_2E_{32}E_1\right)^*
.
\end{align}
The derivation of the coproduct for $P^*$ in \eqref{eq:DP*_1} is as follows
\begin{align}
\copro P^* & =\sum_{n_1,n_2,n_3=0}^{\infty}\left\{ -\left[\prod_{i=1}^3\left(\exta_{i1}+\exta_{i2}\right)^{n_i}\right]
\left(E_{32}\prod_{i=1}^3H_i^{n_i}\right)^*\otimes E_{12}^*
\right.
\nln&\qquad\qquad
+q\left[\prod_{i=1}^3\exta_{i2}^{n_i}\right]
\left(E_{132}\prod_{i=1}^3H_i^{n_i}\right)^*\otimes E_2^*
\nln& \qquad\qquad
+\left[\prod_{i=1}^3\left(\exta_{i1}+2\exta_{i2}\right)^{n_i}\right]
\left(E_3\prod_{i=1}^3H_i^{n_i}\right)^*\otimes\left(E_2E_{12}\right)^*
\nln&\qquad\qquad
 \left.
-\left[\prod_{i=1}^3\exta_{i2}^{n_i}\right]\left(E_{32}E_1\prod_{i=1}^3H_i^{n_i}\right)^*\otimes E_2^*\right\}
\nonumber \\&\qquad
 +P^*\otimes1+1\otimes P^*
\nonumber \\&
 =\sum_{n_1,n_2,n_3=0}^{\infty}\left\{ -E_{32}^*\prod_{i=1}^3
\frac{\left(\exta_{i1}+\exta_{i2}\right)^{n_i}}{n_i!}\left(H_i^*\right)^{n_i}\otimes E_{12}^*
+qE_{132}^*\prod_{i=1}^3\frac{\exta_{i2}^{n_i}}{n_i!}\left(H_i^*\right)^{n_i}\otimes E_2^*\right.
\nonumber \\& \qquad
 \left.-E_3^*\prod_{i=1}^3\frac{\left(\exta_{i1}+2\exta_{i2}\right)^{n_i}}{n_i!}
\left(H_i^*\right)^{n_i}\otimes E_2^*E_{12}^*-E_{32}^*E_1^*
\prod_{i=1}^3\frac{\exta_{i2}^{n_i}}{n_i!}\left(H_i^*\right)^{n_i}\otimes E_2^*\right\}
\nonumber \\&\qquad
 +P^*\otimes1+1\otimes P^*
\nonumber \\
& =P^*\otimes 1+ 1 \otimes P^*
-E_{32}^*e^{\sum_{i=1}^3 \left(\exta_{i1}+\exta_{i2}\right) H_i^*} \otimes E_{12}^*
\nonumber \\ &\qquad
 -E_3^*e^{\sum_{i=1}^3 \left(\exta_{i1}+2\exta_{i2}\right) H_i^*} \otimes E_2^*E_{12}^*
+(qE_{132}^*  -E_{32}^*E_1^*)e^{\sum_{i=1}^3\exta_{i2}H_i^*} \otimes E_2^*
.
\end{align}
The coproducts of powers of simple-root vectors are given by
\begin{align}
\copro E_i^n &= \sum_{k=0}^n \left[ \begin{matrix}
n\\k
\end{matrix}\,;q^{-\exta_{ii}} \right] E_i^{n-k}q^{-k H_i} 
\otimes E_i^k, \quad i=1,3 \,,\\
\copro H_i^n &= \sum_{k=0}^n \begin{pmatrix}
n\\ k
\end{pmatrix} H_i^{n-k}\otimes H_i^{k},\\
\copro P^n &= \sum_{k=0}^n \begin{pmatrix}
n \\ k
\end{pmatrix} P^{n-k}q^{-k2C}\otimes P^k,\label{eq:DPn}
\end{align}
where the q-binomial is defined via the q-numbers
\[
\left[ \begin{matrix}
n\\m
\end{matrix}\,;q \right] =\frac{[n;q]!}{[m;q]![n-m;q]!}\,.
\]

%%%%%%%%%%%%%%%%%%%%%%%%%%%%%%%%%%%%%%%%%%%%%%%%%%%%%%%%%%%%%%%%%%%%%%%%%%%%%%%%
%%%%%%%%%%%%%%%%%%%%%%%%%%%%%%%%%%%%%%%%%%%%%%%%%%%%%%%%%%%%%%%%%%%%%%%%%%%%%%%%

\section{Orthogonality condition}\label{app:Basis}

Consider a Hopf algebra with unit $1$ and $l$ generators $e_i$, $i=1,\cdots,l$ with $\counit (e_i)=0$ for all $i$. 
For any pair of integers $1\leq i \leq j \leq l$, define the sets
\[
\mathcal{B}_{ij}:= \bigrset{e_i^{n_i} e_{i+1}^{n_{i+1}} \cdots e_j^{n_j} }{ n_k \in \Natural_0,\;i \leq k \leq j },
\]
where we understand $e_k^0=1$ as the unit.
Let us assume that $\mathcal{B}_{1l}$ is a PBW basis of $\env_q (\alg{g})$.
Moreover, assume that the Hopf structure of the generators $e_i$ 
satisfies the following conditions regarding the linear spans $\spn{ \mathcal{B}_{ij} }$
\begin{itemize}
\item the product respects the ordering of the basis
\[\label{eq:assumptionprodgen}
e_i e_j \in \spn{ \mathcal{B}_{\min (i,j) \max (i,j)} },
\]
\item the coproduct respects the ordering of the basis
\[\label{eq:assumptioncoprodgen}
\copro e_i \in \spn{ \mathcal{B}_{il} } \otimes
\spn{ \mathcal{B}_{1i} }.
\]
\end{itemize}
We can then prove the following result that was used to compute the R-matrix 
(\foreign{cf.}\ equation \eqref{eq:dualbasis1}) 

%%%%%%%%%%%%%%%%%%%%%%%%%%%%%%%%%%%%%%%%
\begin{prop}\label{prop}
The two natural bases for the dual Hopf algebra $\set{(e_1^*)^{n_1}\ldots(e^*_l)^{n_l}}$ 
and $\set{(e_1^{n_1}\ldots e_l^{n_l})^*}$ are related as follows
\begin{equation}
 e_1^{*n_1} \cdots e_l^{*n_l}
=(-1)^{\sum_{i=1}^l\sum_{j=i+1}^l n_i n_j |e_i||e_j|}
 \pair{ e_1^{*n_1}}{e_1^{n_1} } \cdots \pair{ e_l^{*n_l}}{e_l^{n_l} }
\left(e_1^{n_1}\cdots e_l^{n_l}\right)^*. 
\end{equation}
In other words, 
dualizing is compatible with the product structure of the PBW basis 
satisfying \eqref{eq:assumptionprodgen} and \eqref{eq:assumptioncoprodgen}.
\end{prop}
%

%%%%%%%%%%%%%%%%%%%%%%%%%%%%%%%%%%%%%%%%
\begin{proof}
We will prove this result with four lemmas. 
The proof of \namedref{Proposition}{prop} 
is a direct consequence of \namedref{Lemma}{lemm4}.
\end{proof}

%%%%%%%%%%%%%%%%%%%%%%%%%%%%%%%%%%%%%%%%
\begin{lemma}
Properties \eqref{eq:assumptionprodgen} and \eqref{eq:assumptioncoprodgen} 
 do not just hold for generators, but for any element of the Hopf algebra
\begin{itemize}
\item the product respects the ordering of the basis. 
\[
\label{eq:assumptionprod}
a \in \spn{ \mathcal{B}_{ir} } ,\; b \in \spn{ \mathcal{B}_{js} }
\quad \Rightarrow \quad
ab \in \spn{ \mathcal{B}_{\min (i,j) \max (r,s)} },
\]
\item the coproduct respects the ordering of the basis
\[
\label{eq:assumptioncoprod}
a \in \spn{ \mathcal{B}_{ij} }
\quad \Rightarrow \quad
\copro a \in \spn{ \mathcal{B}_{il} } \otimes
\spn{ \mathcal{B}_{1j} }.
\]
\end{itemize}
\end{lemma}

%%%%%%%%%%%%%%%%%%%%%%%%%%%%%%%%%%%%%%%%
\begin{proof}
Consider two elements $a = e_i^{n_i}\cdots e_r^{n_r}\in \spn{ \mathcal{B}_{ir} } $ 
and $b = e_j^{n_j}\cdots e_s^{n_s}\in \spn{ \mathcal{B}_{js} } $. 
For $r<j$ the concatenation of both words is already in the correct order of the PBW basis and 
we immediately have $a b\in \spn{ \mathcal{B}_{is} } = \spn{ \mathcal{B}_{\min(i,j)\max(r,s)} } $. 
For $j \leq r$ however we need to commute the generators $e_j$ up to $e_{\min(r,s)}$ 
at the beginning of the second word through the generators $e_{\max(i,j)}$ up to $e_r$ at the end of the first word
\[
e_i^{n_i} \cdots e_{\max(i,j)}^{n_{\max(i,j)}} \cdots e_r^{n_r} \; e_j^{n_j} \cdots e_{\min(r,s)}^{n_{\min(r,s)}} \cdots e_s^{n_s}.
\]
Due to \eqref{eq:assumptionprodgen} the commutators satisfy 
$[e_u,e_v] \in \spn{\mathcal{B}_{uv} }$ .
So whatever is created by reordering the generators 
in the product can at most lie in $\spn{ \mathcal{B}_{\min (i,j)\max (r,s)} }$.

The statement for the coproduct follows from the fact that the coproduct is an algebra homomorphism
\[
\copro (e_i^{n_i} \cdots e_j^{n_j})= \copro (e_i)^{n_i} \cdots \copro (e_j)^{n_j}.
\]
Since for each generator the first tensor factor lies in $\spn{ \mathcal{B}_{il} }$ 
also their product lies therein due to \eqref{eq:assumptionprod}. 
Equally since the second tensor factor of the coproduct of each generator 
lies in $\spn{ \mathcal{B}_{1j} }$ also their product lies therein. 
By linearity of the (co)product the lemma follows.
\end{proof}

%%%%%%%%%%%%%%%%%%%%%%%%%%%%%%%%%%%%%%%%
\begin{lemma} \label{lem:copro}
The coproduct of each element of the PBW basis $\mathcal{B}_{1l}$, $\copro e_1^{n_1} e_{2}^{n_{2}} \cdots e_l^{n_l}$, 
contains the terms 
\[
e_1^{n_1} e_{2}^{n_{2}} \cdots e_l^{n_l} \otimes 1 + 1 \otimes e_1^{n_1} e_{2}^{n_{2}} \cdots e_l^{n_l}.
\]
Furthermore these are the only terms containing the identity in one of the tensor factors.
\end{lemma}

%%%%%%%%%%%%%%%%%%%%%%%%%%%%%%%%%%%%%%%%
\begin{proof}
From the multiplicative property of the counit and the requirement $\counit(e_i)=0$, for all $i$ 
we find first of all that the counit is zero on all elements of the PBW basis except on the unit,   
\[
\counit (e_1^{n_1} e_{2}^{n_{2}} \cdots e_l^{n_l} )=
\begin{cases} 
1,&n_1=\cdots=n_l=0,\\
0,& \text{otherwise}.
 \end{cases}
\]
Furthermore by the defining property of the counit we have the identity
\[
(e_1^{n_1} \cdots e_l^{n_l})_{(1)} \counit \bigbrk{(e_1^{n_1} \cdots e_l^{n_l})_{(2)}}
=e_1^{n_1} \cdots e_l^{n_l}= \counit \bigbrk{(e_1^{n_1}  \cdots e_l^{n_l})_{(1)}} (e_1^{n_1}  \cdots e_l^{n_l})_{(2)}.
\]
Subsequently the sum of all left tensor factors that have the unit in the right factor 
has to equal $e_1^{n_1} \cdots e_l^{n_l}$. Since the  words in $\mathcal{B}_{ij}$ 
are linearly independent there can only be the term $e_1^{n_1} \cdots e_l^{n_l} \otimes 1$. 
Equally with left/right exchanged.
\end{proof}

%%%%%%%%%%%%%%%%%%%%%%%%%%%%%%%%%%%%%%%%
\begin{lemma}
\[
\label{eq:otho1}
\pair{ e_i^{*m_i}}{ e_1^{n_1}\cdots e_l^{n_l} } = \pair{ e_i^{*m_i} }{ e_i^{n_i} } \prod_{k\neq i} \delta_{0,n_k}.
\]
\end{lemma}

%%%%%%%%%%%%%%%%%%%%%%%%%%%%%%%%%%%%%%%%
\begin{proof}
Proof by induction. The statement is true by definition of the dual basis for $m_i=0$ and $m_i=1$. 
Now assume \eqref{eq:otho1} holds for some fixed positive integer $m_i$. 
For $m_i+1$ we then find by definition of the pairing \eqref{eq:pairing1}
\[\label{eq:indstep}
\bigpair{ e_i^{*(m_i+1)} }{ e_1^{n_1}\cdots e_l^{n_l} } 
=\bigpair{e_i^{*m_i}\otimes e_i^*}{ \copro (e_1^{n_1}\cdots e_l^{n_l})}.
\]
By the induction hypothesis we know that this only has a chance to evaluate non-trivially, 
if there exists a term of the form $e_i^{k}\otimes e_i $ for some $k\in \Natural_0$ 
in the coproduct $\copro (e_1^{n_1}\cdots e_l^{n_l}) = \copro (e_1^{n_1}\cdots e_{i-1}^{n_{i-1}}) \copro e_i^{n_i} \copro( e_{i+1}^{n_{i+1}} \cdots e_l^{n_l})$. 

Let us consider the first tensor factor. Based on \eqref{eq:assumptionprod} 
we know that 
$\copro ( e_{i+1}^{n_{i+1}}\cdots  e_l^{n_l} )\in \spn{ \mathcal{B}_{i+1l} } \otimes \spn{ \mathcal{B}_{1l} }$ 
so there is no contribution of $e_i$ in the first tensor factor. 
For a non-trivial evaluation of \eqref{eq:indstep} 
only the unit  is therefore allowed in the first tensor factor, namely $1\otimes e_{i+1}^{n_{i+1}}\cdots e_l^{n_l}$.

Analogously, for the second tensor factor we have that by \eqref{eq:assumptionprod} 
$\copro (e_1^{n_1}\cdots  e_{i-1}^{n_{i-1}})\in \spn{ \mathcal{B}_{1l} } \otimes \spn{ \mathcal{B}_{1i-1} }$ 
so there is no $e_i$ in the second tensor factor. 
Thus only the term \linebreak ${e_1^{n_1}\cdots e_{i-1}^{n_{i-1}} \otimes 1}$ contributes.

Summarizing, we find the contributing parts
\[
\arraycolsep=1pt
\begin{array}{rcccl}
\copro (e_1^{n_1} \cdots e_{i-1}^{n_{i-1}})& \cdot 
&\copro e_i^{n_i} & \cdot
& \copro (e_{i+1}^{n_{i+1}} \cdots e_{l}^{n_l})\\
\multicolumn{1}{c}{ \downarrow}&
&\multicolumn{1}{c}{ \downarrow}&
& \multicolumn{1}{c}{ \downarrow} \\
(e_1^{n_1}\cdots e_{i-1}^{n_{i-1}} \otimes 1 )& \cdot 
& \copro e_i^{n_i} &\cdot
& (1\otimes e_{i+1}^{n_{i+1}}\cdots e_l^{n_l}) .
\end{array}
\]
This means that \eqref{eq:indstep} becomes
\[\label{eq:indstep2}
\bigpair{ e_i^{*m_i}}{e_1^{n_1}\cdots e_{i-1}^{n_{i-1}} (e_i^{n_i})_{(1)} }
\bigpair{ e_i^*}{ (e_i^{n_i})_{(2)} e_{i+1}^{n_{i+1}}\cdots e_l^{n_l} }.
\]
Since $\copro e_i^{n_i} \in \spn{ \mathcal{B}_{il}} \otimes \spn{\mathcal{B}_{1i}} $,
 the expressions in \eqref{eq:indstep2} are already ordered, meaning that no new terms are produced. 
Hence, all the $e_i$ terms come from $\copro e_i^{n_i}$ and due to the induction hypothesis, 
we  get that $n_{a\neq i}=0$.  In other words,
\[
\bigpair{ e_i^{*(m_i+1)}}{ e_1^{n_1}\cdots e_l^{n_l}} =
 \bigpair{ e_i^{*(m_i+1)}}{ e_i^{n_i} }
\prod_{k\neq i} \delta_{0,n_k},
\]
which completes the proof.
\end{proof}

%%%%%%%%%%%%%%%%%%%%%%%%%%%%%%%%%%%%%%%%
\begin{lemma}\label{lemm4}
For $1\leq i \leq l$ 
\[\label{eq:ortho2}
\pair{ e_1^{*m_1} \cdots e_i^{*m_i}}{ e_1^{n_1}\cdots e_l^{n_l} } 
=
\pair{ e_1^{*m_1} }{ e_1^{n_1} }
\cdots 
\pair{ e_i^{*m_i}}{e_i^{n_i} }
{\textstyle\prod}_{k>i} \delta_{0,n_k}.
\]
\end{lemma}

%%%%%%%%%%%%%%%%%%%%%%%%%%%%%%%%%%%%%%%%
\begin{proof}
We prove this by induction over $i$. For $i=1$ the result follows from the previous lemma. 
Now assume that for some $i, 1\leq i < l$ the statement \eqref{eq:ortho2} holds. 
For $i+1$ we have from \eqref{eq:pairing1}
\[
\bigpair{ e_1^{*m_1} \cdots e_i^{*m_i} e_{i+1}^{*m_{i+1}}}{ e_1^{n_1}\cdots e_l^{n_l} } 
=\bigpair{ e_1^{*m_1} \cdots e_i^{*m_i} \otimes e_{i+1}^{*m_{i+1}}}{ \copro (e_1^{n_1}\cdots e_l^{n_l})}.
\]
Due to the induction assumption the first tensor factor only evaluates non-trivially 
on $e_1^{k_1} \cdots e_{i}^{k_i}$ for some $k_i$. According to \eqref{eq:assumptioncoprod} 
no such term can appear in the first tensor factor 
of the coproduct 
$\copro (e_{i+1}^{n_{i+1}}\cdots e_l^{n_l} )\in \spn{ \mathcal{B}_{i+1l} } \otimes \spn{ \mathcal{B}_{1l} }$ , 
therefore only the unit is permitted in the first tensor factor of that part of the coproduct. 
\namedref{Lemma}{lem:copro} above tells us there is only one such term $1\otimes e_{i+1}^{n_{i+1}}\cdots e_l^{n_l} $.

Now considering the second tensor factor we know that it only evaluates non-trivially on $e_{i+1}^{k_{i+1}}$ 
for some $k_{i+1}$. Due to \eqref{eq:assumptioncoprod} we know that 
$\copro (e_1^{n_1} \cdots e_i^{n_i}) \in \spn{ \mathcal{B}_{1l}} \otimes \spn{ \mathcal{B}_{1i} }$ 
cannot have such a term in the second tensor factor and therefore must have the unit there. 
Again there is only one such term $e_1^{n_1} \cdots e_i^{n_i} \otimes 1$.

We have now an analogous situation to the proof of the previous lemma. The only contributing terms are
\[
\arraycolsep=1pt
\begin{array}{rcl}
\copro \bigbrk{e_1^{n_1} \cdots e_{i}^{n_{i}}} & \cdot 
& \copro \bigbrk{e_{i+1}^{n_{i+1}} \cdots e_{l}^{n_{l}}} \\
\multicolumn{1}{c}{ \downarrow}&
& \multicolumn{1}{c}{ \downarrow} \\
\bigbrk{e_1^{n_1}\cdots e_{i}^{n_{i}} \otimes 1 }  &  \cdot 
&\bigbrk{1\otimes e_{i+1}^{n_{i+1}}\cdots e_l^{n_l}}.
\end{array}
\]
Thus,
\[
\bigpair{ e_1^{*m_1} \cdots e_i^{*m_i} e_{i+1}^{*m_{i+1}}}{ e_1^{n_1}\cdots e_l^{n_l} }
=
\bigpair{  e_1^{*m_1} \cdots e_i^{*m_i} }{e_1^{n_1}\cdots e_{i}^{n_{i}} }
\bigpair{ e_{i+1}^{*m_{i+1}}}{e_{i+1}^{n_{i+1}}\cdots e_l^{n_l} }.
\]
Now using the previous lemma and the induction hypothesis we complete the proof. 
\end{proof}

%%%%%%%%%%%%%%%%%%%%%%%%%%%%%%%%%%%%%%%%%%%%%%%%%%%%%%%%%%%%%%%%%%%%%%%%%%%%%%%%
%%%%%%%%%%%%%%%%%%%%%%%%%%%%%%%%%%%%%%%%%%%%%%%%%%%%%%%%%%%%%%%%%%%%%%%%%%%%%%%%

\section{Details of the R-matrix calculation}\label{app:r-matrix}

We start by calculating the pairing step by step. 
To facilitate the calculation we will set w.l.o.g.\ $\omega=0$.

%%%%%%%%%%%%%%%%%%%%%%%%%%%%%%%%%%%%%%%%
\begin{lemma}
\begin{align}
\bigpair{ \asymL^{*n}}{P^{k}} & =\delta_{n,0}\delta_{k,0},
\\
\bigpair{ \asymL^{*n}}{\asymL^{k}} & =\delta_{n,k}n!.
\end{align}
\end{lemma}

%%%%%%%%%%%%%%%%%%%%%%%%%%%%%%%%%%%%%%%%
\begin{proof}
For $k=0$ and $k=1$ we have 
\begin{align}
\bigpair{ \asymL^{*n}}{1} 
& =\delta_{n,0}, 
& \bigpair{ \asymL^{*n}}{P} 
& =0, & \bigpair{ \asymL^{*n}}{\asymL}
& =\delta_{n,1},
\end{align}
and for $k>1$ we have 
\begin{align}
\bigpair{ \asymL^{*n}}{P^{k}} & =\sum_{a=0}^{n}\binom{n}{a}
\bigpair{ \asymL^{*n-a}}{P^{k-1}}\bigpair{ \asymL^{*a}}{P}=0,
\\
\bigpair{ \asymL^{*n}}{\asymL^{k}} & =\sum_{a=0}^{n}\binom{n}{a}\bigpair{\asymL^{*n-a}}{\asymL^{k-1}}
\bigpair{ \asymL^{*a}}{\asymL}=n\bigpair{ \asymL^{*n-1}}{\asymL^{k-1}}=\delta_{n,k}n!,
\end{align}
which follows from \eqref{eq:DPn} and
\[
\copro \asymL^{*n} =\sum_{a=0}^{n}\binom{n}{a}\asymL^{*n-a}\otimes e^{-2aH_\aut^{*}}\asymL^{*a}.
\]
\end{proof}

%%%%%%%%%%%%%%%%%%%%%%%%%%%%%%%%%%%%%%%%
\begin{lemma}
We have
\begin{align}
\pair{ \asymL^{*n}}{\asymL^{k}P^{l}} &=\delta_{l,0}\delta_{n,k}n!, 
&\pair{ \asymL^{*n} P^{*m}}{P^l }&=\delta_{n,0}\delta_{m,l}l!,
\end{align}
or equivalently
\begin{align}
\asymL^{*n} & =n!\,(\asymL^{n})^{*}, 
&P^{l}&=l!\,(P^{*l})^{*}. 
\end{align}
\end{lemma}

%%%%%%%%%%%%%%%%%%%%%%%%%%%%%%%%%%%%%%%%
\begin{proof}
\[
\bigpair{ \asymL^{*n}}{\asymL^{k}P^{l}}
=\sum_{a=0}^{n}\binom{n}{a}
 \bigpair{ \asymL^{*n-a}}{\asymL^{k}}
 \underset{\delta_{a,0}\delta_{l,0}}{\underbrace{\bigpair{ \asymL^{*a}}{P^{l}}}}
=\delta_{l,0}\bigpair{ \asymL^{*n}}{\asymL^{k}}=\delta_{l,0}\delta_{n,k}n!.
\]
\end{proof}

%%%%%%%%%%%%%%%%%%%%%%%%%%%%%%%%%%%%%%%%
\begin{lemma}
\label{lem:3}
\[
\bigpair{ \asymL^{*n}P^{*m}}{\asymL^{k}P^{l}}=
\begin{cases}
\displaystyle
\frac{k!}{(k-n)!}\,\frac{m!}{(m-l)!}
\bigpair{ P^{*m-l}}{\asymL^{k-n}}, & k\geq n \; \wedge \; m\geq l ,
\\
0 & k<n \; \vee \; m<l.
\end{cases}
\]
\end{lemma}

%%%%%%%%%%%%%%%%%%%%%%%%%%%%%%%%%%%%%%%%
\begin{proof}
\[
\bigpair{ \asymL^{*n}P^{*m}}{\asymL^{k}P^{l}}
=\bigpair{ \asymL^{*n}\otimes P^{*m}}{\copro (\asymL^{k}P^{l})}
=n!\,\bigpair{(\asymL^{n})^{*}\otimes P^{*m}}{\copro ( \asymL^{k}P^{l})}.
\]
To be non-zero we need exactly $\asymL^{n}$ in the left tensor factor
of $\copro (\asymL^{k}P^{l})=(\copro\asymL)^k (\copro P)^l$. Since $\asymL$ is never produced by any commutator
$\asymL^{n}$ can only come directly from the product of $n$ terms $\asymL\otimes1$, stemming from $n$ factors $\copro \asymL$,
multiplied by terms that have the identity in the left factor, i.e.
only $k-n$ terms $q^{-2C}\otimes \asymL$ from $\copro \asymL$, and $l$ terms
$q^{-2C}\otimes P$ from $\copro P$. In particular $n\leq k$, otherwise
we get zero. There are $\binom{k}{n}$ choices to pick $n$ terms $1\otimes \asymL$
from the $k$ coproducts $\copro \asymL$. Thus 
\[
\bigpair{ \asymL^{*n}\otimes P^{*m}}{\copro \asymL^{k}P^{l}}
=n!\,\binom{k}{n}\bigpair{ P^{*m}}{\asymL^{k-n}P^{l}}.
\]
Similarly, for
\[
\bigpair{ P^{*m}}{\asymL^{k}P^{l} } 
=\bigpair{ \copro P^{*m}}{\asymL^{k}\otimes P^{l}}
=l!\,\bigpair{\copro P^{*m}}{\asymL^{k}\otimes\left(P^{*l}\right)^{*}}
\]
to be non-zero we need exactly $l$ terms $1\otimes P^*$ and $m-l$ terms $P^{*}\otimes e^{-2H_{\aut}^{*}}$. 
There are $\binom{m}{l}$ choices to pick these terms form $\copro P^{*m}$. 
In particular for $l>m$ the pairing will evaluate to zero. 
\[
\pair{ P^{*m}}{\asymL^{k}P^{l}}=l!\,\binom{m}{l}\bigpair{ P^{*m-l}}{\asymL^{k}}.
\]
\end{proof}

To complete the calculation of the pairing we are left with the calculation of 
$\pair{ P^{*m}}{\asymL^n }$:

%%%%%%%%%%%%%%%%%%%%%%%%%%%%%%%%%%%%%%%%
\begin{lemma}
\label{lem:4}
\[
\bigpair{ P^{*n}}{\asymL^{m}}
=\delta_{n,m} n!\,n!\left(q-q^{-1}\right)^{n}f_{n}, 
\]
where $f_n$ is given by the recursion relation  
\begin{align}\label{eq:recursionf}
f_{0}&=1, &nf_{n} & =(n-1)f_{n-1} -\frac{\kappa}{4\hbar}
\sum_{a=0}^{n-2}\frac{f_{a}}{n-a}\,,\quad n\geq 1.
\end{align}
\end{lemma}

%%%%%%%%%%%%%%%%%%%%%%%%%%%%%%%%%%%%%%%%
\begin{proof}
To evaluate the pairing we split it into
\begin{align}
\bigpair{ P^{*n}}{\asymL^{m}} 
& =\bigpair{ P^{*}\otimes P^{*n-1}}{\copro \asymL^{m}}.
\end{align}
For a non-trivial evaluation we need to consider the parts in $\copro \asymL^m=(\copro \asymL)^m$ 
that have exactly a single $P$ in the left tensor factor.
The terms in $\copro \asymL$ that can give rise to such a term are
\[
\copro \asymL=\asymL\otimes1+q^{-2C}\otimes \asymL
-\half(q-q^{-1})P\otimes H_\aut+ \half\kappa (q-q^{-1})Cq^{-2C}\otimes P+\cdots.
\]
Now  $P\otimes\cdot$ can arise by products of these in one of three cases:
\begin{enumerate}
\item There is one term $-\half(q-q^{-1})P\otimes H_\aut$. 
Then it cannot be multiplied by any terms $\asymL\otimes1$ or $Cq^{-2C}\otimes P$, 
because they would lead to higher products $P^n$ or $PC$ on which the pairing $\pair{ P^*}{ \cdot }$ 
would evaluate to zero. Therefore non-zero contributions have to come from
\begin{align}
 &\quad \sum_{k=0}^{m-1}\left(q^{-2C}\otimes \asymL\right)^{k}
\left(-\hbar \frac{q-q^{-1}}{2\hbar}P\otimes H_\aut\right)\left(q^{-2C}\otimes \asymL\right)^{m-1-k}
\nln
 & =-\frac{q-q^{-1}}{2}\sum_{k=0}^{m-1}Pq^{-2(m-1)C}\otimes\left(H_\aut-2k\right)\asymL^{m-1}
\nln
 & \circeq\frac{q-q^{-1}}{2}2\sum_{k=0}^{m-1}k\, Pq^{-2(m-1)C}\otimes \asymL^{m-1}
\nln
 & =\frac{q-q^{-1}}{2}m(m-1)\, Pq^{-2(m-1)C}\otimes \asymL^{m-1},
\end{align}
where $\circeq$ denotes equality up to terms on which the pairing evaluates to zero.

\item There is no $-\frac{q-q^{-1}}{2}P\otimes H_\aut$ term
but one term $\hbar\beta_{2}Cq^{-2C}\otimes P$. Then there needs to be exactly one term $\asymL\otimes 1$
on the right of it to produce a $P$ in the left tensor factor. Thus we get a contribution from
\begin{align}
 & \quad\sum_{k=0}^{m-2}(k+1)\bigbrk{q^{-2C}\otimes \asymL}^{m-2-k}
\lrbrk{\half \kappa (q-q^{-1})Cq^{-2C}\otimes P}\bigbrk{\asymL\otimes1}
\bigbrk{q^{-2C}\otimes \asymL}^{k}
\nln
 & \circeq \kappa \frac{q-q^{-1}}{2}\sum_{k=0}^{m-2}(k+1)
 CLq^{-2(m-1)C}\otimes \asymL^{m-2-k}PL^{k}
\nln
 & \circeq  \frac{\kappa}{2}\sum_{k=0}^{m-2}(k+1)
 CLq^{-2(m-1)C}\otimes \asymL^{m-2-k}\sum_{a=0}^{k}\binom{k}{a}a!\,(q-q^{-1})^{a+1}\asymL^{k-a}P^{1+a}
\nln
 & \circeq -\frac{\kappa}{4\hbar}\sum_{k=1}^{m-1}\sum_{a=1}^{k}\frac{k!}{(k-a)!}(q-q^{-1})^{a+1}
 Pq^{-2(m-1)C}\otimes \asymL^{m-1-a}P^{a}
\nln
 & =-\frac{\kappa}{4\hbar}\sum_{a=1}^{m-1}\sum_{k=a}^{m-1}\frac{k!}{(k-a)!}(q-q^{-1})^{a+1}
 Pq^{-2(m-1)C}\otimes \asymL^{m-1-a}P^{a}
\nln
 & =- \frac{\kappa}{4\hbar}\sum_{a=2}^{m}\frac{m!}{(m-a)!}\,\frac{(q-q^{-1})^{a}}{a}
 Pq^{-2(m-1)C}\otimes \asymL^{m-a}P^{a-1}.
\end{align}

\item Finally if there are no $P\otimes H_\aut$ and no $Cq^{-2C}\otimes P$ terms
then we can only have contributions from
\begin{align}
 & \sum_{k=0}^{m-1}\bigbrk{q^{-2C}\otimes \asymL}^{k}\bigbrk{\asymL\otimes1}
\bigbrk{q^{-2C}\otimes \asymL}^{m-1-k}
\nln
\circeq \mathord{}& \sum_{k=0}^{m-1}q^{-2kC}Lq^{-2(m-1-k)C}\otimes \asymL^{m-1}
\nln
\circeq \mathord{}& \sum_{k=0}^{m-1}k(q-q^{-1})Pq^{-2(m-1)C}\otimes \asymL^{m-1}
\nln
= \mathord{}& \half m\,(m-1)(q-q^{-1})Pq^{-2(m-1)C}\otimes \asymL^{m-1}.
\end{align}
\end{enumerate}
Putting all together we get 
\begin{align}
\bigpair{ P^{*n}}{\asymL^{m}} 
& =m(m-1)(q-q^{-1})\bigpair{ P^{*n-1}}{\asymL^{m-1}}
\nln
&\qquad-\frac{\kappa}{4\hbar}\sum_{a=2}^{m}\frac{m!}{(m-a)!}\,\frac{(q-q^{-1})^{a}}{a}
\bigpair{ P^{*n-1}}{\asymL^{m-a}P^{a-1}}
\nln
 & =m(m-1)(q-q^{-1})\bigpair{ P^{*n-1}}{\asymL^{m-1}} 
\nln
 &\qquad-\frac{\kappa}{4\hbar}\sum_{a=2}^{m}\frac{(n-1)!}{(n-a)!}\,
\frac{m!}{(m-a)!}\, \frac{(q-q^{-1})^{a}}{a}\bigpair{ P^{*n-a}}{\asymL^{m-a}}.
\label{eq:recursion}
\end{align}
A quick induction shows that $\pair{ P^{*n}}{\asymL^{m}}\propto\delta_{n.m}$.
Define $f_{n}$ through 
\[
\bigpair{ P^{*n}}{\asymL^{n}}
=n!\,n!\left(q-q^{-1}\right)^{n}f_{n},
\]
and the recursion \eqref{eq:recursion} leads to \eqref{eq:recursionf}.
\end{proof}

%%%%%%%%%%%%%%%%%%%%%%%%%%%%%%%%%%%%%%%%
\begin{lemma}
\label{lem:5}
The sequence $f_{n}$ is generated by the function
\[\label{eq:generatingf}
f(x)=\sum_{n=0}^{\infty}f_{n}x^{n}=
\exp\lrsbrk{-\frac{\kappa}{4\hbar}\lrbrk{-\Li_2\lrsbrk{\frac{x}{x-1}} +\log(1-x)}}
.
\]
\end{lemma}

%%%%%%%%%%%%%%%%%%%%%%%%%%%%%%%%%%%%%%%%
\begin{proof}
Using the recursion relation 
\begin{align}
\frac{df}{dx}-x\,\frac{df}{dx}  & =f_{1}+\sum_{n=2}^{\infty}\bigbrk{nf_{n}-(n-1)f_{n-1}}x^{n-1}
\nln
 & =-\frac{\kappa}{4\hbar}\sum_{n=0}^{\infty}\sum_{a=0}^{n}\frac{f_{a}}{n-a+2}\,x^{n+1}
\nln
 & =-\frac{\kappa}{4\hbar} \sum_{k=0}^{\infty}\frac{x^{k+1}}{k+2}\sum_{a=0}^{\infty}f_{a}x^{a}
\nln
 & =\frac{\kappa}{4\hbar} \left(\frac{\log\left(1-x\right)}{x}+1\right)f
\end{align}
we get the differential equation
\[
(1-x)\frac{df}{dx}=\frac{\kappa}{4\hbar}
\left(\frac{\log(1-x)}{x}+1\right)f,
\]
which is solved by \eqref{eq:generatingf} for $f_0=1$.
\end{proof}

For each $a\geq 0$ and $0\leq n,m \leq a$ we can write the transformation as 
\[
\bigbrk{\asymL^{a-n}P^{n}}^{*}=\sum_{m=0}^{a}\tilde t^a_{nm}\bigbrk{\asymL^{*a-m}P^{*m}},
\]
where $\tilde t^a= (t^a)^{-1}$ is the inverse matrix of
\begin{align}
t^a_{nm}=\bigpair{ \asymL^{*a-n}P^{*n}}{\asymL^{a-m}P^{m}}
= \charfn_{n\geq m}(a-m)!\,n!\left(q-q^{-1}\right)^{n-m}f_{n-m}.
\end{align}
%

%%%%%%%%%%%%%%%%%%%%%%%%%%%%%%%%%%%%%%%%
\begin{lemma}
\label{lem:6}
The inverse $\tilde t^a_{mk}$ is given by
\[
\tilde t^a_{mk}=\charfn_{m\geq k}\frac{\tilde f_{m-k}}{(a-m)!}\,\frac{1}{k!}\left(q-q^{-1}\right)^{m-k},
\label{eq:Tinverse}
\]
where $\tilde f_{n}$ is generated by 
\[
\sum_{n=0}^{\infty}\tilde f_{n}x^{n}=\frac{1}{f(x)}=
\exp\lrsbrk{\frac{\kappa}{4\hbar}\lrbrk{-\Li_2\lrsbrk{\frac{x}{x-1}}+\log(1-x)}}.
\]
\end{lemma}

%%%%%%%%%%%%%%%%%%%%%%%%%%%%%%%%%%%%%%%%
\begin{proof}
The two series fulfill
\[
1=f(x)\frac{1}{f(x)}=\sum_{n=0}^{\infty}f_{n}x^{n}\sum_{m=0}^{\infty}\tilde f_{m}x^{m}
=\sum_{n=0}^{\infty}\sum_{m=0}^{n}f_{n-m}\tilde f_{m}x^{n},
\]
which yields the identity
\[
\sum_{m=0}^{n}f_{n-m}\tilde f_{m}=\delta_{n,0}.
\]
Now it is straightforward to check that the inverse $\tilde t^a_{mk}$
is given by \eqref{eq:Tinverse}
\begin{align}
\sum_{m=0}^{a}t^a_{nm}\tilde t^a_{mk} & =\sum_{m=0}^{a}\charfn_{n\geq m}\charfn_{m\geq k}
\frac{n!}{k!}\left(q-q^{-1}\right)^{n-k}f_{n-m}\tilde f_{m-k}
\nln
 & =\frac{n!}{k!}\left(q-q^{-1}\right)^{n-k}\sum_{m=k}^{n}f_{n-m}\tilde f_{m-k}
\nln
 & =\frac{n!}{k!}\left(q-q^{-1}\right)^{n-k}\sum_{a=0}^{n-k}f_{n-k-a}\tilde f_{a}
\nln
 & =\delta_{n,k}.
\end{align}
\end{proof}

%%%%%%%%%%%%%%%%%%%%%%%%%%%%%%%%%%%%%%%%
\begin{lemma}
\label{lem:expcombine}
For generators $X$, $Y$ and $Z$ with commutators
\begin{align}
\comm{Z}{X}&=X, 
&
\comm{Z}{Y}&=Y,
&
\comm{X}{Y}&=0.
\end{align}
the following identity holds 
\[
\exp\bigsbrk{X-YZ}
=\exp\lrsbrk{\frac{\log(1+Y)}{Y}\,X}
\exp\bigsbrk{-YZ},
\]
where the logarithmic term is defined by its series expansion 
\[
\frac{\log(1+Y)}{Y}=\sum_{n=0}^\infty \frac{(-1)^n}{n+1}Y^n.
\]
\end{lemma}

%%%%%%%%%%%%%%%%%%%%%%%%%%%%%%%%%%%%%%%%
\begin{proof}
First we derive the commutator of the composite expressions appearing here
\begin{align}
\comm{YZ}{Y^n X}&=Y \comm{Z}{Y^n} X+Y^{n+1} \comm{Z}{X}=(n+1)Y^{n+1}X, \\
\ad(YZ)^k (Y^n X)&=\frac{(k+n)!}{n!}Y^{k+1}X.
\end{align}
Note that $\comm{Y^n X}{\comm{YZ}{Y^k X}}=0$. The Baker--Campbell--Hausdorff formula reduces for this case to 
\begin{align}
&\quad\exp\lrsbrk{\sum_{n=0}^\infty \frac{(-1)^n}{n+1}\,Y^n X}
\exp\bigsbrk{-YZ}
\nln
&=
\exp\lrsbrk{-YZ
+\sum_{n,k=0}^\infty
\frac{\mathrm{B}_k (-1)^n (-1)^k}{(n+1)k!} \ad(YZ)^k(Y^n X)
}
\nln
&=
\exp\lrsbrk{-YZ
+\sum_{n,k=0}^\infty
\frac{\mathrm{B}_k(n+k)!(-1)^{n+k}}{(n+1)!\,k!}\, Y^{n+k}X
}
\nln
&=
\exp\lrsbrk{-YZ
+\sum_{n=0}^\infty \sum_{k=0}^n
\frac{\mathrm{B}_k n!(-1)^{n}}{(n-k+1)!\,k!}\, Y^{n}X
}
\nln
&=
\exp\bigsbrk{-YZ+X}.
\end{align}
Here, we have made use of a defining property 
of the Bernoulli numbers $\mathrm{B}_n$
\[
\sum_{k=0}^n
\frac{n!\,\mathrm{B}_k}{(n-k+1)!\,k!}=\delta_{n,0}.
\]
\end{proof}

%%%%%%%%%%%%%%%%%%%%%%%%%%%%%%%%%%%%%%%%%%%%%%%%
%%%%%%%%%%%%%%%%%%%%%%%%%%%%%%%%%%%%%%%%%%%%%%%%
\begin{bibtex}[\jobname]

@article{BK,
      author         = "Beisert, Niklas and Koroteev, Peter",
      title          = "{Quantum Deformations of the One-Dimensional Hubbard
                        Model}",
      journal        = "J.Phys.",
      volume         = "A41",
      pages          = "255204",
      doi            = "10.1088/1751-8113/41/25/255204",
      year           = "2008",
      eprint         = "0802.0777",
      archivePrefix  = "arXiv",
      primaryClass   = "hep-th",
      reportNumber   = "AEI-2008-003, ITEP-TH-06-08",
      SLACcitation   = "%%CITATION = ARXIV:0802.0777;%%",
}

@article{AdLTuniversal,
      author         = "Arutyunov, Gleb and de Leeuw, Marius and Torrielli,
                        Alessandro",
      title          = "{On Yangian and Long Representations of the Centrally
                        Extended su(2$/$2) Superalgebra}",
      journal        = "JHEP",
      volume         = "1006",
      pages          = "033",
      doi            = "10.1007/JHEP06(2010)033",
      year           = "2010",
      eprint         = "0912.0209",
      archivePrefix  = "arXiv",
      primaryClass   = "hep-th",
      reportNumber   = "ITP-UU-09-55, SPIN-09-45",
      SLACcitation   = "%%CITATION = ARXIV:0912.0209;%%",
}

@article{dLMRQbound,
      author         = "de Leeuw, Marius and Matsumoto, Takuya and Regelskis,
                        Vidas",
      title          = "{The Bound State S-matrix of the Deformed Hubbard Chain}",
      journal        = "JHEP",
      volume         = "1204",
      pages          = "021",
      doi            = "10.1007/JHEP04(2012)021",
      year           = "2012",
      eprint         = "1109.1410",
      archivePrefix  = "arXiv",
      primaryClass   = "math-ph",
      reportNumber   = "AEI-2011-062",
      SLACcitation   = "%%CITATION = ARXIV:1109.1410;%%",
}

@article{dLRTQSecret,
      author         = "de Leeuw, Marius and Regelskis, Vidas and Torrielli,
                        Alessandro",
      title          = "{The Quantum Affine Origin of the AdS/CFT Secret
                        Symmetry}",
      journal        = "J.Phys.",
      volume         = "A45",
      pages          = "175202",
      doi            = "10.1088/1751-8113/45/17/175202",
      year           = "2012",
      eprint         = "1112.4989",
      archivePrefix  = "arXiv",
      primaryClass   = "hep-th",
      reportNumber   = "DMUS-MP-11-02",
      SLACcitation   = "%%CITATION = ARXIV:1112.4989;%%",
}

@article{BGMaffine,
      author         = "Beisert, Niklas and Galleas, Wellington and Matsumoto,
                        Takuya",
      title          = "{A Quantum Affine Algebra for the Deformed Hubbard
                        Chain}",
      journal        = "J.Phys.",
      volume         = "A45",
      pages          = "365206",
      doi            = "10.1088/1751-8113/45/36/365206",
      year           = "2012",
      eprint         = "1102.5700",
      archivePrefix  = "arXiv",
      primaryClass   = "math-ph",
      reportNumber   = "AEI-2011-005",
      SLACcitation   = "%%CITATION = ARXIV:1102.5700;%%",
}

@article{BSclassical,
      author         = "Beisert, Niklas and Spill, Fabian",
      title          = "{The Classical r-matrix of AdS/CFT and its Lie Bialgebra
                        Structure}",
      journal        = "Commun.Math.Phys.",
      volume         = "285",
      pages          = "537-565",
      doi            = "10.1007/s00220-008-0578-2",
      year           = "2009",
      eprint         = "0708.1762",
      archivePrefix  = "arXiv",
      primaryClass   = "hep-th",
      reportNumber   = "AEI-2007-116, HU-EP-07-31",
      SLACcitation   = "%%CITATION = ARXIV:0708.1762;%%",
}

@article{BQClassical,
      author         = "Beisert, Niklas",
      title          = "{The Classical Trigonometric r-Matrix for the
                        Quantum-Deformed Hubbard Chain}",
      journal        = "J.Phys.",
      volume         = "A44",
      pages          = "265202",
      doi            = "10.1088/1751-8113/44/26/265202",
      year           = "2011",
      eprint         = "1002.1097",
      archivePrefix  = "arXiv",
      primaryClass   = "math-ph",
      reportNumber   = "AEI-2010-016",
      SLACcitation   = "%%CITATION = ARXIV:1002.1097;%%",
}

@article{HHMqdefS,
      author         = "Hoare, Ben and Hollowood, Timothy J. and Miramontes, J.
                        Luis",
      title          = "{q-Deformation of the AdS$_5$ $\times$ S$^5$ Superstring S-matrix
                        and its Relativistic Limit}",
      journal        = "JHEP",
      volume         = "1203",
      pages          = "015",
      doi            = "10.1007/JHEP03(2012)015",
      year           = "2012",
      eprint         = "1112.4485",
      archivePrefix  = "arXiv",
      primaryClass   = "hep-th",
      reportNumber   = "IMPERIAL-TP-11-BH-03",
      SLACcitation   = "%%CITATION = ARXIV:1112.4485;%%",
}

@article{HHMbound,
      author         = "Hoare, Ben and Hollowood, Timothy J. and Miramontes, J.
                        Luis",
      title          = "{Bound States of the q-Deformed AdS$_5$ $\times$ S$^5$ Superstring
                        S-matrix}",
      journal        = "JHEP",
      volume         = "1210",
      pages          = "076",
      doi            = "10.1007/JHEP10(2012)076",
      year           = "2012",
      eprint         = "1206.0010",
      archivePrefix  = "arXiv",
      primaryClass   = "hep-th",
      reportNumber   = "IMPERIAL-TP-12-BH-01",
      SLACcitation   = "%%CITATION = ARXIV:1206.0010;%%",
}

@article{HTPohl,
      author         = "Hoare, B. and Tseytlin, A. A.",
      title          = "{Towards the quantum S-matrix of the Pohlmeyer reduced
                        version of AdS$_5$ $\times$ S$^5$ superstring theory}",
      journal        = "Nucl.Phys.",
      volume         = "B851",
      pages          = "161-237",
      doi            = "10.1016/j.nuclphysb.2011.05.016",
      year           = "2011",
      eprint         = "1104.2423",
      archivePrefix  = "arXiv",
      primaryClass   = "hep-th",
      reportNumber   = "IMPERIAL-TP-BH-2011-01",
      SLACcitation   = "%%CITATION = ARXIV:1104.2423;%%",
}

@article{HHMrel,
      author         = "Hoare, Ben and Hollowood, Timothy J. and Miramontes, J. Luis",
      title          = "{A Relativistic Relative of the Magnon S-Matrix}",
      journal        = "JHEP",
      volume         = "1111",
      pages          = "048",
      doi            = "10.1007/JHEP11(2011)048",
      year           = "2011",
      eprint         = "1107.0628",
      archivePrefix  = "arXiv",
      primaryClass   = "hep-th",
      reportNumber   = "IMPERIAL-TP-11-BH-02",
      SLACcitation   = "%%CITATION = ARXIV:1107.0628;%%",
}

@article{BAna,
      author         = "Beisert, Niklas",
      title          = "{The Analytic Bethe Ansatz for a Chain with Centrally
                        Extended su(2$/$2) Symmetry}",
      journal        = "J.Stat.Mech.",
      volume         = "0701",
      pages          = "P01017",
      doi            = "10.1088/1742-5468/2007/01/P01017",
      year           = "2007",
      eprint         = "nlin/0610017",
      archivePrefix  = "arXiv",
      primaryClass   = "nlin.SI",
      reportNumber   = "AEI-2006-074, PUTP-2211",
      SLACcitation   = "%%CITATION = NLIN/0610017;%%",
}

@article{Tdouble,
      author         = "Tjin, T.",
      title          = "{An Introduction to quantized Lie groups and algebras}",
      journal        = "Int.J.Mod.Phys.",
      volume         = "A7",
      pages          = "6175-6213",
      doi            = "10.1142/S0217751X92002805",
      year           = "1992",
      eprint         = "hep-th/9111043",
      archivePrefix  = "arXiv",
      primaryClass   = "hep-th",
      reportNumber   = "PRINT-91-0499 (AMSTERDAM)",
      SLACcitation   = "%%CITATION = HEP-TH/9111043;%%",
}

@article{Bsl3double,
      author         = "Burroughs, Nigel",
      title          = "{The Universal R Matrix for U$_q$ Sl(3) and Beyond!}",
      journal        = "Commun.Math.Phys.",
      volume         = "127",
      pages          = "109",
      doi            = "10.1007/BF02096496",
      year           = "1990",
      reportNumber   = "DAMTP/R-89/4",
      SLACcitation   = "%%CITATION = CMPHA,127,109;%%",
}

@article{DQauntumGroups,
      author         = "Drinfeld, V. G.",
      title          = "{Quantum groups}",
      journal        = "J.Sov.Math.",
      volume         = "41",
      pages          = "898-915",
      doi            = "10.1007/BF01247086",
      year           = "1988",
      SLACcitation   = "%%CITATION = JOSMA,41,898;%%",
}

@article{DMVetaDef,
      author         = "Delduc, Francois and Magro, Marc and Vicedo, Benoit",
      title          = "{An integrable deformation of the AdS$_5$ $\times$ S$^5$
                        superstring action}",
      journal        = "Phys.Rev.Lett.",
      number         = "5",
      volume         = "112",
      pages          = "051601",
      doi            = "10.1103/PhysRevLett.112.051601",
      year           = "2014",
      eprint         = "1309.5850",
      archivePrefix  = "arXiv",
      primaryClass   = "hep-th",
      SLACcitation   = "%%CITATION = ARXIV:1309.5850;%%",
}

@article{ABFqSmat,
      author         = "Arutyunov, Gleb and Borsato, Riccardo and Frolov, Sergey",
      title          = "{S-matrix for strings on $\eta$-deformed AdS$_5$ $\times$ S$^5$}",
      journal        = "JHEP",
      volume         = "1404",
      pages          = "002",
      doi            = "10.1007/JHEP04(2014)002",
      year           = "2014",
      eprint         = "1312.3542",
      archivePrefix  = "arXiv",
      primaryClass   = "hep-th",
      reportNumber   = "ITP-UU-13-31, SPIN-13-23, HU-MATHEMATIK-2013-24,
                        TCD-MATH-13-16",
      SLACcitation   = "%%CITATION = ARXIV:1312.3542;%%",
}

@article{RTwist,
year={1990},
issn={0377-9017},
journal={Lett. Math. Phys.},
volume={20},
number={4},
doi={10.1007/BF00626530},
title={Multiparameter quantum groups and twisted quasitriangular Hopf algebras},
publisher={Kluwer Academic Publishers},
keywords={16A24},
author={Reshetikhin, N.},
pages={331-335},
language={English}
}

@article{BdLHubbAlgebra,
      author         = "Beisert, Niklas and de Leeuw, Marius",
      title          = "{The RTT realization for the deformed {gl}(2$/$2) Yangian}",
      journal        = "J.Phys.",
      volume         = "A47",
      pages          = "305201",
      doi            = "10.1088/1751-8113/47/30/305201",
      year           = "2014",
      eprint         = "1401.7691",
      archivePrefix  = "arXiv",
      primaryClass   = "math-ph",
      SLACcitation   = "%%CITATION = ARXIV:1401.7691;%%",
}

@article{Hubbard,
  author = {Hubbard, J.},
  title = {Electron Correlations in Narrow Energy Bands},
  volume = {A276},
  number = {1365},
  pages = {238-257},
  year = {1963},
  doi = {10.1098/rspa.1963.0204},
  journal = {Proc. R. Soc. Lond.}
}

@article{Shastry,
  year={1988},
  issn={0022-4715},
  journal={J. Stat. Phys.},
  volume={50},
  number={1-2},
  doi={10.1007/BF01022987},
  title={Decorated star-triangle relations and exact integrability of the one-dimensional Hubbard model},
  publisher={Kluwer Academic Publishers-Plenum Publishers},
  keywords={One-dimensional Hubbard model; exactly integrable systems; star-triangle relations},
  author={Shastry, B. Sriram},
  pages={57-79},
  language={English}
}

@Book{HubbBook,
  author =  {Fabian H. L. Essler and Holger Frahm and
                  Frank G{\"o}hmann and Andreas Kl{\"u}mper and
    Vladimir E. Korepin},
  title =   {The One-Dimensional {H}ubbard Model},
  publisher =   {Cambridge University Press},
  year =   {2005},
  address =  {Cambridge, UK},
  doi =          {10.2277/0521802628}
}

@Book{MBook,
  author =  {Shahn Majid},
  title =   {Foundations of quantum group theory},
  publisher =   {Cambridge University Press},
  year =   {1995},
  address =  {Cambridge, UK}
}

@article{Martins:2007hb,
      author         = "Martins, M. J. and Melo, C. S.",
      title          = "{The Bethe ansatz approach for factorizable centrally
                        extended S-matrices}",
      journal        = "Nucl.Phys.",
      volume         = "B785",
      pages          = "246-262",
      doi            = "10.1016/j.nuclphysb.2007.05.021",
      year           = "2007",
      eprint         = "hep-th/0703086",
      archivePrefix  = "arXiv",
      primaryClass   = "hep-th",
      reportNumber   = "UFSCAR-TH-07-03",
      SLACcitation   = "%%CITATION = HEP-TH/0703086;%%",
}

@article{Uglov:1993jy,
      author         = "Uglov, D. B. and Korepin, V. E.",
      title          = "{The Yangian symmetry of the Hubbard model}",
      journal        = "Phys.Lett.",
      volume         = "A190",
      pages          = "238-242",
      doi            = "10.1016/0375-9601(94)90748-X",
      year           = "1994",
      eprint         = "hep-th/9310158",
      archivePrefix  = "arXiv",
      primaryClass   = "hep-th",
      reportNumber   = "ITP-SB-93-66",
      SLACcitation   = "%%CITATION = HEP-TH/9310158;%%",
}

@article{AFP,
      author         = "Arutyunov, Gleb and Frolov, Sergey and Plefka, Jan and
                        Zamaklar, Marija",
      title          = "{The Off-shell Symmetry Algebra of the Light-cone AdS$_5$ $\times$ S$^5$ Superstring}",
      journal        = "J.Phys.",
      volume         = "A40",
      pages          = "3583-3606",
      doi            = "10.1088/1751-8113/40/13/018",
      year           = "2007",
      eprint         = "hep-th/0609157",
      archivePrefix  = "arXiv",
      primaryClass   = "hep-th",
      reportNumber   = "AEI-2006-071, HU-EP-06-31, ITP-UU-06-39, SPIN-06-33,
                        TCDMATH-06-13",
      SLACcitation   = "%%CITATION = HEP-TH/0609157;%%",
}

@article{Bsu22,
      author         = "Beisert, Niklas",
      title          = "{The SU(2$/$2) dynamic S-matrix}",
      journal        = "Adv.Theor.Math.Phys.",
      volume         = "12",
      pages          = "945-979",
      doi            = "10.4310/ATMP.2008.v12.n5.a1",
      year           = "2008",
      eprint         = "hep-th/0511082",
      archivePrefix  = "arXiv",
      primaryClass   = "hep-th",
      reportNumber   = "PUTP-2181, NSF-KITP-05-92",
      SLACcitation   = "%%CITATION = HEP-TH/0511082;%%",
}

@article{BYang,
      author         = "Beisert, Niklas",
      title          = "{The S-matrix of AdS / CFT and Yangian symmetry}",
      journal        = "PoS",
      volume         = "SOLVAY",
      pages          = "002",
      year           = "2006",
      eprint         = "0704.0400",
      archivePrefix  = "arXiv",
      primaryClass   = "nlin.SI",
      reportNumber   = "AEI-2007-019",
      SLACcitation   = "%%CITATION = ARXIV:0704.0400;%%",
}

@article{dLYang,
      author         = "de Leeuw, Marius",
      title          = "{Bound States, Yangian Symmetry and Classical r-matrix
                        for the AdS$_5$ $\times$ S$^5$ Superstring}",
      journal        = "JHEP",
      volume         = "0806",
      pages          = "085",
      doi            = "10.1088/1126-6708/2008/06/085",
      year           = "2008",
      eprint         = "0804.1047",
      archivePrefix  = "arXiv",
      primaryClass   = "hep-th",
      reportNumber   = "ITP-UU-08-18, SPIN-08-17",
      SLACcitation   = "%%CITATION = ARXIV:0804.1047;%%",
}

@article{TClassical,
      author         = "Torrielli, Alessandro",
      title          = "{Classical r-matrix of the su(2$/$2) SYM spin-chain}",
      journal        = "Phys.Rev.",
      volume         = "D75",
      pages          = "105020",
      doi            = "10.1103/PhysRevD.75.105020",
      year           = "2007",
      eprint         = "hep-th/0701281",
      archivePrefix  = "arXiv",
      primaryClass   = "hep-th",
      reportNumber   = "MIT-CTP-3809",
      SLACcitation   = "%%CITATION = HEP-TH/0701281;%%",
}

@article{MTClassical,
      author         = "Moriyama, Sanefumi and Torrielli, Alessandro",
      title          = "{A Yangian double for the AdS/CFT classical r-matrix}",
      journal        = "JHEP",
      volume         = "0706",
      pages          = "083",
      doi            = "10.1088/1126-6708/2007/06/083",
      year           = "2007",
      eprint         = "0706.0884",
      archivePrefix  = "arXiv",
      primaryClass   = "hep-th",
      reportNumber   = "MIT-CTP-3843",
      SLACcitation   = "%%CITATION = ARXIV:0706.0884;%%",
 }

@article{MdLThesis,
      author         = "de Leeuw, Marius",
      title          = "{The S-matrix of the AdS$_5$ $\times$ S$^5$ superstring}",
      year           = "2010",
      eprint         = "1007.4931",
      archivePrefix  = "arXiv",
      primaryClass   = "hep-th",
      reportNumber   = "ITP-UU-10-13, SPIN-10-11",
      SLACcitation   = "%%CITATION = ARXIV:1007.4931;%%",
note = "PhD thesis",
}

@article{MdLClassicalR,
      author         = "de Leeuw, Marius",
      title          = "{Bound States, Yangian Symmetry and Classical r-matrix
                        for the AdS$_5$ $\times$ S$^5$ Superstring}",
      journal        = "JHEP",
      volume         = "0806",
      pages          = "085",
      doi            = "10.1088/1126-6708/2008/06/085",
      year           = "2008",
      eprint         = "0804.1047",
      archivePrefix  = "arXiv",
      primaryClass   = "hep-th",
      reportNumber   = "ITP-UU-08-18, SPIN-08-17",
      SLACcitation   = "%%CITATION = ARXIV:0804.1047;%%",
}

@article{AB,
  author={F. C. Alcaraz and R. Z. Bariev},
  title={Interpolation between Hubbard and supersymmetric t - J models: two-parameter integrable models of correlated electrons},
  journal={J. Phys.},
  volume={A32},
  number={46},
  pages={L483},
  year={1999},
  eprint = {cond-mat/9908265},
  doi = {10.1088/0305-4470/32/46/101}
}

@article{deLeeuw:2011fr,
      author         = "de Leeuw, Marius and Regelskis, Vidas and Torrielli,
                        Alessandro",
      title          = "{The Quantum Affine Origin of the AdS/CFT Secret
                        Symmetry}",
      journal        = "J. Phys.",
      volume         = "A45",
      year           = "2012",
      pages          = "175202",
      doi            = "10.1088/1751-8113/45/17/175202",
      eprint         = "1112.4989",
      archivePrefix  = "arXiv",
      primaryClass   = "hep-th",
      reportNumber   = "DMUS-MP-11-02",
      SLACcitation   = "%%CITATION = ARXIV:1112.4989;%%"
}

@book{QExp,
    author = {Pittner, Ludwig},
    isbn = {3540605878},
    keywords = {ncg, quantum\_groups},
    publisher = {Springer},
    title = {{Algebraic Foundations of Non-Commutative Differential Geometry and Quantum Groups}},
}

@article{LeClair:1991cf,
      author         = "LeClair, Andre and Smirnov, F. A.",
      title          = "{Infinite quantum group symmetry of fields in massive 2-D
                        quantum field theory}",
      journal        = "Int. J. Mod. Phys.",
      volume         = "A7",
      year           = "1992",
      pages          = "2997-3022",
      doi            = "10.1142/S0217751X92001332",
      eprint         = "hep-th/9108007",
      archivePrefix  = "arXiv",
      primaryClass   = "hep-th",
      reportNumber   = "CLNS-91-1056",
      SLACcitation   = "%%CITATION = HEP-TH/9108007;%%"
}

@article{Beisert:2006ez,
      author         = "Beisert, Niklas and Eden, Burkhard and Staudacher,
                        Matthias",
      title          = "{Transcendentality and Crossing}",
      journal        = "J. Stat. Mech.",
      volume         = "0701",
      year           = "2007",
      pages          = "P01021",
      doi            = "10.1088/1742-5468/2007/01/P01021",
      eprint         = "hep-th/0610251",
      archivePrefix  = "arXiv",
      primaryClass   = "hep-th",
      reportNumber   = "AEI-2006-079, ITP-UU-06-44, SPIN-06-34",
      SLACcitation   = "%%CITATION = HEP-TH/0610251;%%"
}

@article{BdLH2,
      author         = "Beisert, Niklas and de Leeuw, Marius and Hecht, Reimar",
      note           = "work in progress",
}
\end{bibtex}

\bibliographystyle{nb}
\bibliography{\jobname}

\end{document}